\documentclass[longauth]{aa}  
\usepackage{amsmath,amssymb,amsfonts}
\usepackage{graphicx}
\usepackage{txfonts}
\usepackage{enumerate} %
\usepackage{colordvi} %
\usepackage{bm}%
\usepackage{epsfig}
\usepackage{hyperref}
\usepackage[normalem]{ulem}

\usepackage[dvipsnames]{xcolor}
\usepackage{color, colortbl}
\usepackage[T1]{fontenc} %
\usepackage{braket}
\RequirePackage{amsmath,amssymb}
\RequirePackage{siunitx}
\RequirePackage{xspace}
\RequirePackage{euclid}
\usepackage{placeins}
\usepackage{tabularx}
\usepackage{csvsimple}
\usepackage[switch]{lineno}
\setcitestyle{citesep={,}}

\newcommand{\bs}[1]{\boldsymbol{#1}}
\newcommand{\dif}{\mathrm{d}}
\newcommand{\mc}[1]{\mathcal{#1}}
\newcommand{\cH}{\mathcal{H}}

\newcommand{\tj}[6]{ \begin{pmatrix}
  #1 & #2 & #3 \\
  #4 & #5 & #6 
\end{pmatrix}^2}
\definecolor{amaranth}{rgb}{0.9, 0.17, 0.31}
\definecolor{forestgreen(web)}{rgb}{0.13, 0.55, 0.13}
\definecolor{lavender(web)}{rgb}{0.6, 0.8, 0.75}
\definecolor{cosmiclatte}{rgb}{1.0, 0.97, 0.91}
\definecolor{jonquil}{rgb}{0.98, 0.85, 0.37}
\definecolor{khaki(x11)(lightkhaki)}{rgb}{0.94, 0.9, 0.55}
\definecolor{thistle}{rgb}{0.85, 0.75, 0.85}

\newcommand{\liger}{\texttt{LIGER} }
\newcommand{\bvr}[1]{\bs{\varv}_{\mathrm{#1}} }

\newcommand{\overc}[1]{{#1\over c}}

\newcommand{\real}{$\mc{R}\,$}
\newcommand{\vrsd}{$\mc{V}\,$}
\newcommand{\grsd}{$\mc{G}\,$}
\newcommand{\obs}{$\mc{O}\,$}

\newcommand{\obsm}{\mc{O}}
\newcommand{\vobstest}{$\mc{O}$-$\mc{G}$ }
\newcommand{\WLtest}{$\mc{O}$-$\mc{V}$ }
\newcommand{\snrtval}[1]{$\text{S/N} = #1$}
\newcommand{\snrt}{$\text{S/N}$\xspace}

\newcommand{\ttt}[1]{\texttt{#1}}
\newcommand{\bshat}[1]{\hat{\bs{#1}}}
\newcommand{\velpot}{\Theta}
\newcommand{\euclidtwopcf}{(Euclid Collaboration: De la Torre et al. in prep.)}
\newcommand{\EuclidAngSystematics}{Euclid Collaboration: Monaco et al. in prep.}
\newcommand{\euclidPower}{(Euclid Collaboration: Salvalaggio et al. in prep.)}

\newcommand{\elkhashabalpha}{(Elkhashab et al. in prep.)}

\usepackage[nameinlink,noabbrev]{cleveref}
\Crefname{equation}{Eq.}{Eqs}
\Crefname{section}{Sect.}{Sects}
\Crefname{figure}{Fig.}{Figs}
\crefname{equation}{Equation}{Equations}
\crefname{section}{Section}{Sections}
\crefname{figure}{Figure}{Figures}

\begin{document}

\title{\Euclid preparation}
\subtitle{The impact of relativistic redshift-space distortions on two-point clustering statistics from the Euclid wide spectroscopic survey}    

\newcommand{\orcid}[1]{} %
\author{Euclid Collaboration: M.~Y.~Elkhashab\orcid{0000-0001-9306-2603}\thanks{\email{mohamed.elkhashab@inaf.it}}\inst{\ref{aff1},\ref{aff2}}
\and D.~Bertacca\orcid{0000-0002-2490-7139}\inst{\ref{aff2},\ref{aff3},\ref{aff1}}
\and C.~Porciani\orcid{0000-0002-7797-2508}\inst{\ref{aff4}}
\and J.~Salvalaggio\orcid{0000-0002-1431-5607}\inst{\ref{aff5},\ref{aff6},\ref{aff7},\ref{aff8}}
\and N.~Aghanim\orcid{0000-0002-6688-8992}\inst{\ref{aff9}}
\and A.~Amara\inst{\ref{aff10}}
\and S.~Andreon\orcid{0000-0002-2041-8784}\inst{\ref{aff11}}
\and N.~Auricchio\orcid{0000-0003-4444-8651}\inst{\ref{aff12}}
\and C.~Baccigalupi\orcid{0000-0002-8211-1630}\inst{\ref{aff7},\ref{aff6},\ref{aff8},\ref{aff13}}
\and M.~Baldi\orcid{0000-0003-4145-1943}\inst{\ref{aff14},\ref{aff12},\ref{aff15}}
\and S.~Bardelli\orcid{0000-0002-8900-0298}\inst{\ref{aff12}}
\and C.~Bodendorf\inst{\ref{aff16}}
\and D.~Bonino\orcid{0000-0002-3336-9977}\inst{\ref{aff17}}
\and E.~Branchini\orcid{0000-0002-0808-6908}\inst{\ref{aff18},\ref{aff19},\ref{aff11}}
\and M.~Brescia\orcid{0000-0001-9506-5680}\inst{\ref{aff20},\ref{aff21},\ref{aff22}}
\and J.~Brinchmann\orcid{0000-0003-4359-8797}\inst{\ref{aff23}}
\and S.~Camera\orcid{0000-0003-3399-3574}\inst{\ref{aff24},\ref{aff25},\ref{aff17}}
\and V.~Capobianco\orcid{0000-0002-3309-7692}\inst{\ref{aff17}}
\and C.~Carbone\orcid{0000-0003-0125-3563}\inst{\ref{aff26}}
\and V.~F.~Cardone\inst{\ref{aff27},\ref{aff28}}
\and J.~Carretero\orcid{0000-0002-3130-0204}\inst{\ref{aff29},\ref{aff30}}
\and R.~Casas\orcid{0000-0002-8165-5601}\inst{\ref{aff31},\ref{aff32}}
\and S.~Casas\orcid{0000-0002-4751-5138}\inst{\ref{aff33}}
\and M.~Castellano\orcid{0000-0001-9875-8263}\inst{\ref{aff27}}
\and G.~Castignani\orcid{0000-0001-6831-0687}\inst{\ref{aff12}}
\and S.~Cavuoti\orcid{0000-0002-3787-4196}\inst{\ref{aff21},\ref{aff22}}
\and A.~Cimatti\inst{\ref{aff34}}
\and C.~Colodro-Conde\inst{\ref{aff35}}
\and G.~Congedo\orcid{0000-0003-2508-0046}\inst{\ref{aff36}}
\and C.~J.~Conselice\orcid{0000-0003-1949-7638}\inst{\ref{aff37}}
\and L.~Conversi\orcid{0000-0002-6710-8476}\inst{\ref{aff38},\ref{aff39}}
\and Y.~Copin\orcid{0000-0002-5317-7518}\inst{\ref{aff40}}
\and F.~Courbin\orcid{0000-0003-0758-6510}\inst{\ref{aff41}}
\and H.~M.~Courtois\orcid{0000-0003-0509-1776}\inst{\ref{aff42}}
\and A.~Da~Silva\orcid{0000-0002-6385-1609}\inst{\ref{aff43},\ref{aff44}}
\and H.~Degaudenzi\orcid{0000-0002-5887-6799}\inst{\ref{aff45}}
\and A.~M.~Di~Giorgio\orcid{0000-0002-4767-2360}\inst{\ref{aff46}}
\and J.~Dinis\orcid{0000-0001-5075-1601}\inst{\ref{aff43},\ref{aff44}}
\and M.~Douspis\orcid{0000-0003-4203-3954}\inst{\ref{aff9}}
\and F.~Dubath\orcid{0000-0002-6533-2810}\inst{\ref{aff45}}
\and C.~A.~J.~Duncan\inst{\ref{aff37}}
\and X.~Dupac\inst{\ref{aff39}}
\and S.~Dusini\orcid{0000-0002-1128-0664}\inst{\ref{aff1}}
\and M.~Farina\orcid{0000-0002-3089-7846}\inst{\ref{aff46}}
\and S.~Farrens\orcid{0000-0002-9594-9387}\inst{\ref{aff47}}
\and S.~Ferriol\inst{\ref{aff40}}
\and P.~Fosalba\orcid{0000-0002-1510-5214}\inst{\ref{aff31},\ref{aff48}}
\and M.~Frailis\orcid{0000-0002-7400-2135}\inst{\ref{aff6}}
\and E.~Franceschi\orcid{0000-0002-0585-6591}\inst{\ref{aff12}}
\and S.~Galeotta\orcid{0000-0002-3748-5115}\inst{\ref{aff6}}
\and B.~Gillis\orcid{0000-0002-4478-1270}\inst{\ref{aff36}}
\and C.~Giocoli\orcid{0000-0002-9590-7961}\inst{\ref{aff12},\ref{aff49}}
\and P.~G\'omez-Alvarez\orcid{0000-0002-8594-5358}\inst{\ref{aff50},\ref{aff39}}
\and A.~Grazian\orcid{0000-0002-5688-0663}\inst{\ref{aff3}}
\and F.~Grupp\inst{\ref{aff16},\ref{aff51}}
\and L.~Guzzo\orcid{0000-0001-8264-5192}\inst{\ref{aff52},\ref{aff11}}
\and S.~V.~H.~Haugan\orcid{0000-0001-9648-7260}\inst{\ref{aff53}}
\and W.~Holmes\inst{\ref{aff54}}
\and F.~Hormuth\inst{\ref{aff55}}
\and A.~Hornstrup\orcid{0000-0002-3363-0936}\inst{\ref{aff56},\ref{aff57}}
\and K.~Jahnke\orcid{0000-0003-3804-2137}\inst{\ref{aff58}}
\and M.~Jhabvala\inst{\ref{aff59}}
\and B.~Joachimi\orcid{0000-0001-7494-1303}\inst{\ref{aff60}}
\and E.~Keih\"anen\orcid{0000-0003-1804-7715}\inst{\ref{aff61}}
\and S.~Kermiche\orcid{0000-0002-0302-5735}\inst{\ref{aff62}}
\and A.~Kiessling\orcid{0000-0002-2590-1273}\inst{\ref{aff54}}
\and M.~Kilbinger\orcid{0000-0001-9513-7138}\inst{\ref{aff47}}
\and T.~Kitching\orcid{0000-0002-4061-4598}\inst{\ref{aff63}}
\and B.~Kubik\orcid{0009-0006-5823-4880}\inst{\ref{aff40}}
\and K.~Kuijken\orcid{0000-0002-3827-0175}\inst{\ref{aff64}}
\and M.~K\"ummel\orcid{0000-0003-2791-2117}\inst{\ref{aff51}}
\and M.~Kunz\orcid{0000-0002-3052-7394}\inst{\ref{aff65}}
\and H.~Kurki-Suonio\orcid{0000-0002-4618-3063}\inst{\ref{aff66},\ref{aff67}}
\and S.~Ligori\orcid{0000-0003-4172-4606}\inst{\ref{aff17}}
\and P.~B.~Lilje\orcid{0000-0003-4324-7794}\inst{\ref{aff53}}
\and V.~Lindholm\orcid{0000-0003-2317-5471}\inst{\ref{aff66},\ref{aff67}}
\and I.~Lloro\inst{\ref{aff68}}
\and G.~Mainetti\orcid{0000-0003-2384-2377}\inst{\ref{aff69}}
\and E.~Maiorano\orcid{0000-0003-2593-4355}\inst{\ref{aff12}}
\and O.~Mansutti\orcid{0000-0001-5758-4658}\inst{\ref{aff6}}
\and O.~Marggraf\orcid{0000-0001-7242-3852}\inst{\ref{aff4}}
\and K.~Markovic\orcid{0000-0001-6764-073X}\inst{\ref{aff54}}
\and N.~Martinet\orcid{0000-0003-2786-7790}\inst{\ref{aff70}}
\and F.~Marulli\orcid{0000-0002-8850-0303}\inst{\ref{aff71},\ref{aff12},\ref{aff15}}
\and R.~Massey\orcid{0000-0002-6085-3780}\inst{\ref{aff72}}
\and E.~Medinaceli\orcid{0000-0002-4040-7783}\inst{\ref{aff12}}
\and S.~Mei\orcid{0000-0002-2849-559X}\inst{\ref{aff73}}
\and Y.~Mellier\inst{\ref{aff74},\ref{aff75}}
\and M.~Meneghetti\orcid{0000-0003-1225-7084}\inst{\ref{aff12},\ref{aff15}}
\and G.~Meylan\inst{\ref{aff41}}
\and M.~Moresco\orcid{0000-0002-7616-7136}\inst{\ref{aff71},\ref{aff12}}
\and L.~Moscardini\orcid{0000-0002-3473-6716}\inst{\ref{aff71},\ref{aff12},\ref{aff15}}
\and S.-M.~Niemi\inst{\ref{aff76}}
\and C.~Padilla\orcid{0000-0001-7951-0166}\inst{\ref{aff77}}
\and S.~Paltani\orcid{0000-0002-8108-9179}\inst{\ref{aff45}}
\and F.~Pasian\orcid{0000-0002-4869-3227}\inst{\ref{aff6}}
\and K.~Pedersen\inst{\ref{aff78}}
\and V.~Pettorino\inst{\ref{aff76}}
\and S.~Pires\orcid{0000-0002-0249-2104}\inst{\ref{aff47}}
\and G.~Polenta\orcid{0000-0003-4067-9196}\inst{\ref{aff79}}
\and M.~Poncet\inst{\ref{aff80}}
\and L.~A.~Popa\inst{\ref{aff81}}
\and L.~Pozzetti\orcid{0000-0001-7085-0412}\inst{\ref{aff12}}
\and F.~Raison\orcid{0000-0002-7819-6918}\inst{\ref{aff16}}
\and R.~Rebolo\inst{\ref{aff35},\ref{aff82}}
\and A.~Renzi\orcid{0000-0001-9856-1970}\inst{\ref{aff2},\ref{aff1}}
\and J.~Rhodes\orcid{0000-0002-4485-8549}\inst{\ref{aff54}}
\and G.~Riccio\inst{\ref{aff21}}
\and E.~Romelli\orcid{0000-0003-3069-9222}\inst{\ref{aff6}}
\and M.~Roncarelli\orcid{0000-0001-9587-7822}\inst{\ref{aff12}}
\and R.~Saglia\orcid{0000-0003-0378-7032}\inst{\ref{aff51},\ref{aff16}}
\and Z.~Sakr\orcid{0000-0002-4823-3757}\inst{\ref{aff83},\ref{aff84},\ref{aff85}}
\and A.~G.~S\'anchez\orcid{0000-0003-1198-831X}\inst{\ref{aff16}}
\and D.~Sapone\orcid{0000-0001-7089-4503}\inst{\ref{aff86}}
\and M.~Schirmer\orcid{0000-0003-2568-9994}\inst{\ref{aff58}}
\and P.~Schneider\orcid{0000-0001-8561-2679}\inst{\ref{aff4}}
\and T.~Schrabback\orcid{0000-0002-6987-7834}\inst{\ref{aff87}}
\and M.~Scodeggio\inst{\ref{aff26}}
\and A.~Secroun\orcid{0000-0003-0505-3710}\inst{\ref{aff62}}
\and E.~Sefusatti\orcid{0000-0003-0473-1567}\inst{\ref{aff6},\ref{aff7},\ref{aff8}}
\and G.~Seidel\orcid{0000-0003-2907-353X}\inst{\ref{aff58}}
\and S.~Serrano\orcid{0000-0002-0211-2861}\inst{\ref{aff31},\ref{aff88},\ref{aff32}}
\and C.~Sirignano\orcid{0000-0002-0995-7146}\inst{\ref{aff2},\ref{aff1}}
\and G.~Sirri\orcid{0000-0003-2626-2853}\inst{\ref{aff15}}
\and L.~Stanco\orcid{0000-0002-9706-5104}\inst{\ref{aff1}}
\and J.~Steinwagner\inst{\ref{aff16}}
\and C.~Surace\orcid{0000-0003-2592-0113}\inst{\ref{aff70}}
\and P.~Tallada-Cresp\'{i}\orcid{0000-0002-1336-8328}\inst{\ref{aff29},\ref{aff30}}
\and A.~N.~Taylor\inst{\ref{aff36}}
\and I.~Tereno\inst{\ref{aff43},\ref{aff89}}
\and R.~Toledo-Moreo\orcid{0000-0002-2997-4859}\inst{\ref{aff90}}
\and F.~Torradeflot\orcid{0000-0003-1160-1517}\inst{\ref{aff30},\ref{aff29}}
\and I.~Tutusaus\orcid{0000-0002-3199-0399}\inst{\ref{aff84}}
\and L.~Valenziano\orcid{0000-0002-1170-0104}\inst{\ref{aff12},\ref{aff91}}
\and T.~Vassallo\orcid{0000-0001-6512-6358}\inst{\ref{aff51},\ref{aff6}}
\and G.~Verdoes~Kleijn\orcid{0000-0001-5803-2580}\inst{\ref{aff92}}
\and A.~Veropalumbo\orcid{0000-0003-2387-1194}\inst{\ref{aff11},\ref{aff19},\ref{aff93}}
\and Y.~Wang\orcid{0000-0002-4749-2984}\inst{\ref{aff94}}
\and J.~Weller\orcid{0000-0002-8282-2010}\inst{\ref{aff51},\ref{aff16}}
\and G.~Zamorani\orcid{0000-0002-2318-301X}\inst{\ref{aff12}}
\and E.~Zucca\orcid{0000-0002-5845-8132}\inst{\ref{aff12}}
\and A.~Biviano\orcid{0000-0002-0857-0732}\inst{\ref{aff6},\ref{aff7}}
\and A.~Boucaud\orcid{0000-0001-7387-2633}\inst{\ref{aff73}}
\and E.~Bozzo\orcid{0000-0002-8201-1525}\inst{\ref{aff45}}
\and C.~Burigana\orcid{0000-0002-3005-5796}\inst{\ref{aff95},\ref{aff91}}
\and M.~Calabrese\orcid{0000-0002-2637-2422}\inst{\ref{aff96},\ref{aff26}}
\and D.~Di~Ferdinando\inst{\ref{aff15}}
\and J.~A.~Escartin~Vigo\inst{\ref{aff16}}
\and R.~Farinelli\inst{\ref{aff12}}
\and F.~Finelli\orcid{0000-0002-6694-3269}\inst{\ref{aff12},\ref{aff91}}
\and J.~Gracia-Carpio\inst{\ref{aff16}}
\and N.~Mauri\orcid{0000-0001-8196-1548}\inst{\ref{aff34},\ref{aff15}}
\and A.~Pezzotta\orcid{0000-0003-0726-2268}\inst{\ref{aff16}}
\and M.~P\"ontinen\orcid{0000-0001-5442-2530}\inst{\ref{aff66}}
\and V.~Scottez\inst{\ref{aff74},\ref{aff97}}
\and M.~Tenti\orcid{0000-0002-4254-5901}\inst{\ref{aff15}}
\and M.~Viel\orcid{0000-0002-2642-5707}\inst{\ref{aff7},\ref{aff6},\ref{aff13},\ref{aff8},\ref{aff98}}
\and M.~Wiesmann\orcid{0009-0000-8199-5860}\inst{\ref{aff53}}
\and Y.~Akrami\orcid{0000-0002-2407-7956}\inst{\ref{aff99},\ref{aff100}}
\and V.~Allevato\orcid{0000-0001-7232-5152}\inst{\ref{aff21}}
\and S.~Anselmi\orcid{0000-0002-3579-9583}\inst{\ref{aff1},\ref{aff2},\ref{aff101}}
\and A.~Balaguera-Antolinez\orcid{0000-0001-5028-3035}\inst{\ref{aff35},\ref{aff82}}
\and M.~Ballardini\orcid{0000-0003-4481-3559}\inst{\ref{aff102},\ref{aff12},\ref{aff103}}
\and A.~Blanchard\orcid{0000-0001-8555-9003}\inst{\ref{aff84}}
\and L.~Blot\orcid{0000-0002-9622-7167}\inst{\ref{aff104},\ref{aff101}}
\and H.~B\"ohringer\orcid{0000-0001-8241-4204}\inst{\ref{aff16},\ref{aff105},\ref{aff106}}
\and S.~Borgani\orcid{0000-0001-6151-6439}\inst{\ref{aff5},\ref{aff7},\ref{aff6},\ref{aff8}}
\and S.~Bruton\orcid{0000-0002-6503-5218}\inst{\ref{aff107}}
\and R.~Cabanac\orcid{0000-0001-6679-2600}\inst{\ref{aff84}}
\and A.~Calabro\orcid{0000-0003-2536-1614}\inst{\ref{aff27}}
\and G.~Canas-Herrera\orcid{0000-0003-2796-2149}\inst{\ref{aff76},\ref{aff108}}
\and A.~Cappi\inst{\ref{aff12},\ref{aff109}}
\and C.~S.~Carvalho\inst{\ref{aff89}}
\and T.~Castro\orcid{0000-0002-6292-3228}\inst{\ref{aff6},\ref{aff8},\ref{aff7},\ref{aff98}}
\and K.~C.~Chambers\orcid{0000-0001-6965-7789}\inst{\ref{aff110}}
\and A.~R.~Cooray\orcid{0000-0002-3892-0190}\inst{\ref{aff111}}
\and S.~Davini\orcid{0000-0003-3269-1718}\inst{\ref{aff19}}
\and B.~De~Caro\inst{\ref{aff26}}
\and S.~de~la~Torre\inst{\ref{aff70}}
\and G.~Desprez\inst{\ref{aff112}}
\and A.~D\'iaz-S\'anchez\orcid{0000-0003-0748-4768}\inst{\ref{aff113}}
\and J.~J.~Diaz\inst{\ref{aff114}}
\and S.~Di~Domizio\orcid{0000-0003-2863-5895}\inst{\ref{aff18},\ref{aff19}}
\and H.~Dole\orcid{0000-0002-9767-3839}\inst{\ref{aff9}}
\and S.~Escoffier\orcid{0000-0002-2847-7498}\inst{\ref{aff62}}
\and A.~G.~Ferrari\orcid{0009-0005-5266-4110}\inst{\ref{aff34},\ref{aff15}}
\and P.~G.~Ferreira\orcid{0000-0002-3021-2851}\inst{\ref{aff115}}
\and I.~Ferrero\orcid{0000-0002-1295-1132}\inst{\ref{aff53}}
\and A.~Finoguenov\orcid{0000-0002-4606-5403}\inst{\ref{aff66}}
\and A.~Fontana\orcid{0000-0003-3820-2823}\inst{\ref{aff27}}
\and F.~Fornari\orcid{0000-0003-2979-6738}\inst{\ref{aff91}}
\and L.~Gabarra\orcid{0000-0002-8486-8856}\inst{\ref{aff115}}
\and K.~Ganga\orcid{0000-0001-8159-8208}\inst{\ref{aff73}}
\and J.~Garc\'ia-Bellido\orcid{0000-0002-9370-8360}\inst{\ref{aff99}}
\and E.~Gaztanaga\orcid{0000-0001-9632-0815}\inst{\ref{aff32},\ref{aff31},\ref{aff116}}
\and F.~Giacomini\orcid{0000-0002-3129-2814}\inst{\ref{aff15}}
\and F.~Gianotti\orcid{0000-0003-4666-119X}\inst{\ref{aff12}}
\and G.~Gozaliasl\orcid{0000-0002-0236-919X}\inst{\ref{aff117},\ref{aff66}}
\and A.~Hall\orcid{0000-0002-3139-8651}\inst{\ref{aff36}}
\and W.~G.~Hartley\inst{\ref{aff45}}
\and H.~Hildebrandt\orcid{0000-0002-9814-3338}\inst{\ref{aff118}}
\and J.~Hjorth\orcid{0000-0002-4571-2306}\inst{\ref{aff119}}
\and A.~Jimenez~Mu\~noz\orcid{0009-0004-5252-185X}\inst{\ref{aff120}}
\and J.~J.~E.~Kajava\orcid{0000-0002-3010-8333}\inst{\ref{aff121},\ref{aff122}}
\and V.~Kansal\orcid{0000-0002-4008-6078}\inst{\ref{aff123},\ref{aff124}}
\and D.~Karagiannis\orcid{0000-0002-4927-0816}\inst{\ref{aff125},\ref{aff126}}
\and C.~C.~Kirkpatrick\inst{\ref{aff61}}
\and F.~Lacasa\orcid{0000-0002-7268-3440}\inst{\ref{aff127},\ref{aff9},\ref{aff65}}
\and J.~Le~Graet\orcid{0000-0001-6523-7971}\inst{\ref{aff62}}
\and L.~Legrand\orcid{0000-0003-0610-5252}\inst{\ref{aff128}}
\and A.~Loureiro\orcid{0000-0002-4371-0876}\inst{\ref{aff129},\ref{aff130}}
\and G.~Maggio\orcid{0000-0003-4020-4836}\inst{\ref{aff6}}
\and M.~Magliocchetti\orcid{0000-0001-9158-4838}\inst{\ref{aff46}}
\and F.~Mannucci\orcid{0000-0002-4803-2381}\inst{\ref{aff131}}
\and R.~Maoli\orcid{0000-0002-6065-3025}\inst{\ref{aff132},\ref{aff27}}
\and C.~J.~A.~P.~Martins\orcid{0000-0002-4886-9261}\inst{\ref{aff133},\ref{aff23}}
\and S.~Matthew\orcid{0000-0001-8448-1697}\inst{\ref{aff36}}
\and L.~Maurin\orcid{0000-0002-8406-0857}\inst{\ref{aff9}}
\and R.~B.~Metcalf\orcid{0000-0003-3167-2574}\inst{\ref{aff71},\ref{aff12}}
\and M.~Migliaccio\inst{\ref{aff134},\ref{aff135}}
\and P.~Monaco\orcid{0000-0003-2083-7564}\inst{\ref{aff5},\ref{aff6},\ref{aff8},\ref{aff7}}
\and C.~Moretti\orcid{0000-0003-3314-8936}\inst{\ref{aff13},\ref{aff98},\ref{aff6},\ref{aff7},\ref{aff8}}
\and G.~Morgante\inst{\ref{aff12}}
\and S.~Nadathur\orcid{0000-0001-9070-3102}\inst{\ref{aff116}}
\and Nicholas~A.~Walton\orcid{0000-0003-3983-8778}\inst{\ref{aff136}}
\and L.~Patrizii\inst{\ref{aff15}}
\and V.~Popa\inst{\ref{aff81}}
\and D.~Potter\orcid{0000-0002-0757-5195}\inst{\ref{aff137}}
\and P.~Reimberg\orcid{0000-0003-3410-0280}\inst{\ref{aff74}}
\and I.~Risso\orcid{0000-0003-2525-7761}\inst{\ref{aff93}}
\and P.-F.~Rocci\inst{\ref{aff9}}
\and M.~Sahl\'en\orcid{0000-0003-0973-4804}\inst{\ref{aff138}}
\and A.~Schneider\orcid{0000-0001-7055-8104}\inst{\ref{aff137}}
\and M.~Sereno\orcid{0000-0003-0302-0325}\inst{\ref{aff12},\ref{aff15}}
\and G.~Sikkema\inst{\ref{aff92}}
\and A.~Silvestri\orcid{0000-0001-6904-5061}\inst{\ref{aff108}}
\and P.~Simon\inst{\ref{aff4}}
\and A.~Spurio~Mancini\orcid{0000-0001-5698-0990}\inst{\ref{aff139},\ref{aff63}}
\and K.~Tanidis\inst{\ref{aff115}}
\and C.~Tao\orcid{0000-0001-7961-8177}\inst{\ref{aff62}}
\and N.~Tessore\orcid{0000-0002-9696-7931}\inst{\ref{aff60}}
\and G.~Testera\inst{\ref{aff19}}
\and R.~Teyssier\orcid{0000-0001-7689-0933}\inst{\ref{aff140}}
\and S.~Toft\orcid{0000-0003-3631-7176}\inst{\ref{aff57},\ref{aff141},\ref{aff142}}
\and S.~Tosi\orcid{0000-0002-7275-9193}\inst{\ref{aff18},\ref{aff19}}
\and A.~Troja\orcid{0000-0003-0239-4595}\inst{\ref{aff2},\ref{aff1}}
\and M.~Tucci\inst{\ref{aff45}}
\and C.~Valieri\inst{\ref{aff15}}
\and J.~Valiviita\orcid{0000-0001-6225-3693}\inst{\ref{aff66},\ref{aff67}}
\and D.~Vergani\orcid{0000-0003-0898-2216}\inst{\ref{aff12}}
\and F.~Vernizzi\orcid{0000-0003-3426-2802}\inst{\ref{aff143}}
\and G.~Verza\orcid{0000-0002-1886-8348}\inst{\ref{aff144},\ref{aff145}}
\and P.~Vielzeuf\orcid{0000-0003-2035-9339}\inst{\ref{aff62}}
\and C.~Hern\'andez-Monteagudo\orcid{0000-0001-5471-9166}\inst{\ref{aff82},\ref{aff35}}}

\institute{INFN-Padova, Via Marzolo 8, 35131 Padova, Italy\label{aff1}
\and
Dipartimento di Fisica e Astronomia "G. Galilei", Universit\`a di Padova, Via Marzolo 8, 35131 Padova, Italy\label{aff2}
\and
INAF-Osservatorio Astronomico di Padova, Via dell'Osservatorio 5, 35122 Padova, Italy\label{aff3}
\and
Universit\"at Bonn, Argelander-Institut f\"ur Astronomie, Auf dem H\"ugel 71, 53121 Bonn, Germany\label{aff4}
\and
Dipartimento di Fisica - Sezione di Astronomia, Universit\`a di Trieste, Via Tiepolo 11, 34131 Trieste, Italy\label{aff5}
\and
INAF-Osservatorio Astronomico di Trieste, Via G. B. Tiepolo 11, 34143 Trieste, Italy\label{aff6}
\and
IFPU, Institute for Fundamental Physics of the Universe, via Beirut 2, 34151 Trieste, Italy\label{aff7}
\and
INFN, Sezione di Trieste, Via Valerio 2, 34127 Trieste TS, Italy\label{aff8}
\and
Universit\'e Paris-Saclay, CNRS, Institut d'astrophysique spatiale, 91405, Orsay, France\label{aff9}
\and
School of Mathematics and Physics, University of Surrey, Guildford, Surrey, GU2 7XH, UK\label{aff10}
\and
INAF-Osservatorio Astronomico di Brera, Via Brera 28, 20122 Milano, Italy\label{aff11}
\and
INAF-Osservatorio di Astrofisica e Scienza dello Spazio di Bologna, Via Piero Gobetti 93/3, 40129 Bologna, Italy\label{aff12}
\and
SISSA, International School for Advanced Studies, Via Bonomea 265, 34136 Trieste TS, Italy\label{aff13}
\and
Dipartimento di Fisica e Astronomia, Universit\`a di Bologna, Via Gobetti 93/2, 40129 Bologna, Italy\label{aff14}
\and
INFN-Sezione di Bologna, Viale Berti Pichat 6/2, 40127 Bologna, Italy\label{aff15}
\and
Max Planck Institute for Extraterrestrial Physics, Giessenbachstr. 1, 85748 Garching, Germany\label{aff16}
\and
INAF-Osservatorio Astrofisico di Torino, Via Osservatorio 20, 10025 Pino Torinese (TO), Italy\label{aff17}
\and
Dipartimento di Fisica, Universit\`a di Genova, Via Dodecaneso 33, 16146, Genova, Italy\label{aff18}
\and
INFN-Sezione di Genova, Via Dodecaneso 33, 16146, Genova, Italy\label{aff19}
\and
Department of Physics "E. Pancini", University Federico II, Via Cinthia 6, 80126, Napoli, Italy\label{aff20}
\and
INAF-Osservatorio Astronomico di Capodimonte, Via Moiariello 16, 80131 Napoli, Italy\label{aff21}
\and
INFN section of Naples, Via Cinthia 6, 80126, Napoli, Italy\label{aff22}
\and
Instituto de Astrof\'isica e Ci\^encias do Espa\c{c}o, Universidade do Porto, CAUP, Rua das Estrelas, PT4150-762 Porto, Portugal\label{aff23}
\and
Dipartimento di Fisica, Universit\`a degli Studi di Torino, Via P. Giuria 1, 10125 Torino, Italy\label{aff24}
\and
INFN-Sezione di Torino, Via P. Giuria 1, 10125 Torino, Italy\label{aff25}
\and
INAF-IASF Milano, Via Alfonso Corti 12, 20133 Milano, Italy\label{aff26}
\and
INAF-Osservatorio Astronomico di Roma, Via Frascati 33, 00078 Monteporzio Catone, Italy\label{aff27}
\and
INFN-Sezione di Roma, Piazzale Aldo Moro, 2 - c/o Dipartimento di Fisica, Edificio G. Marconi, 00185 Roma, Italy\label{aff28}
\and
Centro de Investigaciones Energ\'eticas, Medioambientales y Tecnol\'ogicas (CIEMAT), Avenida Complutense 40, 28040 Madrid, Spain\label{aff29}
\and
Port d'Informaci\'{o} Cient\'{i}fica, Campus UAB, C. Albareda s/n, 08193 Bellaterra (Barcelona), Spain\label{aff30}
\and
Institut d'Estudis Espacials de Catalunya (IEEC),  Edifici RDIT, Campus UPC, 08860 Castelldefels, Barcelona, Spain\label{aff31}
\and
Institute of Space Sciences (ICE, CSIC), Campus UAB, Carrer de Can Magrans, s/n, 08193 Barcelona, Spain\label{aff32}
\and
Institute for Theoretical Particle Physics and Cosmology (TTK), RWTH Aachen University, 52056 Aachen, Germany\label{aff33}
\and
Dipartimento di Fisica e Astronomia "Augusto Righi" - Alma Mater Studiorum Universit\`a di Bologna, Viale Berti Pichat 6/2, 40127 Bologna, Italy\label{aff34}
\and
Instituto de Astrof\'isica de Canarias, Calle V\'ia L\'actea s/n, 38204, San Crist\'obal de La Laguna, Tenerife, Spain\label{aff35}
\and
Institute for Astronomy, University of Edinburgh, Royal Observatory, Blackford Hill, Edinburgh EH9 3HJ, UK\label{aff36}
\and
Jodrell Bank Centre for Astrophysics, Department of Physics and Astronomy, University of Manchester, Oxford Road, Manchester M13 9PL, UK\label{aff37}
\and
European Space Agency/ESRIN, Largo Galileo Galilei 1, 00044 Frascati, Roma, Italy\label{aff38}
\and
ESAC/ESA, Camino Bajo del Castillo, s/n., Urb. Villafranca del Castillo, 28692 Villanueva de la Ca\~nada, Madrid, Spain\label{aff39}
\and
Universit\'e Claude Bernard Lyon 1, CNRS/IN2P3, IP2I Lyon, UMR 5822, Villeurbanne, F-69100, France\label{aff40}
\and
Institute of Physics, Laboratory of Astrophysics, Ecole Polytechnique F\'ed\'erale de Lausanne (EPFL), Observatoire de Sauverny, 1290 Versoix, Switzerland\label{aff41}
\and
UCB Lyon 1, CNRS/IN2P3, IUF, IP2I Lyon, 4 rue Enrico Fermi, 69622 Villeurbanne, France\label{aff42}
\and
Departamento de F\'isica, Faculdade de Ci\^encias, Universidade de Lisboa, Edif\'icio C8, Campo Grande, PT1749-016 Lisboa, Portugal\label{aff43}
\and
Instituto de Astrof\'isica e Ci\^encias do Espa\c{c}o, Faculdade de Ci\^encias, Universidade de Lisboa, Campo Grande, 1749-016 Lisboa, Portugal\label{aff44}
\and
Department of Astronomy, University of Geneva, ch. d'Ecogia 16, 1290 Versoix, Switzerland\label{aff45}
\and
INAF-Istituto di Astrofisica e Planetologia Spaziali, via del Fosso del Cavaliere, 100, 00100 Roma, Italy\label{aff46}
\and
Universit\'e Paris-Saclay, Universit\'e Paris Cit\'e, CEA, CNRS, AIM, 91191, Gif-sur-Yvette, France\label{aff47}
\and
Institut de Ciencies de l'Espai (IEEC-CSIC), Campus UAB, Carrer de Can Magrans, s/n Cerdanyola del Vall\'es, 08193 Barcelona, Spain\label{aff48}
\and
Istituto Nazionale di Fisica Nucleare, Sezione di Bologna, Via Irnerio 46, 40126 Bologna, Italy\label{aff49}
\and
FRACTAL S.L.N.E., calle Tulip\'an 2, Portal 13 1A, 28231, Las Rozas de Madrid, Spain\label{aff50}
\and
Universit\"ats-Sternwarte M\"unchen, Fakult\"at f\"ur Physik, Ludwig-Maximilians-Universit\"at M\"unchen, Scheinerstrasse 1, 81679 M\"unchen, Germany\label{aff51}
\and
Dipartimento di Fisica "Aldo Pontremoli", Universit\`a degli Studi di Milano, Via Celoria 16, 20133 Milano, Italy\label{aff52}
\and
Institute of Theoretical Astrophysics, University of Oslo, P.O. Box 1029 Blindern, 0315 Oslo, Norway\label{aff53}
\and
Jet Propulsion Laboratory, California Institute of Technology, 4800 Oak Grove Drive, Pasadena, CA, 91109, USA\label{aff54}
\and
Felix Hormuth Engineering, Goethestr. 17, 69181 Leimen, Germany\label{aff55}
\and
Technical University of Denmark, Elektrovej 327, 2800 Kgs. Lyngby, Denmark\label{aff56}
\and
Cosmic Dawn Center (DAWN), Denmark\label{aff57}
\and
Max-Planck-Institut f\"ur Astronomie, K\"onigstuhl 17, 69117 Heidelberg, Germany\label{aff58}
\and
NASA Goddard Space Flight Center, Greenbelt, MD 20771, USA\label{aff59}
\and
Department of Physics and Astronomy, University College London, Gower Street, London WC1E 6BT, UK\label{aff60}
\and
Department of Physics and Helsinki Institute of Physics, Gustaf H\"allstr\"omin katu 2, 00014 University of Helsinki, Finland\label{aff61}
\and
Aix-Marseille Universit\'e, CNRS/IN2P3, CPPM, Marseille, France\label{aff62}
\and
Mullard Space Science Laboratory, University College London, Holmbury St Mary, Dorking, Surrey RH5 6NT, UK\label{aff63}
\and
Leiden Observatory, Leiden University, Einsteinweg 55, 2333 CC Leiden, The Netherlands\label{aff64}
\and
Universit\'e de Gen\`eve, D\'epartement de Physique Th\'eorique and Centre for Astroparticle Physics, 24 quai Ernest-Ansermet, CH-1211 Gen\`eve 4, Switzerland\label{aff65}
\and
Department of Physics, P.O. Box 64, 00014 University of Helsinki, Finland\label{aff66}
\and
Helsinki Institute of Physics, Gustaf H{\"a}llstr{\"o}min katu 2, University of Helsinki, Helsinki, Finland\label{aff67}
\and
NOVA optical infrared instrumentation group at ASTRON, Oude Hoogeveensedijk 4, 7991PD, Dwingeloo, The Netherlands\label{aff68}
\and
Centre de Calcul de l'IN2P3/CNRS, 21 avenue Pierre de Coubertin 69627 Villeurbanne Cedex, France\label{aff69}
\and
Aix-Marseille Universit\'e, CNRS, CNES, LAM, Marseille, France\label{aff70}
\and
Dipartimento di Fisica e Astronomia "Augusto Righi" - Alma Mater Studiorum Universit\`a di Bologna, via Piero Gobetti 93/2, 40129 Bologna, Italy\label{aff71}
\and
Department of Physics, Institute for Computational Cosmology, Durham University, South Road, DH1 3LE, UK\label{aff72}
\and
Universit\'e Paris Cit\'e, CNRS, Astroparticule et Cosmologie, 75013 Paris, France\label{aff73}
\and
Institut d'Astrophysique de Paris, 98bis Boulevard Arago, 75014, Paris, France\label{aff74}
\and
Institut d'Astrophysique de Paris, UMR 7095, CNRS, and Sorbonne Universit\'e, 98 bis boulevard Arago, 75014 Paris, France\label{aff75}
\and
European Space Agency/ESTEC, Keplerlaan 1, 2201 AZ Noordwijk, The Netherlands\label{aff76}
\and
Institut de F\'{i}sica d'Altes Energies (IFAE), The Barcelona Institute of Science and Technology, Campus UAB, 08193 Bellaterra (Barcelona), Spain\label{aff77}
\and
Department of Physics and Astronomy, University of Aarhus, Ny Munkegade 120, DK-8000 Aarhus C, Denmark\label{aff78}
\and
Space Science Data Center, Italian Space Agency, via del Politecnico snc, 00133 Roma, Italy\label{aff79}
\and
Centre National d'Etudes Spatiales -- Centre spatial de Toulouse, 18 avenue Edouard Belin, 31401 Toulouse Cedex 9, France\label{aff80}
\and
Institute of Space Science, Str. Atomistilor, nr. 409 M\u{a}gurele, Ilfov, 077125, Romania\label{aff81}
\and
Departamento de Astrof\'isica, Universidad de La Laguna, 38206, La Laguna, Tenerife, Spain\label{aff82}
\and
Institut f\"ur Theoretische Physik, University of Heidelberg, Philosophenweg 16, 69120 Heidelberg, Germany\label{aff83}
\and
Institut de Recherche en Astrophysique et Plan\'etologie (IRAP), Universit\'e de Toulouse, CNRS, UPS, CNES, 14 Av. Edouard Belin, 31400 Toulouse, France\label{aff84}
\and
Universit\'e St Joseph; Faculty of Sciences, Beirut, Lebanon\label{aff85}
\and
Departamento de F\'isica, FCFM, Universidad de Chile, Blanco Encalada 2008, Santiago, Chile\label{aff86}
\and
Universit\"at Innsbruck, Institut f\"ur Astro- und Teilchenphysik, Technikerstr. 25/8, 6020 Innsbruck, Austria\label{aff87}
\and
Satlantis, University Science Park, Sede Bld 48940, Leioa-Bilbao, Spain\label{aff88}
\and
Instituto de Astrof\'isica e Ci\^encias do Espa\c{c}o, Faculdade de Ci\^encias, Universidade de Lisboa, Tapada da Ajuda, 1349-018 Lisboa, Portugal\label{aff89}
\and
Universidad Polit\'ecnica de Cartagena, Departamento de Electr\'onica y Tecnolog\'ia de Computadoras,  Plaza del Hospital 1, 30202 Cartagena, Spain\label{aff90}
\and
INFN-Bologna, Via Irnerio 46, 40126 Bologna, Italy\label{aff91}
\and
Kapteyn Astronomical Institute, University of Groningen, PO Box 800, 9700 AV Groningen, The Netherlands\label{aff92}
\and
Dipartimento di Fisica, Universit\`a degli studi di Genova, and INFN-Sezione di Genova, via Dodecaneso 33, 16146, Genova, Italy\label{aff93}
\and
Infrared Processing and Analysis Center, California Institute of Technology, Pasadena, CA 91125, USA\label{aff94}
\and
INAF, Istituto di Radioastronomia, Via Piero Gobetti 101, 40129 Bologna, Italy\label{aff95}
\and
Astronomical Observatory of the Autonomous Region of the Aosta Valley (OAVdA), Loc. Lignan 39, I-11020, Nus (Aosta Valley), Italy\label{aff96}
\and
Junia, EPA department, 41 Bd Vauban, 59800 Lille, France\label{aff97}
\and
ICSC - Centro Nazionale di Ricerca in High Performance Computing, Big Data e Quantum Computing, Via Magnanelli 2, Bologna, Italy\label{aff98}
\and
Instituto de F\'isica Te\'orica UAM-CSIC, Campus de Cantoblanco, 28049 Madrid, Spain\label{aff99}
\and
CERCA/ISO, Department of Physics, Case Western Reserve University, 10900 Euclid Avenue, Cleveland, OH 44106, USA\label{aff100}
\and
Laboratoire Univers et Th\'eorie, Observatoire de Paris, Universit\'e PSL, Universit\'e Paris Cit\'e, CNRS, 92190 Meudon, France\label{aff101}
\and
Dipartimento di Fisica e Scienze della Terra, Universit\`a degli Studi di Ferrara, Via Giuseppe Saragat 1, 44122 Ferrara, Italy\label{aff102}
\and
Istituto Nazionale di Fisica Nucleare, Sezione di Ferrara, Via Giuseppe Saragat 1, 44122 Ferrara, Italy\label{aff103}
\and
Kavli Institute for the Physics and Mathematics of the Universe (WPI), University of Tokyo, Kashiwa, Chiba 277-8583, Japan\label{aff104}
\and
Ludwig-Maximilians-University, Schellingstrasse 4, 80799 Munich, Germany\label{aff105}
\and
Max-Planck-Institut f\"ur Physik, Boltzmannstr. 8, 85748 Garching, Germany\label{aff106}
\and
Minnesota Institute for Astrophysics, University of Minnesota, 116 Church St SE, Minneapolis, MN 55455, USA\label{aff107}
\and
Institute Lorentz, Leiden University, Niels Bohrweg 2, 2333 CA Leiden, The Netherlands\label{aff108}
\and
Universit\'e C\^{o}te d'Azur, Observatoire de la C\^{o}te d'Azur, CNRS, Laboratoire Lagrange, Bd de l'Observatoire, CS 34229, 06304 Nice cedex 4, France\label{aff109}
\and
Institute for Astronomy, University of Hawaii, 2680 Woodlawn Drive, Honolulu, HI 96822, USA\label{aff110}
\and
Department of Physics \& Astronomy, University of California Irvine, Irvine CA 92697, USA\label{aff111}
\and
Department of Astronomy \& Physics and Institute for Computational Astrophysics, Saint Mary's University, 923 Robie Street, Halifax, Nova Scotia, B3H 3C3, Canada\label{aff112}
\and
Departamento F\'isica Aplicada, Universidad Polit\'ecnica de Cartagena, Campus Muralla del Mar, 30202 Cartagena, Murcia, Spain\label{aff113}
\and
Instituto de Astrof\'isica de Canarias (IAC); Departamento de Astrof\'isica, Universidad de La Laguna (ULL), 38200, La Laguna, Tenerife, Spain\label{aff114}
\and
Department of Physics, Oxford University, Keble Road, Oxford OX1 3RH, UK\label{aff115}
\and
Institute of Cosmology and Gravitation, University of Portsmouth, Portsmouth PO1 3FX, UK\label{aff116}
\and
Department of Computer Science, Aalto University, PO Box 15400, Espoo, FI-00 076, Finland\label{aff117}
\and
Ruhr University Bochum, Faculty of Physics and Astronomy, Astronomical Institute (AIRUB), German Centre for Cosmological Lensing (GCCL), 44780 Bochum, Germany\label{aff118}
\and
DARK, Niels Bohr Institute, University of Copenhagen, Jagtvej 155, 2200 Copenhagen, Denmark\label{aff119}
\and
Univ. Grenoble Alpes, CNRS, Grenoble INP, LPSC-IN2P3, 53, Avenue des Martyrs, 38000, Grenoble, France\label{aff120}
\and
Department of Physics and Astronomy, Vesilinnantie 5, 20014 University of Turku, Finland\label{aff121}
\and
Serco for European Space Agency (ESA), Camino bajo del Castillo, s/n, Urbanizacion Villafranca del Castillo, Villanueva de la Ca\~nada, 28692 Madrid, Spain\label{aff122}
\and
ARC Centre of Excellence for Dark Matter Particle Physics, Melbourne, Australia\label{aff123}
\and
Centre for Astrophysics \& Supercomputing, Swinburne University of Technology,  Hawthorn, Victoria 3122, Australia\label{aff124}
\and
School of Physics and Astronomy, Queen Mary University of London, Mile End Road, London E1 4NS, UK\label{aff125}
\and
Department of Physics and Astronomy, University of the Western Cape, Bellville, Cape Town, 7535, South Africa\label{aff126}
\and
Universit\'e Libre de Bruxelles (ULB), Service de Physique Th\'eorique CP225, Boulevard du Triophe, 1050 Bruxelles, Belgium\label{aff127}
\and
ICTP South American Institute for Fundamental Research, Instituto de F\'{\i}sica Te\'orica, Universidade Estadual Paulista, S\~ao Paulo, Brazil\label{aff128}
\and
Oskar Klein Centre for Cosmoparticle Physics, Department of Physics, Stockholm University, Stockholm, SE-106 91, Sweden\label{aff129}
\and
Astrophysics Group, Blackett Laboratory, Imperial College London, London SW7 2AZ, UK\label{aff130}
\and
INAF-Osservatorio Astrofisico di Arcetri, Largo E. Fermi 5, 50125, Firenze, Italy\label{aff131}
\and
Dipartimento di Fisica, Sapienza Universit\`a di Roma, Piazzale Aldo Moro 2, 00185 Roma, Italy\label{aff132}
\and
Centro de Astrof\'{\i}sica da Universidade do Porto, Rua das Estrelas, 4150-762 Porto, Portugal\label{aff133}
\and
Dipartimento di Fisica, Universit\`a di Roma Tor Vergata, Via della Ricerca Scientifica 1, Roma, Italy\label{aff134}
\and
INFN, Sezione di Roma 2, Via della Ricerca Scientifica 1, Roma, Italy\label{aff135}
\and
Institute of Astronomy, University of Cambridge, Madingley Road, Cambridge CB3 0HA, UK\label{aff136}
\and
Department of Astrophysics, University of Zurich, Winterthurerstrasse 190, 8057 Zurich, Switzerland\label{aff137}
\and
Theoretical astrophysics, Department of Physics and Astronomy, Uppsala University, Box 515, 751 20 Uppsala, Sweden\label{aff138}
\and
Department of Physics, Royal Holloway, University of London, TW20 0EX, UK\label{aff139}
\and
Department of Astrophysical Sciences, Peyton Hall, Princeton University, Princeton, NJ 08544, USA\label{aff140}
\and
Cosmic Dawn Center (DAWN)\label{aff141}
\and
Niels Bohr Institute, University of Copenhagen, Jagtvej 128, 2200 Copenhagen, Denmark\label{aff142}
\and
Institut de Physique Th\'eorique, CEA, CNRS, Universit\'e Paris-Saclay 91191 Gif-sur-Yvette Cedex, France\label{aff143}
\and
Center for Cosmology and Particle Physics, Department of Physics, New York University, New York, NY 10003, USA\label{aff144}
\and
Center for Computational Astrophysics, Flatiron Institute, 162 5th Avenue, 10010, New York, NY, USA\label{aff145}}

\date{\today}

\authorrunning{Euclid Collaboration}

\titlerunning{\Euclid preparation. Relativistic RSD in the two-point statistics from the EWSS}

  \abstract
   {Measurements of galaxy clustering are affected by redshift-space distortions (RSD). Peculiar velocities, gravitational lensing, and
   other light-cone projection effects
   modify the observed redshifts, fluxes, and sky positions of distant light sources. We determine which of these effects
   leave a detectable imprint on 
   several two-point clustering statistics extracted from the Euclid Wide Spectroscopic Survey (EWSS) on large scales. We generate 140 mock galaxy catalogues with the survey geometry and selection function of the EWSS and make use of the LIGER (LIght cones with GEneral Relativity) method to account for a variable number of relativistic RSD to linear order in the cosmological perturbations.
We estimate different two-point clustering statistics from the mocks and use the likelihood-ratio test to calculate the statistical significance with which the EWSS could reject the
null hypothesis that certain relativistic projection effects can be
neglected in the theoretical models. 
We find that the combined effects of lensing magnification and convergence imprint characteristic signatures on several clustering observables.
Their signal-to-noise ratio (\snrt) ranges between 2.5 and 6 (depending on the adopted summary statistic) for the highest-redshift galaxies in the EWSS.
The corresponding feature due to the peculiar velocity of the Sun is measured with a \snrt of order one or two. The multipoles of the power spectrum from the catalogues that include all relativistic effects reject the null hypothesis that RSD are only generated by the variation of the peculiar velocity along the line of sight with a significance of 2.9 standard deviations.
As a byproduct of our study, we demonstrate that the mixing-matrix formalism to model finite-volume effects in the multipole moments of the power spectrum can be
robustly applied to surveys made of several disconnected patches. Our results indicate that relativistic RSD, the contribution from weak gravitational lensing in particular, cannot be disregarded when modelling two-point clustering statistics extracted from the EWSS.}


   \maketitle

\tableofcontents
\section{Introduction}

The primary science goal of the recently launched \Euclid space mission 
\citep{euclidcollaboration2024euclidiovervieweuclid} is to test whether the cosmological constant can be ruled out as the driver of the accelerated expansion of the Universe. To that end, \Euclid is carrying out a wide-angle survey covering nearly $15\,000$ deg$^2$ of the extragalactic sky. The \Euclid mission is optimized for the combination of two cosmological probes -- weak gravitational lensing and galaxy clustering -- and relies on two instruments. The visual imager \citep[VIS,][]{Borlaf_2022} operates in the 550 to 900 nm pass-band and produces high-quality galaxy images to perform measurements of cosmic shear. The near-infrared spectrometer and photometer \citep[NISP,][]{Maciaszek_2022,Schirmer_2022} carries out imaging photometry (as an input for the estimation of photometric redshifts) and slitless spectroscopy to precisely measure the redshift of the H$\alpha$ emission line in the range $0.9<z<1.8$. Over six years of observations, the Euclid Wide Spectroscopic Survey (EWSS) will measure redshifts and positions of nearly 30 million emission-line galaxies, while its photometric counterpart will measure positions and shapes of approximately 1.5 billion galaxies.

Galaxy clustering (the only cosmological probe discussed in this paper) sets constraints on the cause of the accelerated expansion of the Universe in two ways. First, the expansion history of the Universe can be reconstructed by locating the characteristic scale imprinted by baryonic acoustic oscillations on the galaxy power spectrum (or the two-point correlation function, 2PCF) as a function of redshift \citep[e.g.][]{Cole+2005,Eisenstein+2005}. Second, the growth-rate of structure can be determined by studying the {anisotropy}
of the clustering signal \citep[e.g.][]{Peacock+2001}.
The latter option is generally known as the study of
redshift-space distortions (RSD) which arise when galaxy redshifts are mapped into distances by assuming an unperturbed Friedmann--Lema\^{\i}tre--Robertson--Walker (FLRW) metric. The observed redshift of a galaxy does not coincide with its cosmological
component. The dominant (but not sole) correction is due to the relative peculiar velocity between the galaxy and the observer along the line of sight. In a seminal work, \cite{Kaiser87} used linear perturbation theory to calculate how peculiar velocities distort
the galaxy power spectrum \citep[see also ][for extensions to configuration space]{Hamilton-Culhane,Hamilton_review, Hamilton00}.
The derivation relies on a simplifying assumption that the size of the surveyed region is negligibly small compared to its distance from the observer, so that the lines of sight to all galaxies are effectively parallel. This is nowadays known as the `global plane-parallel' (GPP) approximation and implies that the galaxy power spectrum depends
on the cosine of the angle between the wavevector and the fixed line of sight. When decomposed in Legendre polynomials, this functional
dependence only includes multipoles of degree 0, 2, and 4.

The EWSS provides us with the opportunity to study galaxy clustering on unprecedentedly large scales.  This possibility, however, brings forth new challenges. First of all, galaxy pairs with large angular separations contribute to the clustering signal and might cause systematic deviations of the observations from models based on the GPP approximation. Wide-angle corrections have been investigated both for the galaxy 2PCF \citep[see e.g. ][]{Matsubara_2000_2PCF,Szapudi_2004,P_pai_2008,Raccanelli_2010,Samushia_2012,raccanelli2016doppler} and the power spectrum \citep[see e.g.][]{Zaroubi-Hoffman96,Reimberg+2016,Castorina-white2017,Castorina_2019}.
The current leading opinion is that they should not be an hindrance for conducting cosmological studies with surveys of similar size and depth to the EWSS \citep{Castorina_2022, Noorikuhani_Scoccimarro2023}. The second complication 
arises from the fact that Kaiser's RSD should be complemented
with additional corrections. The light bundles from distant galaxies to us propagate through the inhomogeneous Universe and are thus subject to effects like gravitational lensing or aberration. Hence, the observed galaxy positions on the sky, redshifts, and fluxes differ from their
analogues obtained in the corresponding unperturbed FLRW model. However, we construct maps of the galaxy distribution by assuming such a homogeneous model
to convert the observed properties into three-dimensional positions and luminosities.
This step introduces a number of artefacts in the galaxy overdensity
field \citep{Yoo:2009au,Bonvin-Durrer2011,Challinor:2011bk,Jeong:2011as} that we refer to as `relativistic\footnote{We assume that gravitation is described by the theory of general relativity. } RSD' and that are also known in the literature as `relativistic effects' or `projection effects' (since
we observe the projection of the actual Universe on our past light cone). These alterations can be studied perturbatively. The leading
term coincides with Kaiser's RSD due to the peculiar-velocity
gradient along the line of sight. Nevertheless, there exist several additional corrections that can potentially influence
galaxy-clustering statistics on very large scales. A number of investigations have characterised the impact of these terms
on the angular power spectrum \citep[][]{Dio_2013,MIKO_2017}, the 2PCF \citep[e.g.][]{Bertacca:2015,Raccanelli+2016_GR_CORRECTIONS,Tansella_2018,Bertacca_2020_Rocket_effect,Jelic_Cizmek_2021,Breton+2022,BRETON_2022_cross_correlation}, and the {(3D)} power spectrum \citep{ Elkhashab_2021, Castorina_2022,Foglieni_2023,Noorikuhani_Scoccimarro2023}.

Which of the relativistic RSD will leave a detectable imprint on the two-point summary statistics measured from the EWSS?
Answering this question is the main goal of this study. \footnote{Related work based on the \Euclid photometric sample and its cross-correlation with measurements of cosmic shear is presented in \citet{EUCLID_PHOTOMETRIC_WP9} and \citet{EUCLID_PHOTOMETRIC_TANIDIS}.} 
We investigate four different suites of mock galaxy catalogues 
built with the \liger (LIght cones
with GEneral Relativity) method \citep{MIKO_2017,Elkhashab_2021},
which allows us to self-consistently correct
the output of Netwonian $N$-body simulations and introduce
relativistic RSD to linear order in the cosmological perturbations.
All the mock catalogues we generate match the survey geometry and selection function of the EWSS but the four kinds we consider differ in the number of RSD terms they include.
This helps us isolate  the contributions from various effects (e.g. gravitational lensing or the peculiar velocity of the observer). We make use of popular estimators to measure clustering summary statistics from the mock catalogues and we build unbiased models that exactly account for wide-angle effects by averaging the measurements over a large number of realisations. Finally, we employ the likelihood-ratio test to quantify the signal-to-noise ratio (\snrt) with which certain effects can be detected and to determine the fraction of realisations in which models that do not account for these effects could be ruled out with a given statistical
significance.

The paper is structured as follows.
We introduce the
relativistic RSD in Sect.~\ref{Sec:Relat-RSD} and our mock
EWSS galaxy catalogues in Sect.~\ref{sec:gal_mocks}. We describe how we assess the detectability of 
various relativistic RSD in
Sect.~\ref{sec:chisq}. Our results for the angular power spectrum,
the multipole moments of the 2PCF, and the multipole moments of the power spectrum are presented in
Sects.~\ref{sec:ang_power_all}, \ref{Sec:2PCF_ALL}, and
\ref{Sec:Power_Spectrum}, respectively.
Moreover, in Sect.~\ref{sec:Window_analysis_sec}, we compare the
power-spectrum multipoles extracted from the mocks to the predictions
from Kaiser's model after accounting 
for the window function of the survey and the integral constraint.
Eventually, in Sect.~\ref{sec:summary} we summarise our
findings and conclude.

Throughout this paper, we adopt Einstein's summation convention and  define the space-time metric tensor to have the  
signature $(-,+,+,+)$. Greek indices refer to space-time components (i.e. run from 0 to 3) while Latin indices label spatial components (i.e. run 
from 1 to 3).  Furthermore, the Dirac delta and 
the Kronecker delta functions are denoted by the symbols $\delta_{\rm D}$ and $\delta_{\rm K}$, respectively. Our Fourier-transform convention is $\tilde{f}(\bs{k}) = \int f(\bs{x})\,\mathrm{e}^{-i\bs{k}\cdot \bs{x}}\,\dif ^3 x $. 
Finally, the symbol $c$ denotes the speed of light in vacuum. 

\section{Relativistic RSD}\label{Sec:Relat-RSD}

In order to  build three-dimensional maps of the galaxy distribution, it is usually assumed that the light bundles emitted by the galaxies propagate in an unperturbed FLRW model universe and that their observed redshift, $z_\mathrm{obs}$, coincides with the cosmological one. 
This implies that their comoving distance in the so-called `redshift space' is 
\be \label{xr}
{x} = \int_0^{{z_\mathrm{obs}}} \frac{c}{H( z)}\;\dif  z\,,
\ee
where $H(z)$ {denotes} the  Hubble parameter in the model universe as a function of redshift.
This procedure, however, neglects the fact that inhomogeneities
in the Universe alter the observed redshifts and angular positions
of the galaxies. Therefore,  the reconstructed  galaxy maps in redshift space are not faithful \citep{Sargent_Turner_1977}. A number of effects, collectively 
called redshift-space distortions, artificially shift the reconstructed positions of galaxies in both the radial and tangential directions
with respect to their actual (hereafter, real-space) location. 

The pioneering work by \cite{Kaiser87} investigated the 
relationship between galaxy densities in real and redshift space 
 at linear order in {the} cosmological perturbations, focusing
 on the impact of peculiar velocities generated by gravitational
 instabilities \citep[see also][]{Hamilton-Culhane, Hamilton_review, Matsubara_2000}. More recently, this subject has been revisited using a 
fully general-relativistic approach and accounting for additional effects like gravitational lensing, the Sachs--Wolfe effects, and
the Shapiro delay
\citep{Yoo:2009au, Bonvin-Durrer2011,Challinor:2011bk,Jeong:2011as}. 
In this latter case, the goal is to compute the geodesics of photons emitted from a source galaxy {in the presence of linear} cosmological perturbations.
This is sufficient to address the large spatial scales considered in this paper. In the remainder of this section, we 
summarise the main results obtained within the general-relativistic
framework.

By definition, the coordinates of a distant galaxy in redshift space can be
trivially expressed as
\begin{equation} \label{eq:LCcoordinates}
x^\mu=(c\,\eta_0- x, \;  x \, {\bs n})\;, 
\end{equation} 
where $\eta_0$ is the present-day value of conformal time (i.e. at observation), $x$ denotes the comoving distance from the observer (see Eq.~\ref{xr}), 
and $\bs{ n}$ is the observed galaxy position {(pointing towards the galaxy)} on the sky. The mapping between the real- and redshift-space coordinates of a galaxy
can be generically written as $x_{\rm r}^\mu=x^\mu+\Delta x^{\mu}$.
In order to compute the coordinate transformation {explicitly}, we need to specify a gauge. We express the space-time metric in the Poisson gauge,  
assuming a flat cosmology while neglecting vector and
tensor perturbations\footnote{{The scalar-restricted Poisson gauge is also known as the conformal Newtonian gauge. }}, %
\be
\label{eq:metric}
\dif s^2 = a^2(\eta)\left[-(1+2\Psi)\,c^2\,\dif\eta ^2 +(1-2{\Phi})\,\delta_{{\rm K}\,ij}\, \dif x^i\;\dif x^j\;\right]\,,
\ee
where $\Psi$ and $\Phi$ are the dimensionless Bardeen potentials, $\eta$ is the conformal time, and $a$ is the cosmic scale factor. 
From this choice, it follows that
\citep{Hui_2006_perturbation, Yoo:2009au, Bonvin-Durrer2011, Challinor:2011bk,Jeong:2011as}
\begin{subequations}\label{eq:shift_lig}
\begin{align}
    \Delta x^{0} &= \frac{c}{\mc{H}}\,\delta \ln a\,, \\
    \Delta x^{i} &= -\left( \,\Phi_{\rm{o}}+\Psi_{\rm{o}} + \; {\bvr{e}\over c }\cdot {\bs n}  \right)x^i
    -x\, {\varv^i_{\rm{o}}\over c} - \frac{c}{\mc{H}}\,n^i\, \delta \ln a   \nonumber \\ \nonumber
    & +  n^i\int^{x}_0 (x -\tilde{x})\,\overc{(\Phi'+\Psi')}\;\dif\tilde{x} -\int^{x}_0 (x -\tilde{x})\,\delta_{\rm K}^{ij}\,\tilde{\partial}_j(\Phi+\Psi)\;\dif\tilde{x}\\
    &+ 2 n^i \int^{x}_0 (\Phi+\Psi) \;\dif \tilde{x}\,,
\end{align}
\end{subequations}
where 
 \be \label{eq:deltalna}
 \delta \ln a := \left[{(\bs{\varv}_\mathrm{e}-\bs{\varv}_\mathrm{o})\over c }\cdot {\bs {n}} - (\Phi_\mathrm{e}-\Phi_\mathrm{o}) - \int_0^x {(\Phi' + \Psi')\over c}\;\dif \tilde{x}\right]\,,
 \ee 
represents the fractional redshift change due to the perturbations, i.e. $-\delta z/(1+z)$. Here,
$\bs{\varv}$ denotes peculiar 3-velocities, the subscripts `e' and `o' specify whether the functions
are evaluated at the source or observer locations, respectively, 
$\tilde{\partial}_j=\partial / \partial  \tilde x^j$, $\mc{H} = a(z_{\rm obs})\,H(z_{\rm obs})$ and the prime superscript 
denotes the partial derivative w.r.t. $\eta$. %
It is worth mentioning that the equations above assume that
the peculiar velocity of a galaxy coincides with that of the matter at the same location, i.e. that there is no velocity bias.\footnote{In Eqs.\ (\ref{eq:shift_lig}) and (\ref{eq:deltalna}), we are {neglecting} the fact that the  coordinate time does not coincide with the proper time of the observer in an inhomogenous universe. %
This only affects the mean number density and has no impact on the density contrast \citep{Bertacca_2020,Grimm_2021}.}

Cosmological perturbations also alter the solid angle under which galaxies are seen by distant observers, thus enhancing, or decreasing their apparent flux \citep[e.g.][]{Broadhurst:1994qu}. 
In terms of the luminosity distance $D_{\rm L}$, the magnification of a galaxy is defined as
\begin{equation} 
 {\mathcal M}=\left( \frac{D_{\rm L}}{\bar{D}_{\rm L}}\right)^{-2}\;, 
\end{equation} 
where $\bar{D}_{\rm L}$ denotes the luminosity distance in the background model universe evaluated at $z_\mathrm{obs}$.
At linear order  \citep[e.g.][]{Challinor:2011bk,Bertacca:2015}, 
\begin{align}
\label{eq:Mag_lig}
\mc{M} &= 1 + 2\Phi_{\rm e} - \,2\left(1-\frac{c}{\cH x} \right)\,\delta \ln a - 2\,\overc{\bvr{o}}\cdot \bs{n} \\\nonumber
&+ {2\,\kappa} - \frac{2}{x}\int^{x}_0 (\Phi+\Psi)\; \dif \tilde{x}\,\,,
\end{align}
where the weak lensing convergence is
\be 
\kappa := {1\over 2}\int^x_0\, (x-\tilde{x})\,\frac{\tilde{x}}{x} \left[\nabla^2 +(\bs{n}\cdot \bs{\nabla})^2 - {2\over x} \bs{n}\cdot \bs{\nabla} \right] (\Phi + \Psi)\; \dif \tilde{x}\,.
\ee 

The next step is to understand how the local number density of galaxies in a survey responds to redshift perturbations and magnification.
To first approximation, the EWSS is flux limited as it only selects galaxies above a given observed H$\alpha$ flux,\footnote{Strictly speaking, also other factors like the source size determine whether a galaxy is selected or not.}
corresponding to a redshift-dependent luminosity limit, $L_\mathrm{lim}(z)$.
Let us indicate by $n(L_\textrm{min},z)$ the mean number density of the
target population of galaxies with luminosity $L>L_\textrm{min}$ at
redshift $z$. Then, the so-called evolution bias, 
\be
\label{eq:EV_bias}
    \mc{E}(z)=-\left.\frac{\partial \ln n(L_\textrm{min},z)}{\partial \ln (1+z)}\right|_{L_\mathrm{min}=L_\mathrm{lim}(z)} \;,
\ee 
quantifies {how rapidly the number density of the selected galaxies changes} with redshift. Similarly, the magnification bias, 
\be 
\label{eq:mag_bias}
    \mc{Q}(z)=-\left.\frac{\partial \ln n(L_\textrm{min},z)}{\partial \ln L_\textrm{min}}\right|_{L_\mathrm{min}=L_\mathrm{lim}(z)}\,,
\ee
gives the slope of the cumulative luminosity function evaluated
at the luminosity limit of the survey. Taking into account all linear-order corrections {to} the ``observed'' galaxy density {(in redshift space)}, $n_{\rm g}$, and its angular average {at fixed $z_\mathrm{obs}$}, $\bar{n}_{\rm g}$, 
{it is possible to express} the overdensity $\delta_{\rm g, s}:= n_{\rm g}/\bar{n}_{\rm g}-1$ {in terms of the cosmological perturbations}
\citep{Yoo:2009au, Challinor:2011bk, Jeong:2011as, Bertacca:2015} %
\begin{align}
\label{eq:Deltag}
\delta_{\rm g, s}(\bs{x}) 
 &=  \delta^{\rm com}_{\rm g}-\frac{1}{\cH}\frac{\partial (\bs{\varv}_\mathrm{e}\cdot \bs{n})}{\partial x}+ 2\,(\mc{Q}-1) \,{\kappa}\nonumber\\
 \nonumber&+
 \left[ \mc{E}-2\mc{Q}  - \frac{\cH'}{\cH^2} - \frac{2(1-\mc{Q})\,c}{x\, \cH}\right]\\&%
 \times\left[\overc{\bs{\varv}_\mathrm{e}}\cdot \bs{n} - (\Phi_\mathrm{e}-\Phi_\mathrm{o}) - \int_0^x {(\Phi' + \Psi')\over c}\;\dif \tilde{x}\right]\nonumber 
 \\\nonumber
 & -2\,(1-\mc{Q})\,\Phi_\mathrm{e}+\Psi_{\rm e} + {\Phi_{\rm e}'\over \mc{H}} +  \left(3- \mc{E} \right)\frac{\mc{H}\velpot}{{c}^2} \nonumber\\
 & +\frac{2\,(1-\mc{Q})}{x} \int_0^x(\Phi+\Psi)\;\dif \tilde{x}\,\nonumber\\
 &+\left[2- \mc{E}  + \frac{\cH'}{\cH^2} + \frac{2\,(1-\mc{Q})\,c}{x\, \cH}\right]\overc{\bs{\varv}_\mathrm{o}}\cdot \bs{n}\;,
\end{align}
where $\velpot$ is the linear velocity potential (i.e., $\bs{\varv} = \nabla \velpot$). 
We note that the real-space galaxy overdensity $\delta^{\rm com}_{\rm g}$ is defined in the synchronous comoving gauge while all the rest is set in the Poisson gauge.\footnote{{The synchronous comoving galaxy overdensity is related to its Poisson gauge counterpart, $\delta^{\rm P}_{\rm g}$, via $\delta^{\rm P}_{\rm g} = \delta^{\rm com}_{\rm g} + \left(3-\mc{E}\right)\,\mc{H} \velpot/c^2$ \citep[e.g.][]{Jeong:2011as}.} 
}
{At linear order, $\delta^{\rm com}_{\rm g}$ is related to the underlying matter density fluctuation through the linear bias parameter $b$, i.e. $\delta^{\rm com}_{\rm g} = b\,\delta^{\rm com}_{\rm m}$ \citep{Challinor:2011bk,Jeong:2011as}. }

Equation~\eqref{eq:Deltag} defines what is meant by `relativistic (linear) RSD' and forms the starting point for our study.
In brief, it says that the galaxy overdensities in real and redshift space differ because of a number of physical effects.
The second term on the rhs is the classic Kaiser correction due to the variation of peculiar velocities along the line of sight \citep{Kaiser87}. 
The third term is the weak lensing contribution due to volume and magnification corrections which is expected to have an effect on different clustering statistics on large scales \citep[e.g.][]{Matsubara_2000, Hui_2007, Hui_2008,Yoo:2009au, Challinor:2011bk,CAMERA_RELATIVISTIC, Raccanelli+2016_GR_CORRECTIONS,MIKO_2017}.
There are then several additional corrections that depend on the gravitational potentials and the peculiar velocities.
Previous studies have shown that those proportional to the peculiar velocity of the
observer -- i.e. the last term in Eq.~\eqref{eq:Deltag} --
could generate observable features
in the 2PCF at wide angles \citep{Bertacca_2020_Rocket_effect} as well as superimpose an oscillatory signal to the power-spectrum monopole at very large scales \citep[][see also \citealt{Bahr-Kalus:2021jvu}]{Elkhashab_2021}, dubbed the finger-of-the-observer effect. 
\begin{figure*}
    \centering
    \includegraphics[width=1\textwidth]{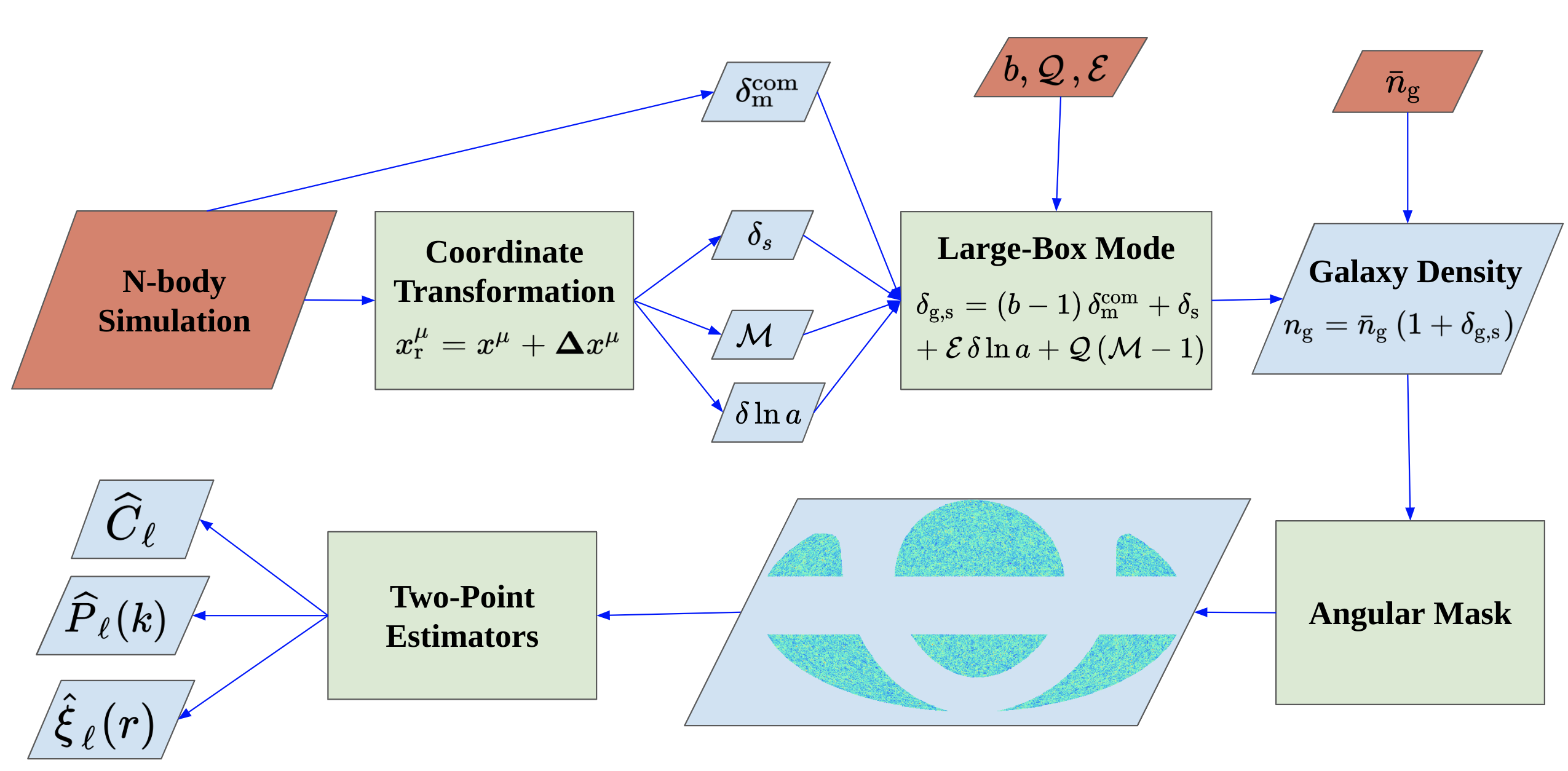}
    \caption{Flow chart of the \liger method in the `large-box' mode (top row) and of the clustering analysis performed in this paper (bottom row).
    Input and output are displayed as red and blue parallelograms, respectively, while processes are shown as green rectangles. }
    \label{fig:liger_schematic}
\end{figure*}

\section{Galaxy mocks}\label{sec:gal_mocks}
{In this paper, we use a suite of realistic mock galaxy catalogues to determine which of the corrections appearing in Eq.~\eqref{eq:Deltag} should be accounted for in the analysis of two-point clustering statistics extracted from
the EWSS. 
The main steps to generate the mock catalogues are described below
\cite[for further details, see][]{Elkhashab_2021}.
}

\subsection{The \liger method}
\label{sec:liger_theory}

{\liger \citep{MIKO_2017, Elkhashab_2021} is a numerical tool for building mock realizations of the %
galaxy distribution on the past light cone of an observer. \footnote{
A code implementation in C is publicly available at \url{https://astro.uni-bonn.de/~porciani/LIGER/}.}
As an input, it takes a Newtonian cosmological simulation. 
This can be either a hydrodynamic simulation or a $N$-body simulation combined with a semi-analytic model of galaxy formation to which one applies the selection criteria of a given survey.
However, resolving individual galaxies within very large comoving volumes is extremely challenging and time consuming with current software and facilities. Therefore, a special `large-box' mode has been developed
in which a Newtonian $N$-body simulation is used in combination with a set of functions describing the galaxy population under study (i.e.
their number density, linear bias parameter, magnification bias, and evolution bias).
The key argument underlying this approach is that, at linear order and for a pressureless fluid in a $\Lambda$CDM background, $\Psi=\Phi$ in the Poisson gauge. The potentials satisfy the standard Poisson equation and can be computed starting from the matter overdensity in the Newtonian simulations, which is equivalent to its counterpart in the synchronous comoving gauge, $\delta_{\rm m}^{\rm com}$ \citep[for more details, see Sect. 2.1.4 in][]{MIKO_2017}. }

We use \liger's large-box framework to produce the mock \Euclid catalogues.
A schematic diagram representing our workflow is presented
in Fig.~\ref{fig:liger_schematic}. 
{In this case,} \liger computes both the coordinate maps (Eq.~\ref{eq:shift_lig}) and the magnification (Eq.~\ref{eq:Mag_lig}) starting from the real-space position of {the particles} in the input $N$-body simulation. 
Then the code identifies the snapshots within which the backward light cone of the observer intersects the world lines {of the particles} after adding the displacement {given in} Eq.~(\ref{eq:shift_lig}). {We save} the intersection position {together with} the corresponding magnification and redshift change. 
\begin{figure}
    \centering
    \includegraphics[width=1\linewidth]{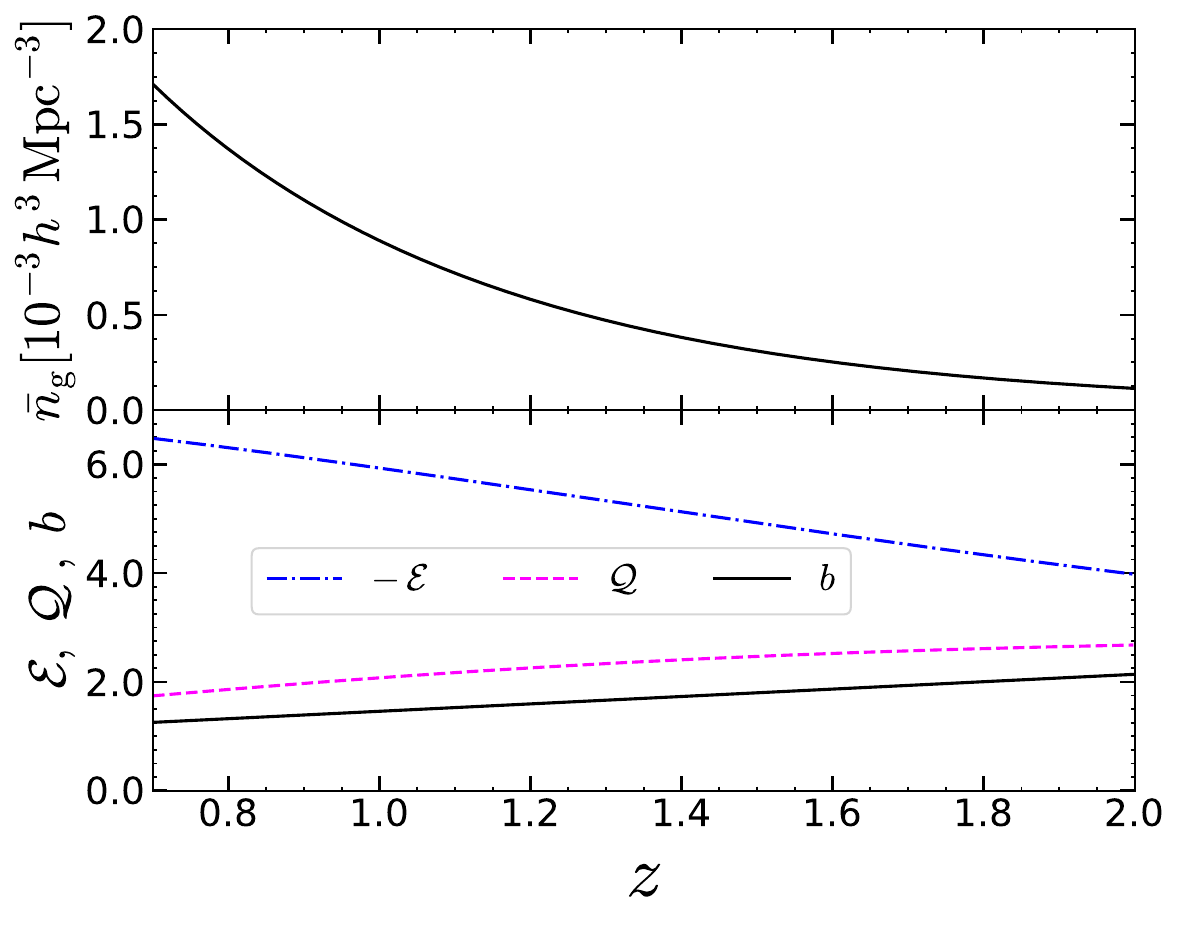}
    \caption{Top: The background number density of galaxies in the EWSS %
    as a function of redshift. Bottom: The corresponding evolution, magnification, and linear-bias parameters.}  
    \label{fig:SURVFUNCS}
\end{figure}

\begin{figure*}
    \centering
\includegraphics[width=0.9\textwidth]{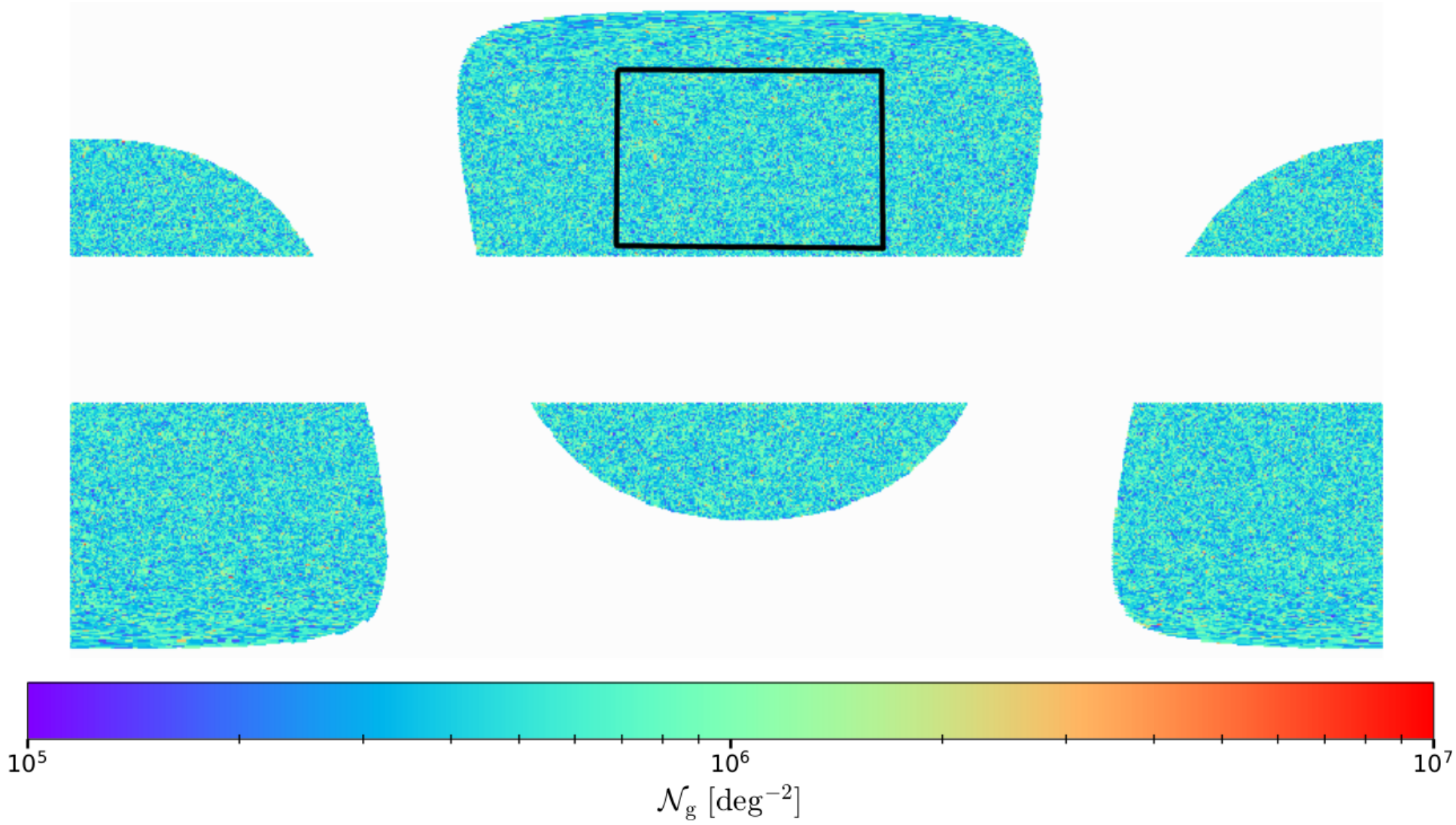}
    \caption{Projected galaxy number counts for one of the \vrsd mocks  in the $z\in(0.9,1.1)$ redshift bin. The black lines enclose the region used in Sect.~\ref{sec:disconnected_regions} as an example of a simply connected domain. }
    \label{fig:proj_den}
\end{figure*}
{The updated particle positions are used to compute the matter overdensity in redshift space,
$\delta_{\mathrm s}$.
Eventually, the galaxy distribution in redshift space is obtained using}
\begin{equation}
\label{eq:Buildcone}
    \delta_{\rm g,\rm{s}}=
    (b-1) \,\delta_{\rm m}^{\rm com}+\delta_{\mathrm s}+\mathcal{E}\, \delta \ln a+\mathcal{Q}\,(\mathcal{M}-1)\;,
\end{equation}
which matches Eq.~\eqref{eq:Deltag} under two assumptions.
Namely, 
({\it i}) $|{\cal H}\velpot|/{c^2}\ll|\delta\ln a|$;
({\it ii}) we can neglect the linear perturbation of $\sqrt{- g}$, where $g$ denotes the determinant of the metric tensor.\footnote{{At linear order, the perturbation of $\sqrt{- g}$ corresponds to the trace of the metric given in Eq.~\eqref{eq:metric}.} 
}
The {neglected} terms are only relevant at scales larger than {the Hubble radius} \citep[for more details, see][]{MIKO_2017}. Aside from these minor contributions, Eq.~\eqref{eq:Buildcone}  recovers the theoretical results obtained in Eq.~\eqref{eq:Deltag}.

\subsection{$N$-body simulations}\label{sec:sims}
We consider a flat $\Lambda$CDM background cosmological model based on the results from the \textit{Planck} mission \citep{planck18} with
matter density parameter $\Omega_{\rm m} = 0.3158$, baryon density parameter $\Omega_{\rm b} = 0.0508$, and dimensionless Hubble constant $h = 0.673$. 
We also assume that primordial scalar perturbations form a Gaussian
random field with a linear power spectrum of power-law shape
characterized by the spectral index $n_\mathrm{s}=0.966$ and the amplitude
$A_\mathrm{s}=2.1\times 10^{-9}$ (defined at the wavenumber $k_*=0.05$ Mpc$^{-1}$). 
We compute the matter transfer function using the \texttt{CAMB} code \citep{CAMBS}. 

In order to encompass the full EWSS within each of our simulation boxes, we study structure formation within periodic
cubic volumes with a comoving side length of $L_{\rm{box}} = 12 \,h^{-1}$Gpc.
As we are only interested in quasi-linear scales,
we use second-order Lagrangian perturbation theory (2LPT) to build the dark-matter distribution that forms the input to \liger.
For this step, we apply the \texttt{MUSIC} code \citep{music} to
$1024^3$ equal-mass particles which, initially, form a regular Cartesian grid.
Overdensities are computed with the classical cloud-in-cell scheme
using the same grid. The gravitational potential is obtained by solving
the Poisson equation with spectral methods \citep{Hockney-Eastwood}.

We run 35 independent $N$-body simulations and
extract four non-overlapping lightcones from each of them, resulting in a total of $N_{\rm mocks} = 140$ mock catalogues. 

\subsection{\Euclid H$\alpha$ galaxies}\label{sec:bias_funcs}
Applying Eq.~\eqref{eq:Buildcone} to \Euclid requires knowledge of the functions $b, \mc{E}$, and $\mc{Q}$ for the H$\alpha$ galaxies targeted by the EWSS.
In the absence of accurate data, we assume that model 3 in \cite{pozzetti16} provides an accurate description of the luminosity function as suggested by recent observations \citep{bagley20}.
Considering a flux limit of $F_{\rm lim} = 2\times10^{-16} \,{\rm erg}\, \rm{cm}^{-2}\, \rm{s}^{-1}$ \citep{Euclid_Wide_survey}, 
we compute $\bar{n}_{\rm g}$ by integrating the luminosity function 
and assuming a (uniform) completeness factor of $70\%$.
We derive the evolution and magnification bias factors using Eqs.~(\ref{eq:EV_bias}) and~(\ref{eq:mag_bias}).
Finally, for the linear bias coefficient, we adopt the linear relation $b(z)=1.46+0.68\,(z-1)$ obtained by fitting the data from Table 3 in \citet[][]{EuclidVII}.
All our results are presented in Fig.~\ref{fig:SURVFUNCS}.

\subsection{Building the mock catalogues}\label{sec:catalog_types}

For each of the 140 light cones, we build four different mock catalogues that progressively include an increasing number of relativistic RSD.
We first generate the galaxy distribution in real space (hereafter denoted by \real). Second, we include the RSD due to the peculiar velocities of
the distant galaxies -- i.e. the terms depending on $\bvr{e}$ in Eqs.~(\ref{eq:shift_lig}) and (\ref{eq:Mag_lig}) -- setting however $\mc{Q}$ and $\mc{E}$ to zero in Eq.~(\ref{eq:Buildcone}) and we dub the corresponding catalogues \vrsd.
Next, we consider all relativistic RSD except those due to the observer peculiar velocity $\bvr{o}$ (from now on the \grsd mocks).
Finally, we include all terms to generate the \obs catalogues. 
In the latter set, we assume that $\bvr{o}$ coincides with the peculiar
velocity of the Sun as derived from the CMB dipole \citep{planck-dipole-18}.

To produce catalogues of discrete galaxies, we proceed as follows.
Based on $\delta_\mathrm{g,s}$ and $\bar{n}_\mathrm{g}$,
we first compute the expected number of galaxies $N_\mathrm{g}$ in each volume element of the lightcone. We then draw %
{from a Poisson distribution} with mean $N_\mathrm{g}$ and randomly distribute the corresponding number of galaxies within the cell.

After taking all these steps, we obtain a full-sky galaxy catalogue covering the redshift range $0.9<z<1.8$. In order to mimic the
expected angular distribution of the EWSS, we mask $20^\circ$ around the Galactic and Ecliptic planes {as shown in Fig.~\ref{fig:proj_den}}.\footnote{{We do not simulate direction-dependent incompleteness (for instance due to Galactic extinction, \EuclidAngSystematics) which {affect the clustering summary statistics at large scales.} These effects can be mitigated when estimating the clustering statistics
\citep[e.g.][]{Burden_2017,Paviot_2022}}} We measure clustering statistics in four tomographic redshift bins
with boundaries $z\in\{0.9,1.1,1.3,1.5,1.8\}$  and, particularly for the angular power spectrum, in a broader bin that covers the entire redshift range covered by the mock catalogues. The  binning strategy used in this work has been chosen to reproduce that in \citet[][]{EuclidVII}.

\subsection{Random catalogues}
\label{sec:catalog}
Estimating two-point statistics requires generating unclustered distributions of points with the same angular footprint and radial selection function
as the actual galaxy data (see Sect.~\ref{sec:2pcf_est} {and} Sect.~\ref{sec:power_est}). We build a `random catalogue' for each of our mocks in three steps. We first measure the mean galaxy number counts %
within radial shells of $20\,\hMpcc$ width. We then interpolate the results with a cubic spline to obtain the cumulative
redshift distribution of the galaxies. Finally, we use the inverse transform method to pick a redshift for the unclustered points, which are also assigned a random line-of-sight direction. The random catalogues contain five times more galaxies than the original light cones. %
 These catalogues are smaller than those typically employed in power-spectrum estimation. Increasing the size of the random catalogue reduces the shot noise contribution to  the statistical error of the power spectrum \citep{FKP}. However, at the large scales considered in this work, the statistical error is dominated by the sample variance. Thus, using a smaller catalogue has minimal impact on our results while significantly reducing computational overhead.

\section{Statistical methods}
\label{sec:chisq}

Given a clustering statistic $S$, we want to understand 
whether the contribution of specific relativistic RSD
to the measured signal is detectable or not with the EWSS.
For instance, the impact of the peculiar velocity of the observer
can be quantified by comparing the clustering statistic extracted
from our \obs and \grsd mock catalogues. Similarly, by comparing the \obs and \vrsd light cones we can also study the relevance of the
the weak lensing contribution. 

Let us denote by $\bs{D}_i^{a}$ the $n$-dimensional (column) data vector containing
the measurements of the clustering statistics in a particular mock (characterized by the index $i$) of type $a\in\{$\real, \vrsd, \grsd, \obs$\}$. By taking the expectation over the 140 realisations, we compute
the mean signal and the covariance matrix %
in the measurements,

\begin{align}
 \bs{\mu}_a &= \mathbb{E}[\bs{D}_i^{a}]\,,\\
 \tens{C}_a &= \mathbb{E}[(\bs{D}_i^{a}-\bs{\mu}_a)(\bs{D}_i^{a}-\bs{\mu}_a)^{\mathrm{T}}]\,\label{Eq:COV_MODEL}.
 \end{align}
Based on the Fisher-information matrix, we can then estimate the signal-to-noise ratio for the detection of the RSD that are not included in the $a$ mocks by using \citep[see e.g. Sect. 3.3.3 in][]{MIKO_2017}
\begin{equation}
(\mathrm{S}/\mathrm{N})^2=    (\bs{\mu}_\mathcal{O}-\bs{\mu}_a)^{\mathrm{T}}\,\tens{C}_\mathcal{O}^{-1} \,(\bs{\mu}_\mathcal{O}-\bs{\mu}_a)=: m_0\,,
\label{eq:Fisher}
\end{equation}
where we correct for the bias of the inverse covariance matrix
due to the finite number of realisations used to estimate it \citep{Kaufman67,Hartlap+2007}.

Classical hypothesis testing based on the likelihood function provides another possibility to quantify the detectability of the different RSD terms. Assuming Gaussian errors, the likelihood that the dataset $\bs{D}_i^{\obsm}$ is drawn from a model $M_a$ with signal $\bs{\mu_a}$ is
  \be
 \label{eq:likeli}
 L(M_a|\bs{D}_i^{\obsm})  \propto {\exp\left[-(\vec{D}_i^{\obsm}-\bs{\mu}_a)^{\rm T} \,\tens{C}^{-1}_a\,(\bs{D}_i^{\obsm}-\bs{\mu}_a)/2\right]\over (2\,\pi)^{n/2}\det \tens{C}_a }\,.
 \ee
  We use the term `model' to indicate a prediction for $S$ with fixed values of the parameters that describe the galaxy population and the underlying cosmology. No fitting of model parameters is considered here. Basically, a model corresponds to an infinite ensemble of mocks all including the
same RSD terms (for instance, the \grsd mocks) and is described by the corresponding signal and noise covariance.

 Let us now formulate the null hypothesis $\mc{H}_0$ that the \Euclid data
 are a realisation of model $M_a$ that does not include all RSD terms
 present in $M_\mathcal{O}$. We want to test this assumption against the alternative hypothesis $\mc{H}_1$ that the data are drawn from $M_\mathcal{O}$.
 The Neyman--Pearson lemma states that the likelihood-ratio test statistic 
 \begin{align}
\lambda_i = 2\ln\left[{L(M_a|\bs{D}_i^{\obsm})\over L(M_\obsm|\bs{D}_i^{\obsm})}\right]\,%
\end{align}
provides the most powerful test for two simple hypotheses (i.e. with fixed model parameters). The null hypothesis is rejected with confidence level $\alpha$ if $\lambda_i<\omega_\alpha$, where $\omega_\alpha$
is a real number such that the probability $\mc{P}(\lambda_i<\omega_\alpha|\mc{H}_0)=1- \alpha$.
Under $\mc{H}_0$, $\mc{P}(\lambda_i|\mc{H}_0)$ is Gaussian with
mean $m_0\geq 0$ 
and variance $s_0^2=4m_0$ \citep[see Appendix A in][]{MIKO_2017}. Adopting a 95\% confidence level, we thus obtain $\omega_{95}=m_0-3.29 \sqrt{m_0}$.
Similarly,
under $\mc{H}_1$, $\lambda_i$ follows a Gaussian distribution with
mean $m=-(\mu_\obsm-\mu_a)^{\rm T}\tens{C}^{-1}_a (\mu_\obsm-\mu_a)\leq 0$ and variance $s^2=4 |m|$. Therefore, $\mc{H}_0$ is rejected in a fraction
\be 
f_{95}={\frac{1}{2}}\left[1+\mathrm{erf}\left(\frac{\omega_{95}-m}{2\sqrt{2|m|}}\right)\right]
\ee
of the realisations. 
If we neglect the small difference
between the covariance matrices, we find that $m=-m_0$ and $s=s_0=4m_0$.\footnote{This is an excellent approximation given that the RSD corrections we are considering are
relatively small.} 
In this case, the separation between $m$ and $m_0$ expressed in units
of the standard deviation of the distributions is
$(m_0-m)/(2\sqrt{m_0})=\sqrt{m_0}$, which coincides with the signal-to-noise ratio given in Eq.~(\ref{eq:Fisher}). The fraction of 
realisations in which the data reject $\mc{H}_0$ is thus $f_{95}=\{1+\mathrm{erf}[(\sqrt{m_0}-1.645)/\sqrt{2}]\}/2$.
Note that \snrtval{1} gives $f_{95}=0.259$, while
$f_{95}=0.5$ and $0.9$ correspond to \snrtval{1.645} and $2.93$, respectively.

\begin{figure}
    \centering
    \includegraphics[width=0.9\linewidth]{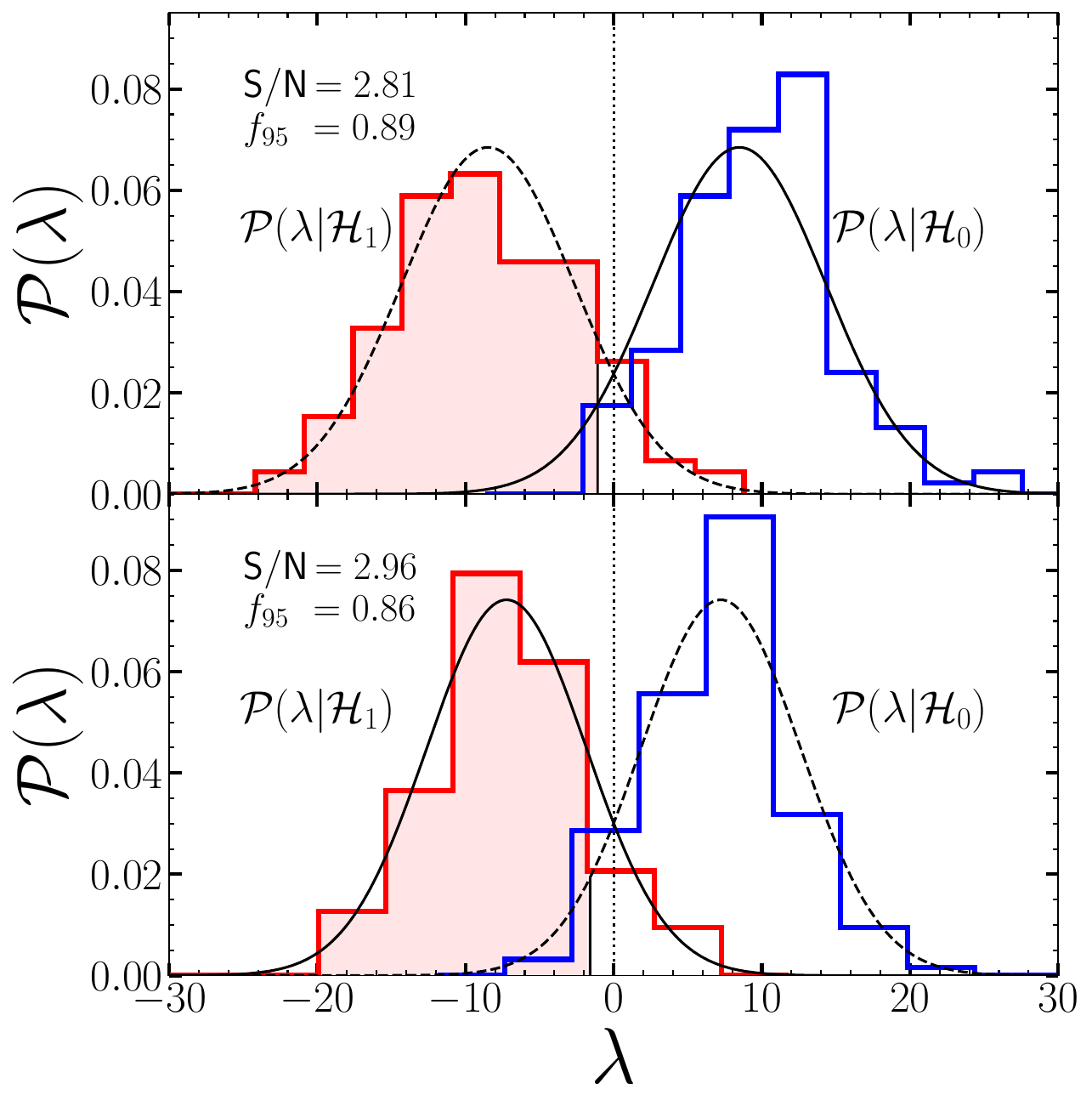}
    \caption{{An elucidatory example of the statistical test
    outlined in Sect.~\ref{sec:chisq}. In the top panel, the blue histogram on the right-hand
    side shows the distribution of the likelihood-ratio
    test statistic $\lambda$ evaluated from the mock catalogues that do not include all relativistic RSD (i.e. under the null hypothesis $\mc{H}_0$ that we try to rule out using the observed data). The red histogram on the left-hand side, instead, displays the distribution
    of $\lambda$ in the mock light cones that account for all effects (i.e. under the alternative hypothesis $\mc{H}_1$). The solid and dashed
    curves represent Gaussian models for the histograms as described in
    the main text. The \snrt is a measure of 
    the separation between the two histograms in units of their RMS scatter.
    The shaded region highlights 
    the realisations in which $\mc{H}_0$ is ruled out at the 95\% confidence level. The bottom panel only differs from the top one in the fact that the covariance matrix $\tens{C}_\mathcal{O}$ has been used
    to compute all likelihood functions. We use this approximation    in the remainder of this paper.}}
    \label{fig:Examble_lam}
\end{figure}
In the remainder of this paper, we apply these statistical tests
to our mock \Euclid light cones. As anticipated at the beginning of this section, by comparing the \obs and \grsd sets
(hereafter \vobstest test) we assess the detectability of the RSD generated by the peculiar velocity of the Sun. The goal here is to
determine if and how often we manage to reject the null hypothesis that the \obs data are drawn from an ensemble with $\bs{\varv}_\mathrm{o}=0$.
In addition, we apply the likelihood-ratio test to the \obs and
\vrsd mocks (hereafter \WLtest test). In this case, we aim to quantify
whether the EWSS can reject the null hypothesis that the influence of the integrated terms (dominated by the weak lensing contribution) on the clustering signal is negligible.

In order to perform these tests, we directly compute histograms of the values of $\lambda_i$ derived from the 140 \obs mock catalogues. %
In all cases,
we only use the $\tens{C}_\mathcal{O}$ covariance. This operation mimics what is usually done in the analysis of actual surveys, where the covariance
is estimated from suites of mock catalogues and not changed with the theoretical models for which the likelihood is evaluated.
We compute the signal-to-noise ratio as
\be
\label{eq:SN}
\text{S$/$N}=2\hat{m}_0/\hat{s}_0\;,
\ee
where the hat denotes estimates derived from the 140 mocks and determine
$f_{95}$ by counting the number of realisations in which $\lambda_i<\omega_{95}$. In order to facilitate the understanding of the likelihood-ratio test for two simple hypotheses, we present an illustrative example in Fig.~\ref{fig:Examble_lam}.

\section{Angular power spectrum}
\label{sec:ang_power_all}
{In this section, we investigate the}
impact of relativistic RSD on the angular power spectrum, $C_\ell$.

\subsection{Estimator}

We make use of the Hierarchical Equal Area isoLatitude Pixelisation  \citep[\texttt{HEALPix}][]{HEALpix,Zonca2019} algorithm\footnote{The \ttt{HEALPix/healpy} software package is available in \url{http://healpix.sourceforge.net}.} to {partition} the sky into $N_{\rm pix} = 12\,\times\,(1024)^2$ pixels.
{After dividing}
our mock light cones into multiple redshift bins, {we compute the projected number counts of galaxies in each bin (indicated by the superscript $i$) and pixel,} $\mc{N}^i_{\rm g}(\Omega)$.
A sample sky map with the \Euclid mask is shown in Fig.~\ref{fig:proj_den} using a cylindrical Cartesian coordinate system. 
The projected density contrast is then
\be
\Sigma_{\rm g}^i (\Omega) = {\mc{N}^i_{\rm g}(\Omega)\over \bar{\mc{N}}^i_{\rm g}}-1\,,
\ee
where $\bar{\mc{N}}^i_{\rm g}$
denotes the average of $\mc{N}^i_{\rm g}(\Omega)$ over the solid angle subtended by the survey.
\begin{figure}
    \centering
    \includegraphics[width=0.8\linewidth]{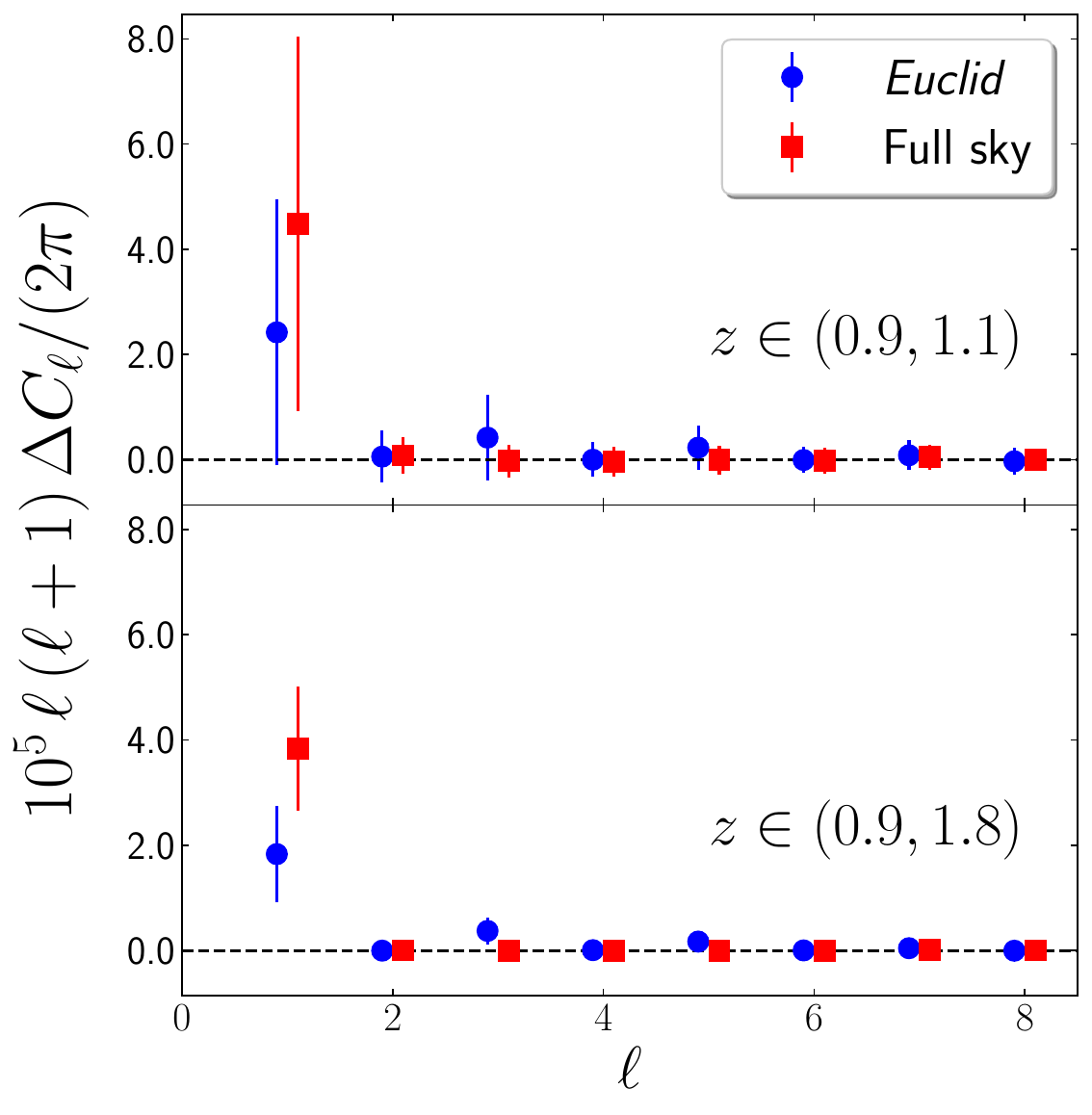}
    \caption{{The difference between the angular power spectra of the \obs and  \grsd mocks for a full-sky survey (red) and the EWSS (blue).
    The symbols indicate the mean signal while the errorbars show the standard deviation over the 140 realisations.} }
    \label{fig:DIFFCLSVOBS}
\end{figure}
\begin{figure*}
    \centering
    \includegraphics[width=0.95\linewidth]{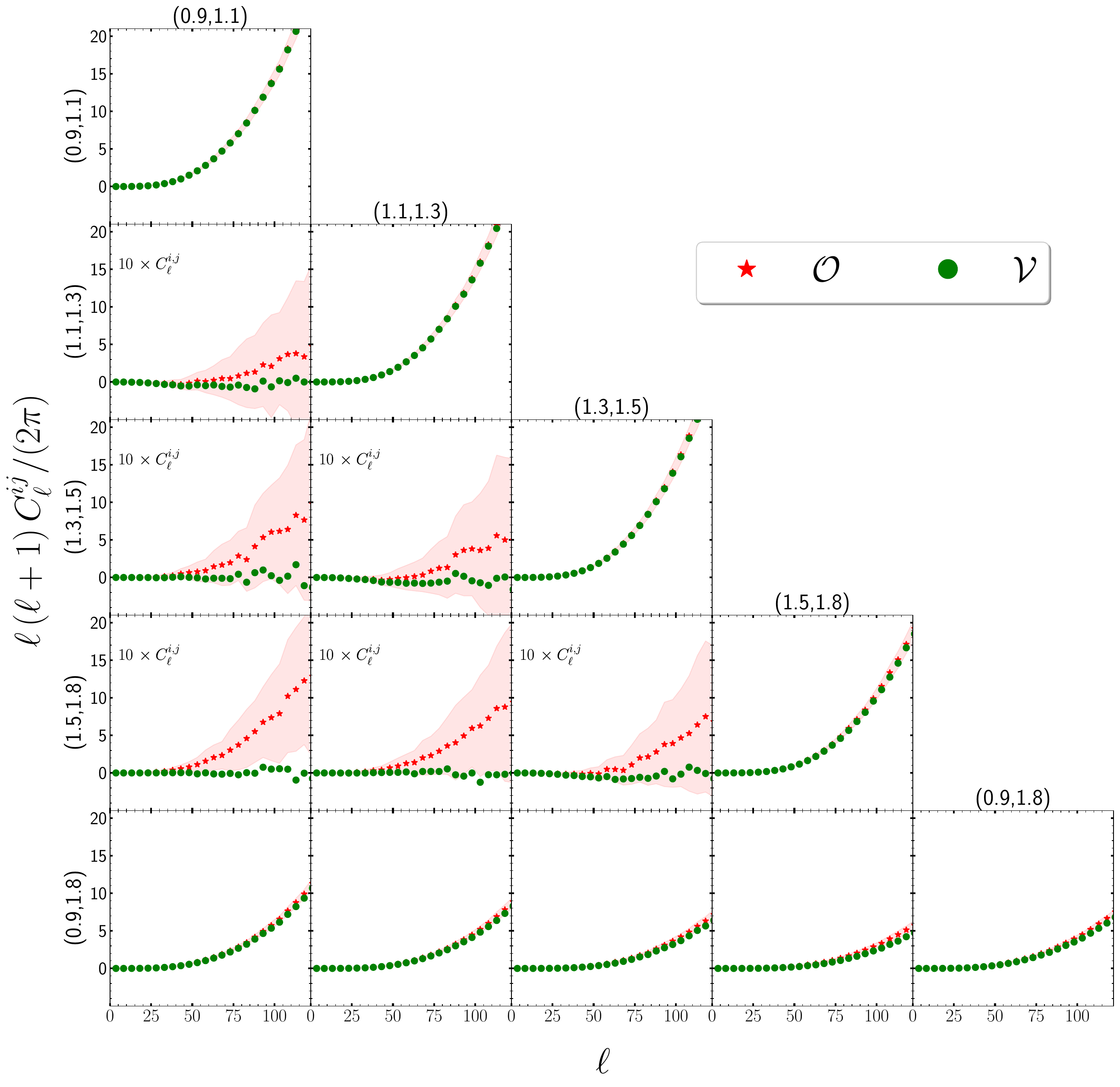}
    \caption{{Mean auto-and cross-angular power spectra extracted
    from the \obs (red stars) and \vrsd (green circles) mocks. The shaded region indicates the RMS scatter of the \obs spectra. The panels are ordered as the entries in Tables~\ref{tab:S_N_TABLE_CLS_O_V} and \ref{tab:S_N_TABLE_CLS_G_V}. The cross-spectra of the tomographic redshift bins are multiplied by ten to improve the readability of the figure.}}
    \label{fig:CLS_FULL}
\end{figure*}

We {expand} the projected overdensities in spherical harmonics 
\begin{equation}
\label{EQ:SPH_HARMON}
    a^i_{\ell m } = \int \Sigma^i_{\rm g}(\Omega)\, Y^*_{\ell m} (\Omega) \;\dif^2 \Omega\,
\end{equation}
{and measure}
the angular {auto- and cross}-power spectra using the pseudo-$C_{\ell}$ (PCL) estimator \citep{Peebles-1973,Loureiro+2019}
\be
\label{eq:estimator_cl}
\hat{C}^{\,i\,j}_{\ell} = \frac{1}{w^2_{\rm p}\,(2\ell + 1)\,f_{\rm sky}} \sum ^\ell_{m=-\ell} a^i_{\ell m}\;{a^j_{\ell m}}^* -{\delta_{{\rm K}\,ij}\over \bar{\mc{N}}^{i}_{\rm g}}\,,
\ee
where $f_{\rm sky}$ {denotes} the fraction of the sky covered by the survey and $w_{\rm p}$ is a correction factor due to the finite pixelisation  of the sphere (see {the} \ttt{HEALPix} documentation for more details). 
In App.~\ref{Sec:Validation},  we present a validation test of our pipeline for producing the mock catalogues and measuring the angular power spectra.

\subsection{Results} 

\label{Sec:ang-corr-results}
\subsubsection{Peculiar velocity of the observer}
{In Eq.~\eqref{eq:Deltag}, a dipolar pattern is superimposed
to the galaxy density contrast whenever the observer is not comoving
with the cosmic expansion traced by matter \citep[see also][]{Gibelyou_2012}. %
Figure~\ref{fig:DIFFCLSVOBS} displays how this kinematic dipole alters
the angular power spectrum. Shown are the average and root-mean-square (RMS) scatter
over the 140 mocks of the}
difference between the \obs and \grsd spectra, i.e. $\Delta \hat{C}_\ell :=\hat{C}_{\ell,\text{\obs}}-\hat{C}_{\ell,\text{\grsd}}$, 
for two different redshift bins. 
It is evident that, in a full-sky survey (red symbols),  only the dipole ($\ell=1$) is affected by $\bs{\varv}_\mathrm{o}$. On the other hand, the correction   
spreads over all the odd multipoles ($\ell=3,5,\dots$) in our \Euclid mocks due to the partial sky coverage of the catalogues. Expanding the projected overdensity in spherical harmonics (Eq.~\ref{EQ:SPH_HARMON}) on a partial sky, where the base functions are no longer orthogonal, results in mode-mixing in the estimated angular power spectrum \citep{Peebles-1973}. Consequently, the dipole signal leaks into higher odd multipoles. 

\begingroup

\setlength{\tabcolsep}{2pt} %
\renewcommand{\arraystretch}{1.3} %
\begin{table}
{
	\begin{center}
	\caption{{\snrt from the \vobstest test applied to $\hat{C}_\ell^{ij}$ with $\ell\in[1,10] $.}}
	\label{tab:S_N_TABLE_CLS_O_V}
  \begin{tabular}{cccccc}
  								
$(z_\text{min},z_\text{max})$
            &$(0.9,1.1)$&$(1.1,1.3)$&$(1.3,1.5)$&$(1.5,1.8)$&$(0.9,1.8)$\\
$(0.9,1.1)$ &0.9        &          &           &           &         \\
$(1.1,1.3)$ &1.3        & 0.7     &           &           &         \\
$(1.3,1.5)$ &1.1        & 1.1     &  0.8     &           &         \\ 
$(1.5,1.8)$ &1.2        & 1.1     &  1.2     &  1.0     &         \\
\hline
$(0.9,1.8)$ &1.4        & 1.4     &  1.5     &  1.6     & 2.1   \\
\hline
  \end{tabular}
  \end{center}
  }
  \vspace{-.4cm}
\end{table}
\endgroup

{In order to quantify how well the modification due to $\bs{\varv}_\mathrm{o}$ can be detected with the EWSS, we apply the \vobstest test introduced in Sect.~\ref{sec:chisq}
to the first ten multipoles.}
The resultant \snrt values for the  auto- and cross-spectra 
are shown in Table~\ref{tab:S_N_TABLE_CLS_O_V}.
For the tomographic redshift bins, the excess clustering induced by $\bs{\varv}_\mathrm{o}$ can be barely identified with a signal-to-noise ratio of order one. This characteristic signature becomes much more discernible in the auto-correlation function evaluated after projecting the galaxies from the broad redshift interval $z\in(0.9,1.8)$. In this case, we obtain \snrtval{2.1}.%

\subsubsection{Weak lensing}
\begingroup
\setlength{\tabcolsep}{2pt} %
\renewcommand{\arraystretch}{1.3} %
\begin{table}
{
	\begin{center}
	\caption{{As in Table~\ref{tab:S_N_TABLE_CLS_O_V} but for the \WLtest test with $\ell \in [1,120]$.}}
	\label{tab:S_N_TABLE_CLS_G_V}
      \begin{tabular}{cccccc}
             $(z_\text{min},z_\text{max})$
            &$(0.9,1.1)$&$(1.1,1.3)$&$(1.3,1.5)$&$(1.5,1.8)$&$(0.9,1.8)$\\
             $(0.9,1.1)$ &0.9        &          &           &          &\\
             $(1.1,1.3)$ &1.7        & 0.8     &            &           &\\
             $(1.3,1.5)$ &3.0        & 2.0     &  1.1      &            &\\ 
             $(1.5,1.8)$ &4.5        & 3.4     &  2.8     &  2.3       &\\
            \hline
             $(0.9,1.8)$ &3.3        & 2.9     &  4.1     &  5.9     & 5.4   \\
\hline
  \end{tabular}
  \end{center}
  }
  \vspace{-.4cm}
\end{table}
\endgroup

{Figure~\ref{fig:CLS_FULL} displays the angular auto- and cross-power spectra extracted from the \vrsd and \obs mocks and re-binned with $\Delta \ell=5$. The symbols show the average signal
over the 140 realisations and the shaded region indicates the RMS
scatter for the \obs measurements. Since 2LPT at $z\simeq 1$ underestimates the non-linear matter power spectrum for wavenumbers $k \gtrsim 0.05\,h\,$Mpc$^{-1}$ \citep[e.g.][]{Taruya+2018},
we only consider harmonics of degree $\ell\leq120$.
The positioning
of the panels is as in Tables~\ref{tab:S_N_TABLE_CLS_O_V} and \ref{tab:S_N_TABLE_CLS_G_V}. 
The cross-spectra between non-overlapping redshift bins are consistent with zero for the \vrsd mocks and show a positive clustering signal for the \obs catalogues. This difference}
is due to the integral terms in Eq.~(\ref{eq:Deltag}), in particular to the dominant  weak lensing  
contribution. The auto-spectra also show enhanced clustering for the \obs mocks.
This is particularly evident in the bins that include galaxies
with the highest redshifts.

{The \snrt values obtained with the \WLtest test are reported in  
Table~\ref{tab:S_N_TABLE_CLS_G_V}. The detection of the lensing term
is highly significant in the cross-correlations of well separated tomographic redshift bins and in all statistics involving the wide
bin $z\in(0.9,1.8)$.}

\begin{figure*}
    \centering
    \includegraphics[width=0.9\linewidth]{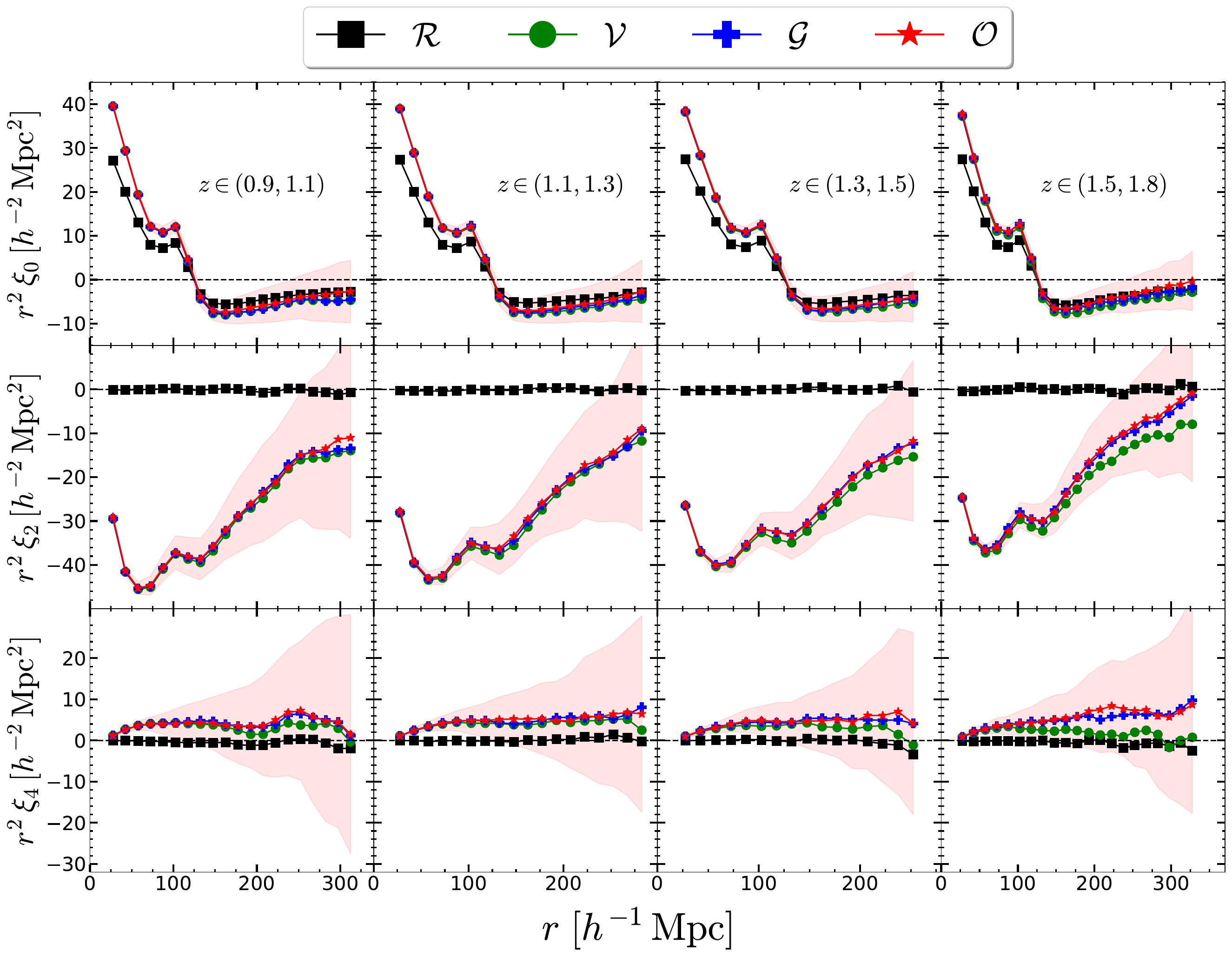}
    \caption{{Mean $\ell=0, 2$, and 4 multipoles of the 2PCF measured from
    the \real (black squares), \vrsd (green circles), \grsd (blue crosses), and \obs (red stars) mocks in the four tomographic redshift bins. The shaded areas highlight
    the RMS scatter among the \obs light cones.
    }}
    \label{fig:XI_CORR_NOV}
\end{figure*}

\section{Two-point correlation function}\label{Sec:2PCF_ALL}

The 2PCF is one of the
most employed summary statistics to extract cosmological information from the large-scale structure of the Universe.
In terms of the galaxy overdensity, it can be defined as
\citep{Peebles1980}
\be
\label{eq:xi_rsd}\left\langle{\delta_{\rm g}(\bs{x}_1)\,\delta_{\rm g}(\bs{x}_2)}\right\rangle=\xi_{\rm g}(\bs{x}_1,\bs{x}_2)\,,
\ee
where the brackets denote the average over an ensemble of realisations.
In real space, we expect that
$\delta_{\rm g}$ is a statistically homogeneous  and isotropic random field, so that
$\xi_{\rm g}$ depends only on the magnitude of the separation between the points at which it is evaluated. 
{RSD, however, break}
the translational and rotational symmetry of the 2PCF (as several terms in Eq.~\ref{eq:Deltag} depend on the
line-of-sight direction with respect to the observer)
into an azimuthal symmetry with respect to the line of sight.
The 2PCF in redshift space thus depends on the shape and size of
the (possibly non-Euclidean) triangle formed by the observer and the galaxy pair \citep{Szalay+1998, Matsubara_2000_2PCF}.

\begin{figure*}
    \centering
    \includegraphics[width=0.95\linewidth]{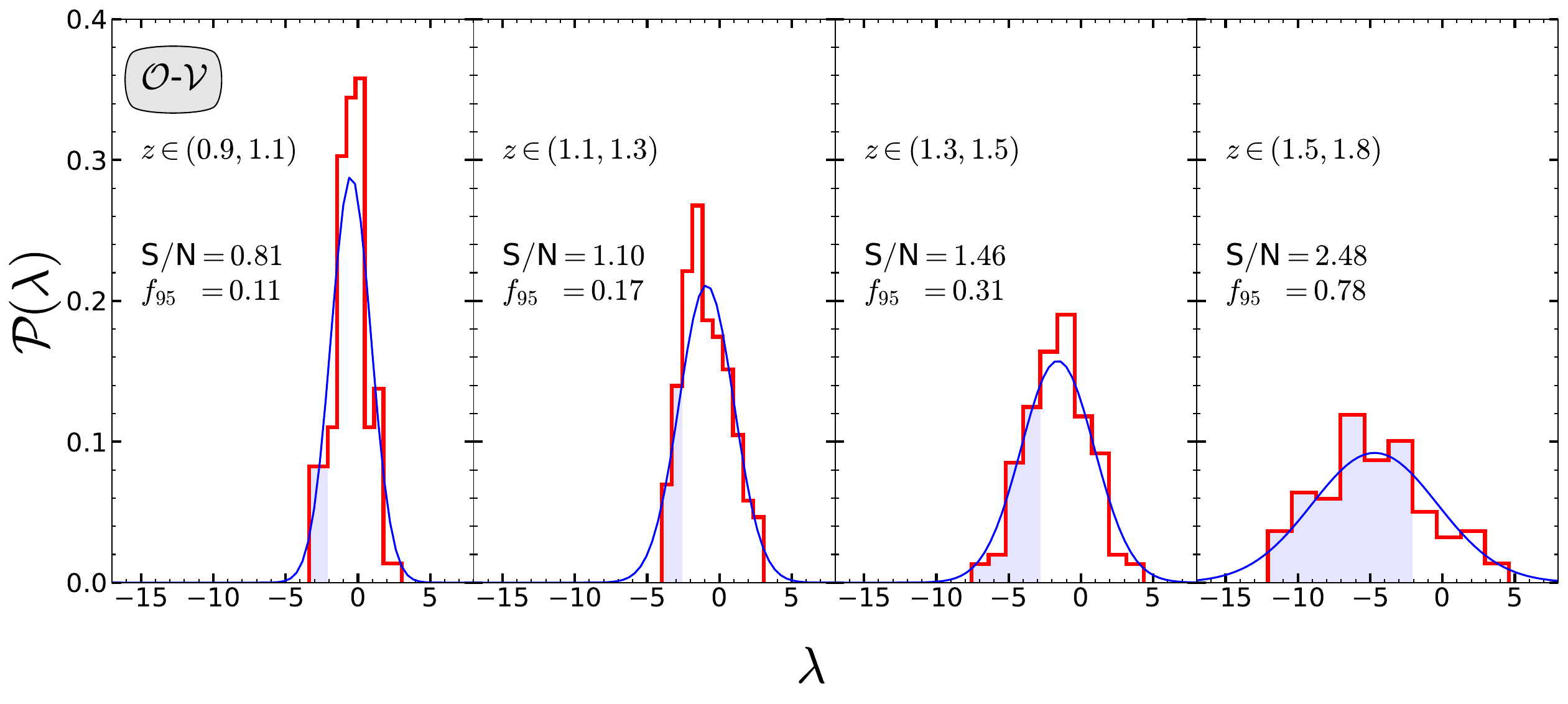}
    \caption{{PDF of the likelihood-ratio test statistic for
    the \WLtest test applied to the multipoles of the 2PCF.}}
    \label{fig:XI_CORR_HIST}
\end{figure*}

\subsection{Estimator}\label{subsection:conf-Estimator}
\label{sec:2pcf_est}
{We use the Landy--Szalay (LS) estimator \citep{Landy-Szalay98} as
implemented in the official \Euclid code \euclidtwopcf~which adopts}
the midpoint coordinate system
\be
\bs{r} = \bs{x}_2 - \bs{x}_1\,\,\,,\,\,\,
\bs{x}_{\mathrm{m}} = {\bs{x}_2 + \bs{x}_1\over 2}\,,
\ee 
and 
defines the pair-orientation angle $\varphi$ {through}
\begin{equation}
\mu =\cos{\varphi}= \hat{\bs{r}}\cdot \hat{ \bs{x}}_{\mathrm{m}}\,,      
\end{equation}
where the hat denotes the unit vector, i.e. $\bshat{r}:=\bs{r}/r $. \footnote{
Many studies of wide-angle effects %
use a different convention %
which defines the line of sight as the direction bisecting the angle formed by the two galaxies \citep[e.g. ][]{Szalay+1998,Matsubara_2000_2PCF,Szapudi_2004,Raccanelli_2010,Samushia_2012, Bertacca_2012}. For small angles, this differs from the midpoint method
at $\mathcal{O}(\varphi^2)$.}
After averaging over $\bs{x}_{\mathrm{m}}$ within 
{the volume of interest,} the code outputs
an estimate of the 2PCF as a function of the galaxy separation and 
orientation with respect to the local line of sight, $\hat{\xi}(r,\mu)$. 
It also computes the  2PCF Legendre multipoles 
\be 
\label{eq:multi_LS}
\hat{\xi}_{\ell}(r)  :={2\ell +1 \over 2} \int_{-1}^{1}  \,\hat{\xi}(r,\mu)\,\mc{L}_{\ell}(\mu)\,\dif \mu\,,
\ee
where $\mc{L}_{\ell}(\mu)$ denotes the Legendre polynomial of degree $\ell$. 
In what follows, we only consider the monopole ($\ell=0$),
quadrupole ($\ell=2$), and hexadecapole ($\ell=4$) moments.
These are the only non-vanishing moments in Kaiser's GPP model and dominate the signal in general. 

We originally estimate them in 500 equally spaced bins covering
the range $r\in [20\,\hMpcc,\, \Delta x]$ where $\Delta x$ denotes the comoving radial width of each redshift bin. This choice prevents that
the clustering signal is dominated by galaxy pairs with particular angular separations and thus limits window-function effects.
We re-bin our results in different ways depending on our applications. For instance, we use 10 equally separated bins in the range $r\in[35\,\hMpcc,\,\Delta x]$ for each multipole to perform the \vobstest and \WLtest tests.

\subsection{Results}
\begin{table}
\label{tab:2PCF}
	\begin{center}
	\caption{%
 {\snrt from the \WLtest and \vobstest tests for the multipoles
 of the 2PCF.}}
	\label{tab:S_N_TABLE_2PCF}
  \begin{tabular}{ccc}
   $(z_\text{min},z_\text{max})$&{\WLtest }&{\vobstest}\\
       \hline%
$(0.9,1.1)$  & 0.8 & 0.6    \\
$(1.1,1.3)$  & 1.1 & 0.7    \\
$(1.3,1.5)$  & 1.5 & 0.6   \\
$(1.5,1.8)$  & 2.5 & 1.0   \\
\hline
\hline
  \end{tabular}
  \end{center}
\end{table}

\label{sec:2PCF_results}

{Figure~\ref{fig:XI_CORR_NOV} shows the mean multipoles of the 2PCF obtained from the different sets of mock catalogues. The shaded areas indicate the scatter for the \obs light cones.
The real-space monopole moment is positive at small separations, present the baryonic-oscillation feature at $r\simeq 100 \, \hMpcc$, and crosses zero at $r\simeq 125 \, \hMpcc$ while $\hat{\xi}_{2}$ and $\hat{\xi}_{4}$ vanish as expected. RSD enhance the clustering signal in $\hat{\xi}_{0}$ and generate a negative $\hat{\xi}_{2}$ and a positive $\hat{\xi}_{4}$.
}
\subsubsection{Peculiar velocity of the observer}
{The clustering signal extracted from the \obs (red stars) and \grsd (blue crosses) mocks is hardly distinguishable at all scales in
all tomographic redshift bins. This visual impression is confirmed
by the \vobstest test which consistently gives values of  $\text{S/N}\leq 1$ (see Table ~\ref{tab:S_N_TABLE_2PCF}) for all redshift bins. At first sight, this appears to be at %
odds with the results by \cite{Bertacca_2020} who predict stronger corrections due to $\bs{\varv}_\mathrm{o}$. However, this study pushes the analysis to
larger separations than ours and uses the bisector convention to define the line of sight to a galaxy pair \citep[which changes the multipoles, e.g.,][]{Raccanelli_2012,Reimberg+2016}.
\subsubsection{Weak lensing}
{Comparing the mean signal from the \grsd (blue crosses) and \vrsd (green circles) {multipoles}, we notice that their difference increases with
redshift and the pair separation. The contribution of the integral
terms always enhances the clustering signal in the quadrupole and hexadecapole moments but by an amount which is relatively small compared to the scatter in the measurements.

Performing the \WLtest test, we find that the \snrt steadily grows
from 0.81 to 2.48 from the lowest to the highest redshift bin (see Fig.~\ref{fig:XI_CORR_HIST} and
Table ~\ref{tab:S_N_TABLE_2PCF}). 
In the latter, the likelihood-ratio test manages to reject the velocity-only model for RSD at the 95\% confidence level in 78\% of our mock catalogues. }

\section{Power Spectrum}
\label{Sec:Power_Spectrum}

The galaxy power spectrum is the workhorse of cosmological-parameter inference. By analogy with Sect.~\ref{Sec:2PCF_ALL}, we introduce the covariance between two Fourier modes of the galaxy overdensity $\langle \tilde{\delta}_\mathrm{g}(\bs{k})\,\tilde{\delta}_\mathrm{g}(\bs{k}')\rangle=(2\,\pi)^3\,C(\bs{k},\bs{k}')$.
If, in real space, $\delta_\mathrm{g}(\bs{x})$ is a statistically homogeneous and isotropic field, then only the diagonal part of the covariance does not vanish,
and the galaxy power spectrum $P_\mathrm{g}(k)$ can be introduced using the relation  $\langle \tilde{\delta}_\mathrm{g}(\bs{k})\,\tilde{\delta}_\mathrm{g}(\bs{k}')\rangle=(2\,\pi)^3\,
\delta_\mathrm{D}(\bs{k}+\bs{k}')\,P_\mathrm{g}(k)$.
However, in redshift-space, where statistical homogeneity is lost, the covariance is not diagonal and the definition above does not apply \citep{Zaroubi-Hoffman96}. A different approach is thus needed.

The `local' power spectrum, $P_{\rm loc}$, is obtained by  Fourier transforming the 2PCF with respect to $\bs{r}$ \citep{scoccimarro_fast_2015}, 
\begin{align}
\label{eq:lps}
P_{\rm loc} (\bs{x}_{\rm m},\bs{k}) &:= \int \left\langle\delta_{\rm g}(\bs{x}_{\rm m} +\bs{r}/ 2)\,\delta_{\rm g}(\bs{x}_{\rm m} -{\bs{r}/ 2})\right\rangle \,\mathrm{e}^{-{\rm i}\bs{k}\cdot\bs{r}}\;\dif^3r \\&
=
\int C(-\bs{k}+\bs{q}/2,\bs{k}+\bs{q}/2)\,\mathrm{e}^{{\rm i}\bs{q}\cdot\bs{x}_{\rm m}}\;\dif^3q \,.
\end{align} 
In order to compress the clustering information into
a set of functions of the wavenumber $k$, it is convenient
to expand $P_{\rm loc}$ in Legendre polynomials of $\hat{\bs{k}}\cdot \hat{\bs{x}}_{\rm m}$ and average over $\bs{x}_{\rm m}$. 
These operations yield the so-called multipole moments of the power spectrum, 
\be 
\label{eq:multpoles_lps}
P_{\ell} (k) := (2\ell + 1)\iint P_{\rm loc} (\bs{x}_{\rm m},\bs{k})\,\mc{L}_{\ell}(\hat{\bs{k}}\cdot \hat{\bs{x}}_{\rm m})\;  {\dif^2 \Omega_k\over 4\pi}\;{\dif^3 x_{\rm m}\over V}\,, 
\ee 
where $V$ denotes the volume under consideration. In this section, we investigate the impact of relativistic RSD on the power spectrum multipoles measured by the \Euclid survey.

\begin{figure*}
    \centering
    \includegraphics[width=\linewidth]{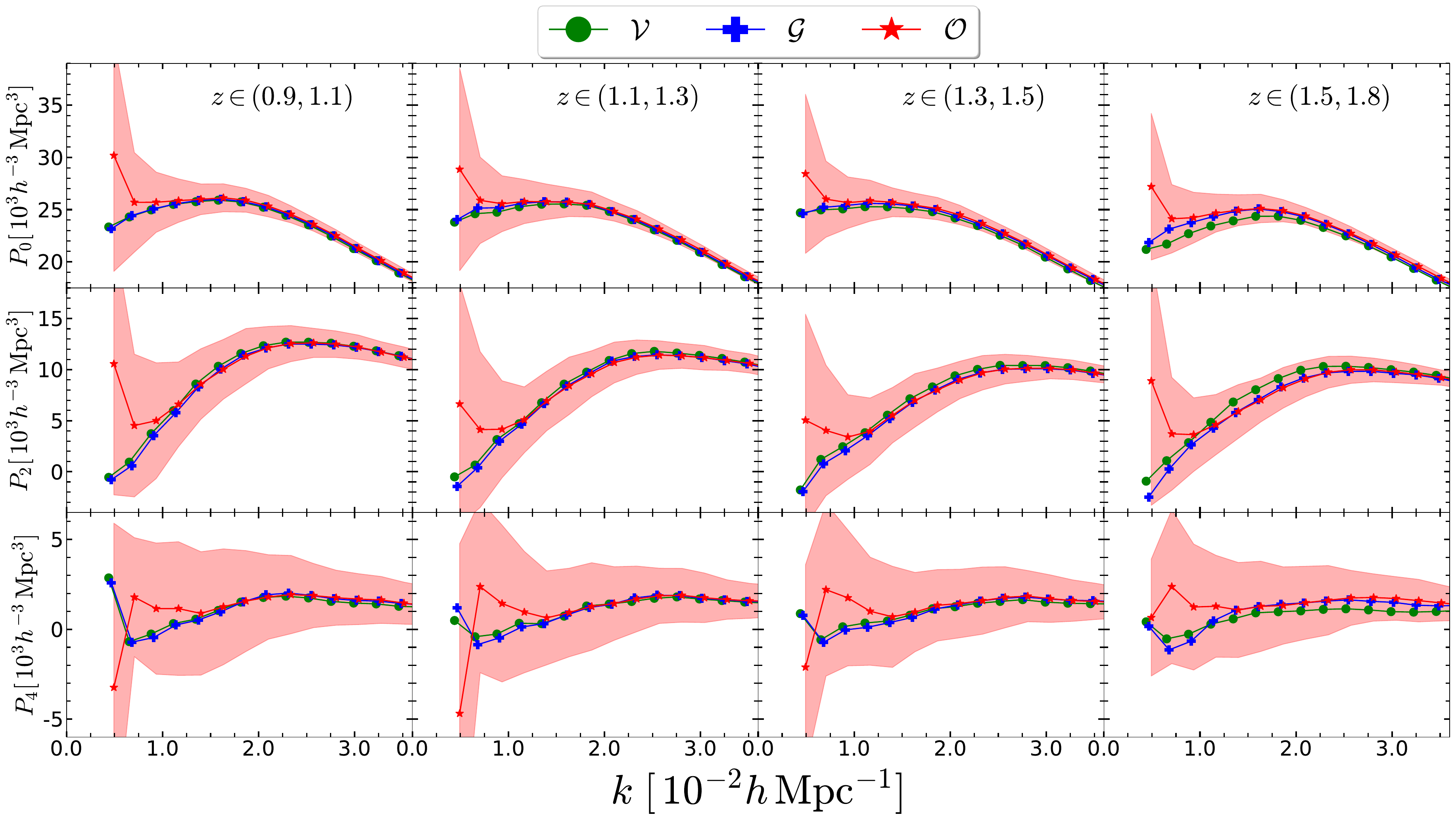}
    \caption{
    {Mean $\ell=0, 2,$ and 4 multipoles of the power spectrum measured from
    the \vrsd (green circles), \grsd (blue crosses), and \obs (red stars) mocks in the four tomographic redshift bins. The shaded areas highlight
    the RMS scatter among the \obs light cones.}
    }
    \label{fig:POWER_SPEC}
\end{figure*}
\subsection{Estimator}
\label{sec:power_est}
{In order to measure the multipoles of the power spectrum
from a galaxy redshift survey, 
the ensemble average in Eq.~(\ref{eq:lps}) 
is replaced with a mean over a 
set of Fourier modes.
We use the Yamamoto--Bianchi \citep[][]{bianchi_measuring_2015,scoccimarro_fast_2015} estimator as implemented
in the official \Euclid code \euclidPower.}%

{Following 
\citet[][hereafter FKP]{FKP}, we first build 
the weighted galaxy overdensity
\begin{align}
\label{eq:FKP_FIELD}
    F \left (\bs{x} \right ) &=  \frac{w \left (\bs{x} \right )}{\sqrt{A}}   \left [\,\widehat{n}_{\rm g} \left (\bs{x} \right ) - \alpha \,\widehat{n}_{\rm r} ( \bs{x})\,\right ]\,,
\end{align}
by comparing the observed number density of galaxies
$\widehat{n}_{\rm g}(\bs{x})$ and its counterpart in the corresponding
random catalogue, $\widehat{n}_{\rm r}(\bs{x})$, introduced in Sect.~\ref{sec:catalog}.
In order to minimize the variance of the measured multipoles,
we adopt the weight function 
\be w(\bs{x}) = \mc{I}(\bs{x}){\left[1+ {\overline{n}_{\rm g}}(x) \,\mc{P}_{0}\right]^{-1}}\,,
\ee where  $\overline{n}_{\rm g}$ denotes the mean density of galaxies,  $ \mc{I}(\bs{x})$ is an indicator function that is one inside the 
volume under study 
and zero elsewhere, and the parameter $\mc{P}_{0}=2\times 10^4\, h^{-3}$ Mpc$^{3}$ gives an approximate value for the galaxy power spectrum
at the scales of interest. 
The normalization factor $A$ is determined through the integral $A = \int w^2\, {\bar{n}^{\,2}}_{\rm g}\, \dif^{3} x$. Finally, the rescaling factor $\alpha$ is calculated using
\begin{equation}
\alpha = \frac{{\int w(\bs{x}) \,\widehat{n}_{\rm g}(\bs{x})\; \dif^{3} x}}{{\int w(\bs{x})\,\widehat{n}_{\rm r}(\bs{x}) \;\dif^{3} x}}.
\end{equation}
}

The multipoles of the galaxy power spectrum with respect to the local line-of-sight direction could be directly estimated by computing
\citep{yamamoto2000effect} 
\begin{align}
    \hat{P}_\ell(k) = (2\ell + 1)\iiint
    \left[F(\bs{x}_1)\,F(\bs{x}_2)\,\mathrm{e}^{-{\rm i}\bs{k}\cdot(\bs{x_1}-\bs{x_2})}\right.\nonumber\\\left.\mc{L}_{\ell}\left(\bshat{k}\cdot {\bshat{x}_{\rm m}}\right)\;\dif^3 x_1\;\dif^3 x_2\;\frac{\dif^2 \Omega_k}{4\pi}\right]-\hat{P}^{\rm SN}_\ell(k)\label{eq:True_estimator}\,,
\end{align}
where $\Omega_k$ denotes the solid angle in Fourier space and the shot noise contribution $\hat{P}^{\rm SN}_\ell(k)$ is given by 
\be
\hat{P}^{\rm SN}_\ell(k) = {(1+\alpha)\over A} \int w^2(\bs{x})\, {\widehat{n}}_{\rm g}(\bs{x})\,\mc{L}_{\ell}(\bshat{k}\cdot \bshat{x}_{\rm m}) \, \dif^{3} x\,.
\ee 
\cite{Yamamoto_2006} noticed that 
replacing the factor $\mc{L}_{\ell}(\bshat{k}\cdot \bshat{x}_{\rm m})$
with either $\mc{L}_{\ell}(\bshat{k}\cdot \bshat{x}_{1})$ 
or $\mc{L}_{\ell}(\bshat{k}\cdot \bshat{x}_{2})$ in Eq.~(\ref{eq:True_estimator}) results in a tremendous speed-up of the estimator. With this substitution -- nowadays known as the `local plane-parallel' (LPP) approximation -- in fact,  $\hat{P}_\ell(k)$ can
be written as the product of two Fourier transforms that can be
conveniently evaluated using the FFT algorithm \citep{Beutler+14, bianchi_measuring_2015,scoccimarro_fast_2015}, i.e.
\begin{equation}
\label{eq:Power_spectrum}
    \hat{P}_\ell(k) = (2\ell + 1)\int F_\ell(\bs{k})F_0(-\bs{k}) \frac{\dif ^2\Omega_k}{4\pi}-\hat{P}^{\rm SN}_\ell(k)\,,
\end{equation}
where
\begin{equation}
    F_\ell(\bs{k}) := \int F(\bs{x})\, {\rm e}^{-{\rm i} \bs{k}\cdot \bs{x}}\mc{L}_{\ell}(\bshat{k}\cdot \bshat{x}) \,\;{\dif^3 x } \,.
\end{equation}
The official \Euclid code we use implements the faster estimator.
It is worth stressing that Eqs.~(\ref{eq:True_estimator})
and (\ref{eq:Power_spectrum}) define two different statistics
which generate different outputs when applied to wide-angle surveys
\citep[e.g.][]{Samushia_2015}. For example, it has been demonstrated that the end-point convention distorts the values of the multipoles when compared to those evaluated using the midpoint convention \citep{Reimberg+2016, Castorina-white2017}.

As shown in Figs.~\ref{fig:XI_CORR_NOV} and \ref{fig:POWER_SPEC}, the effects we investigate here
leave their imprints on extremely large scales. For this reason, we 
compute FFTs within (periodic) cubic boxes with a side length of $L_{\rm FFT} = 16\,015\,\hMpcc$. That choice %
allows us to measure the power-spectrum
multipoles down to the fundamental frequency $k_{\rm F} = 4 \times 10^{-4} \,h\,$Mpc$^{-1}$. 
{We originally measure the multipoles with $\ell=0, 2,$ and 4 within
bins of size $\Delta k=k_{\rm F}$ and we re-bin them in different
ways depending on usage. For the \vobstest and \WLtest tests we
employ 12 equally spaced bins in the range $k\in[0.4,36]\times 10^{-3}\,h\,$Mpc$^{-1}$.} 

\begin{figure*}
    \centering
    \includegraphics[width=\linewidth]{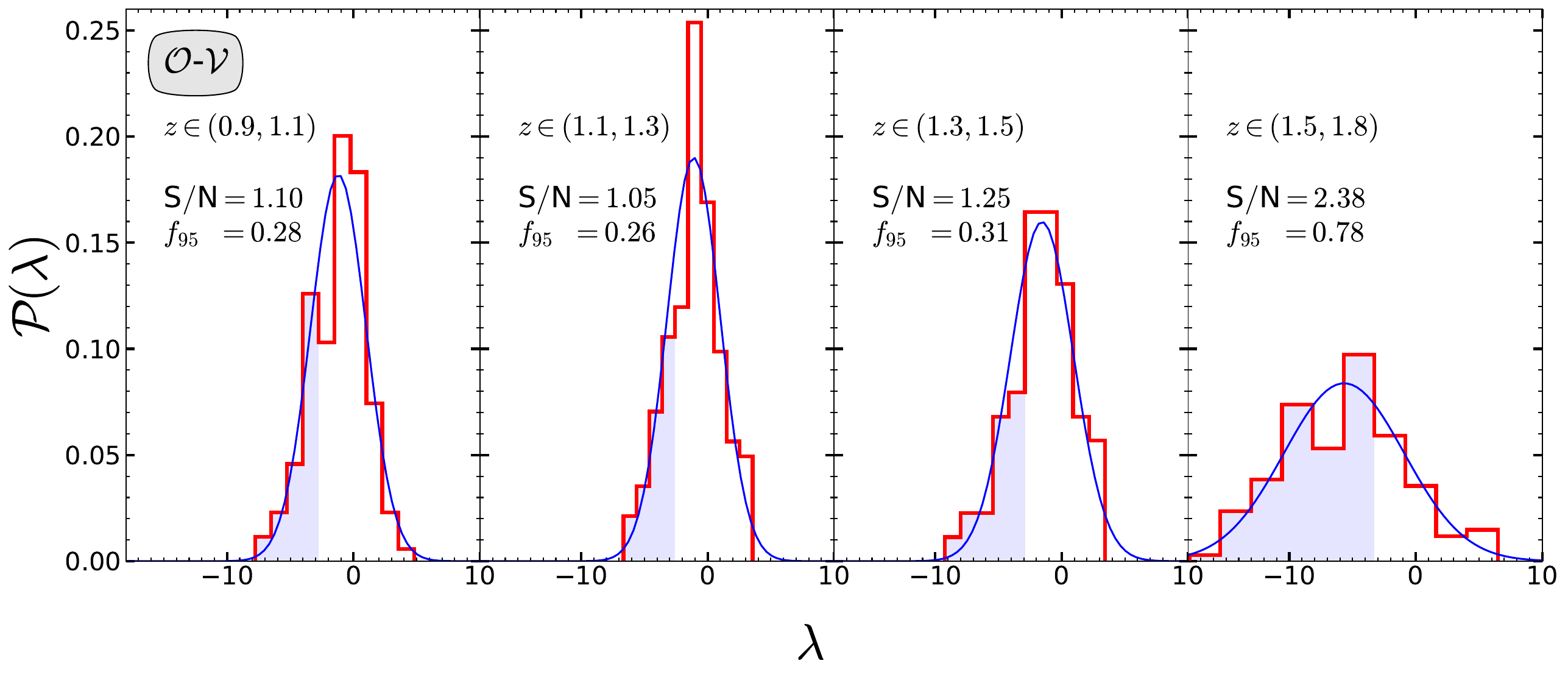}
    \caption{ Similar to Fig.~\ref{fig:XI_CORR_HIST} using the power spectrum multipoles as the data vector.
    }
    \label{fig:POWER_SHIST}
\end{figure*}

\subsection{Results}\label{Sec:p(k)results}

{The mean multipoles of the power spectrum measured from the
different sets of mock catalogues are shown in Fig.~\ref{fig:POWER_SPEC}
together with the RMS scatter from the \obs set.
The monopole moment shows the largest clustering amplitude and is measured with a high signal-to-noise ratio, particularly for $k>2\times 10^{-2}\,h$ Mpc$^{-1}$. At the opposite extreme, the hexadecapole
moment is suppressed by an order of magnitude with respect to $P_0$ and its measurements are very noisy at all scales. The quadrupole moment has intermediate properties between the other two.}

\subsubsection{Peculiar velocity of the observer}
By comparing the \obs and \grsd spectra, we observe that  the peculiar velocity of the observer modifies all multipoles at extremely
large scales. For the monopole, this is consistent with the results presented by \cite{Elkhashab_2021} who showed that a non-vanishing $\bs{\varv}_\mathrm{o}$ adds an oscillatory signal (damped with increasing $k$) to $P_0$ with an oscillation frequency that increases with the characteristic redshift of the galaxy population. The coarse $k$-binning we adopt in this work does not reveal the details of the oscillations that then appear as a large-scale boost of the clustering amplitude in Fig.~\ref{fig:POWER_SPEC}. Similar distortions are clearly noticeable also in $P_2$ and $P_4$. \footnote{A derivation of the impact of $\bs{\varv}_\mathrm{o}$ on all multipoles is presented in \elkhashabalpha} For the latter, in the first three
tomographic bins, the corrected signal becomes negative  at the largest scales probed here.
The chances to detect these signatures with the EWSS are meagre, however, given the large scatter in the measurements at small wavenumbers. The \snrt obtained with the \vobstest test is 1.3 at best (see Table~\ref{tab:S_N_TABLE_P_K}) since the corrections due to $\bs{\varv}_\mathrm{o}$ are localised on very large scales where the measurement noise is large. Still, the enhanced clustering could bias measurements of the local-non-Gaussianity parameter, $f_\mathrm{NL}$, based on $P_0$.
\begin{table}
\label{tab:surveys}
	\begin{center}
	\caption{\snrt from the \WLtest and \vobstest tests for the multipoles of the power spectrum. }
 \label{tab:S_N_TABLE_P_K}
  \begin{tabular}{ccc}
   $(z_\text{min},z_\text{max})$&{\WLtest}&{\vobstest}\\
       \hline%
$(0.9,1.1)$  & 1.1 & 1.1   \\
$(1.1,1.3)$  & 1.0 & 0.9  \\
$(1.3,1.5)$  & 1.3 & 0.9  \\
$(1.5,1.8)$  & 2.4 & 1.3  \\
\hline
\hline
  \end{tabular}
  \end{center}
\end{table}
\subsubsection{Weak lensing}
\label{sec:PK_WL_RESULTS}

The difference between the multipoles extracted from the \grsd and \vrsd mocks is minimal in the first three tomographic redshift bins but becomes more pronounced in the last one, particularly for $\ell=2$ and $4$. This is consistent with the expectation that weak gravitational lensing should have a larger impact on the clustering of high-redshift galaxies.
The \WLtest test closely mirrors the results we obtained for the 2PCF multipoles: the \snrt for the detection of the integrated RSD is always around one in the first three bins and jumps to 2.4 in the last one, for which the velocity-only model for RSD is ruled out with 95\% confidence in 78\% of the realisations.

\section{Perturbative models and survey window function}
\label{sec:Window_analysis_sec}
\begin{figure}
    \centering \includegraphics[width=\linewidth]{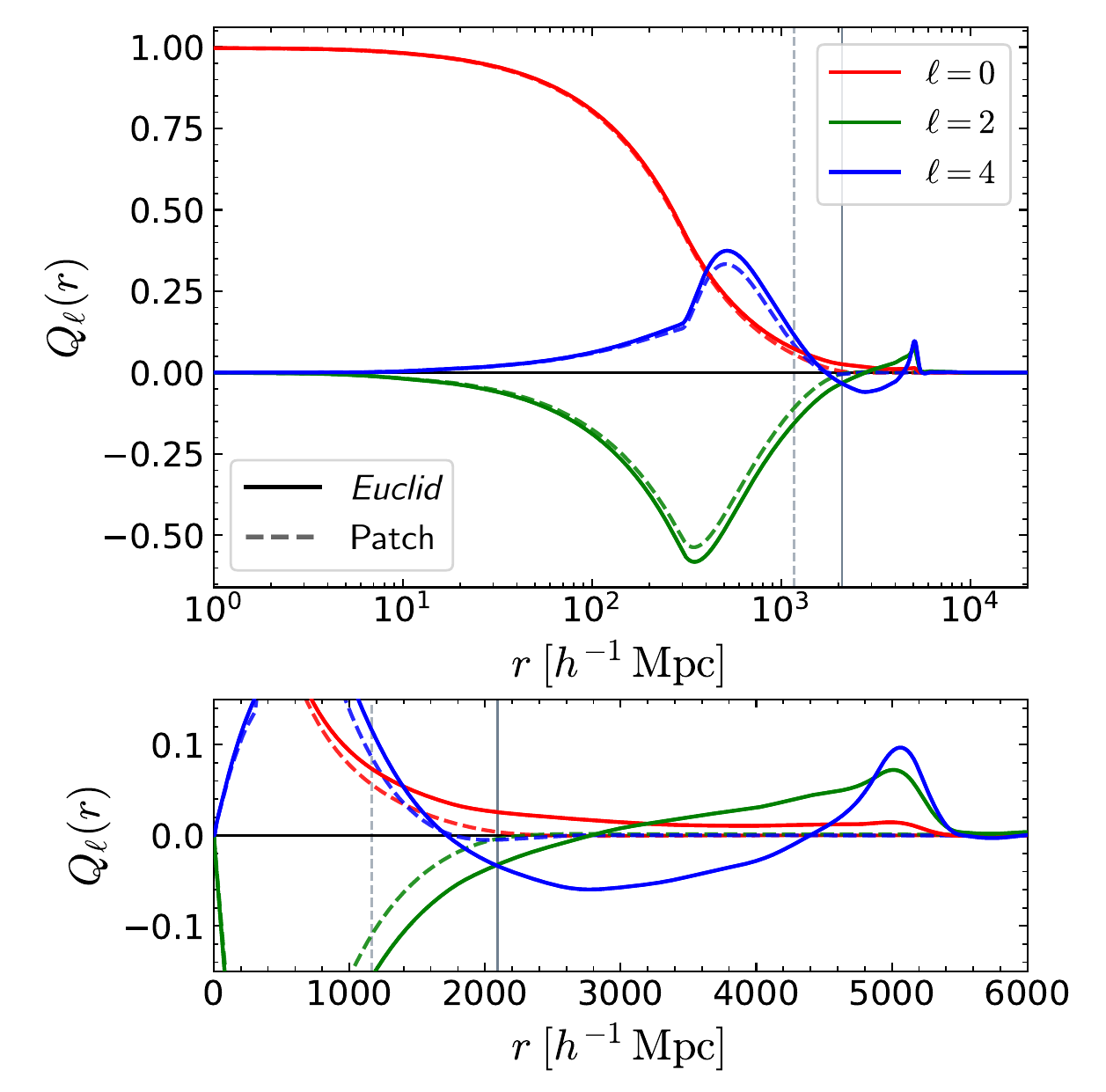}
    \caption{2PCF multipoles of the window function for the
    EWSS (solid) in the redshift bin $z\in(1.1,1.3)$  and for the single patch (dashed) shown in Fig.~\ref{fig:proj_den} in the same redshift range. The vertical lines indicate the characteristic length scale $V^{1/3}$ for the two surveys. The bottom panel offers a more detailed view of the large-scale behaviour of the functions.}
    \label{fig:convolution-ql}
\end{figure}

\begin{figure*}
    \centering
    \includegraphics[width=\linewidth]{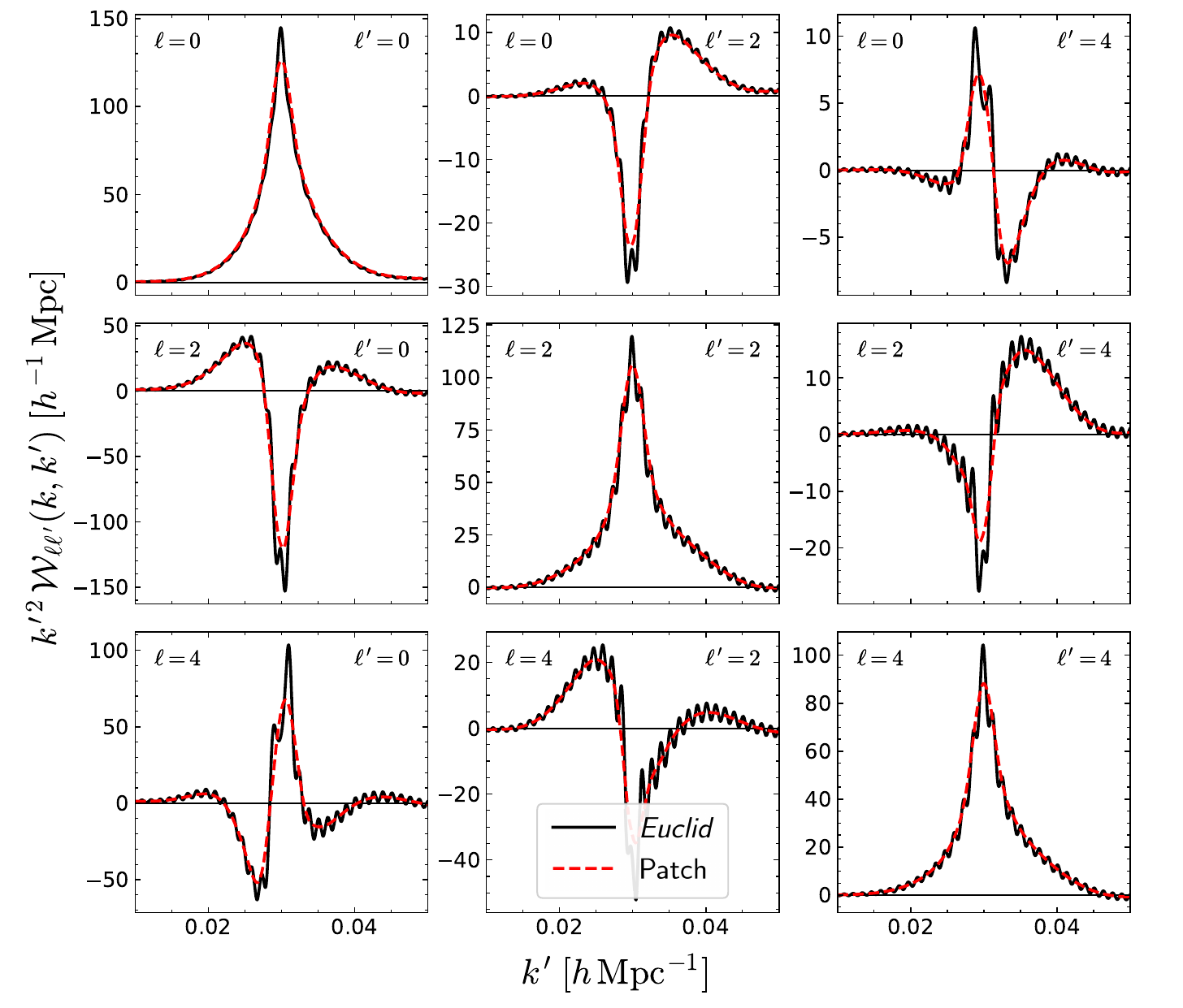}
    \caption{{Elements of the mixing matrix for the two surveys discussed in Fig.~\ref{fig:convolution-ql} evaluated at $k=0.03\, h \, \mathrm{Mpc}^{-1}$.}}
    \label{fig:mixing_kernels}
\end{figure*}
Up to this point, we have estimated the relative importance of various types of RSD in the forthcoming \Euclid data
by comparing the outputs of our different suites of mock catalogues. %
This approach deviates from what is usually done
to interpret clustering measurements from redshift surveys. The standard
procedure is to compare the observed summary statistics to analytical models (based
on some flavour of perturbation theory) after accounting for the window function of the survey. In this section, we pursue this approach 
and assess whether Kaiser's model for RSD is accurate enough to
describe the large-scale limit of the power-spectrum multipoles that \Euclid will measure. We also present some interesting findings about the properties of the \Euclid window function and the possibility of using disjoint patches of the sky in the same
measurement of $\hat{P}_\ell(k)$.

\subsection{The Kaiser model}
In a seminal paper, \cite{Kaiser87} presented a theoretical model for the galaxy power spectrum in redshift space based on linear perturbation theory. The model only considers RSD arising from the peculiar velocity gradient in Eq.~\eqref{eq:Deltag}. It also
relies on the GPP
approximation according to which the lines of sight to all galaxies are parallel (as expected for a small volume located at large distance from the observer and which is thus seen under a narrow solid angle, i.e. under the distant-observer approximation). 
At fixed cosmological redshift, it gives\begin{align}
    \label{Eq:Kaiser_P_L}
    P_{\rm g}(\bs{k})= \sum_{\ell=0,2,4}\mc{F}_\ell\, b^2\,D^2_+\,P_{\rm m}(k)\,\mc{L}_\ell(\bshat{k}\cdot \bshat{x}_{\rm g})\,,  
\end{align}
{where $D_+$ denotes the linear growth factor for matter
perturbations, $P_{\rm m}(k)$ is their linear power spectrum
at $z=0$, {$\hat{\bs{x}}_\mathrm{g}$ is the line-of-sight direction,} and }
\begin{align}
    \mc{F}_0 &= 1 + {2\over 3} \beta + {1\over5} \beta^2\,, \\
    \mc{F}_2 &= {4\over 3} \beta+ {4\over7} \beta^2\,, \\
    \mc{F}_4 &= {8\over35}\beta^2\,,
\end{align}
{in terms of the (redshift-dependent)} linear RSD parameter
\be
\beta= {1\over b}\left.{\dif \ln D_+ \over \dif \ln a}\right|_{a=(1+z)^{-1}}\,.
\ee

{This result can be generalised to model observations taken
on a section of the past lightcone (with volume $V_{\rm s}$) of an observer obtaining 
\citep[to first approximation, e.g.][]{Yamamoto_1999,Pryer_2022}
\begin{align}
    \label{Eq:Kaiser_P_L_LC}
    P_{\rm g}&(\bs{k}) = \sum_{\ell=0,2,4}\overline{\mc{F}}_{\ell}\,P_{\rm m}(k)\,\mc{L}_\ell(\bshat{k}\cdot \bshat{x}_{\rm g})\,,
\end{align}
where 
\begin{align}
\overline{\mc{F}}_\ell={\int_{V_{\rm s}} \mc{F}_\ell\, b^2  \, D^2_+ \,\bar{n}_{\rm g}^2 \,\dif V \over  \int_{V_{\rm s}} \bar{n}_{\rm g}^2  \,\dif V}\,.
\end{align}
}
Going beyond {this} model requires dropping the  GPP approximation and/or accounting for all the RSD terms appearing in Eq.~\eqref{eq:Deltag}. In order to correct the GPP predictions, \cite{Castorina-white2017} proposed to expand the multipoles of the wide-angle power spectrum in the parameter $(k\,x_{\rm m})^{-1}$.  %
In this framework, the additional relativistic RSD can then be treated perturbatively \citep{Beutler+2019,Castorina_2022,Noorikuhani_Scoccimarro2023}. The expansion in $(k\,x_{\rm m})^{-1}$, %
however, might become inaccurate for large angular separations.  
Another possible approach  is to use the “spherical-Fourier-Bessel” formalism which was introduced by \cite{Peebles-1973}, extended to redshift space by \cite{Heavens_1995}, and applied to survey data in \cite{Percival_2004}. The inclusion of GR effects in this formalism is discussed in
\citet{Yoo_2013} and \citet{Bertacca_2018}.

\subsection{The window convolution and integral constraint }
\label{sec:P_L_WINDOW_CONV}

In order to compare theoretical models
to the multipoles estimated from a survey, one needs to account for the fact that only a finite volume is observed.
We can gain some insight into this issue by first considering
the FKP power-spectrum estimator $\hat{P}_\mathrm{obs}(\bs{k})=|\tilde{F}(\bs{k})|^2$ derived under the
GPP approximation. In this case, we obtain \citep{Peacock_1991}
\begin{align}
    {P}_{\mathrm{obs}}(\bs{k}) =  \int  P_{\rm g}(\bs{k}')\,& | \widetilde{W}(\bs{k} - \bs{k}') |^2\,\dif^3 k' \nonumber \\
    & - \frac{| \widetilde{W}(\bs{k}) |^2}{| \widetilde{W}(0) |^2} \int  P_{\rm g}(\bs{k}')\, | \widetilde{W}(\bs{k}') |^2\,\dif^3 k'\,,\label{eq:P_convo_full}
\end{align}
where $P_{\mathrm{obs}}(\bs{k})=\langle \hat{P}_{\mathrm{obs}}(\bs{k}) \rangle$, and $\widetilde{W}(\bs{k})$ is the Fourier transform of the survey window function
\begin{equation}
    W(\bs{x}) = \frac{\alpha \, w(\bs{x}) \, \,\widehat{n}_{\rm r} ( \bs{x})}{\sqrt{A }}\,. \label{eq:W}
\end{equation}
The first term {in Eq.~(\ref{eq:P_convo_full}) shows
that the power-spectrum  estimator mixes the contributions from Fourier modes that
differ by less than the characteristic width of $\widetilde{W}(\bs{k})$
which is of the order of $V^{-1/3}$ (in accordance with the uncertainty relation between conjugate Fourier variables). This has two main consequences: ({\it i}) the power spectrum is substantially distorted on large scales and ({\it ii}) additional anisotropy is generated on top of RSD
and the Alcock-Paczynski effect because of the spherical asymmetry of the window function.} 
The second term {in Eq.~(\ref{eq:P_convo_full}) gives the so-called} (global) integral constraint which {arises from} the assumption that the average density within the survey {coincides with} the {actual mean} density of the Universe. {This term subtracts the actual power at $\bs{k}=0$ which leaks to larger wavenumbers because of the convolution with the window function, thus
enforcing that $P_{\rm obs}(\bs{k})$ is zero for $\bs{k}=0$. }

The considerations above can be generalised to the estimator
for the power-spectrum multipoles introduced in Eq.~(\ref{eq:Power_spectrum}) based on the
LPP approximation. This gives \citep{Beutler+14, Wilson+17,Beutler_2021}
\begin{align}
    P_{\mathrm{obs},\,\ell}(k) = & \sum_{\ell'=0,2,4}\int_{0}^{\infty}\, {k'}^2 \, \mathcal{W}_{\ell\ell'}(k,k') \, P_{{\rm g}\,,\ell'}(k') \;\dif k' \, \nonumber \\
    & - \frac{P_{\ell}^{W}(k)}{P_{0}^{W}(0)} \sum_{\ell'=0,2,4}\int_{0}^{\infty}  \, {k'}^2 \, \mathcal{W}_{\ell\ell'}(0,k') \,P_{{\rm g}\,,\ell'}(k')\;\dif k'\,,\label{eq:wind_fucn}
\end{align}
where $P_{{\rm g}\,,\ell'}(k')$ is computed under
the GPP approximation, $P_{\ell}^{W}$ denotes the multipole spectral moments of the 
window function and $\mathcal{W}_{\ell\ell'}(k,k')$ are the elements
of the so-called mixing matrix.
The latter can be obtained from the multipoles $Q_{\ell}(r)$ of the 2PCF of $W(\bs{x})$ using
\begin{align}
\label{Eq:Mixing_mat}
    \mathcal{W}_{\ell \ell'}(k, k') =  \frac{{\rm i}^{2\ell' - \ell}}{2 \pi^2} (2\ell + 1) & \sum_{L=0}^{\infty} \tj{\ell}{\ell'}{L}{0}{0}{0}  \,  \nonumber \\
    & \times\,\int_0^{\infty} r^2 j_{\ell}(k\,r) \, j_{\ell'}(k'\,r)\, Q_{L}(r)\,\dif r\,,
\end{align}
where %
{the brackets denote the Wigner 3j symbols and  $j_{\ell}$ is the spherical Bessel function of order $\ell$. }
The functions
$Q_{\ell}(r)$ could be directly estimated by counting  pairs in the random catalogue \citep{Wilson+17}.
However, it is computationally faster to obtain them using a Hankel transform 
\begin{equation}
\label{eq:Q_L_X}
    Q_{\ell}(r) = \frac{{\rm i}^{\ell}}{2\,\pi^2} \int_0^{\infty} k^2\, P_{\ell}^{W}(k) \,j_{\ell}(k\,r)\,\dif k \,,
\end{equation}
evaluated with the \ttt{FFTLog} method \citep{Hamilton00}. %

\subsubsection{Disconnected patches}
\label{sec:disconnected_regions}
The EWSS avoids the ecliptic and Galactic planes and is thus composed of four disconnected regions (see Fig.~\ref{fig:proj_den}). Considering all of them together would allow us to measure galaxy clustering on
the largest possible scales. This is what we did in Sect.~\ref{Sec:Power_Spectrum} to test the importance of the different RSD terms. However, this procedure differs from what is regularly done in ground-based surveys
where Northern and Southern Galactic caps are analysed separately 
since each of them is subject to different angular systematics. %

{In order to contrast these approaches, in Fig.~\ref{fig:convolution-ql}, we compare the functions $Q_{0},\,  Q_2$, and $Q_4$ derived for the full \Euclid footprint (solid lines) and for the simply connected patch covering $2565 \,\mathrm{deg}^2$ displayed in Fig.~\ref{fig:proj_den} 
(dashed lines). In both cases, we consider the redshift bin $z\in(1.1,1.3)$. }
For separations of a few $\hMpcc$, the monopole moment approaches
one while $Q_2$ and $Q_4$ are close to zero, reflecting the fact
that $Q(\bs{r})$ is nearly constant and isotropic for spatial
lags that are well 'contained' within the survey window function. On the other hand, $Q(\bs{r})$ vanishes when $r$ is larger than the maximum distance between two galaxies in the patch. {In between these two asymptotic regimes, the 2PCF of the random catalogue becomes
highly anisotropic with $|Q_2|$ and $Q_4$ that assume values
larger than $Q_0$. The difference between the two survey footprints
becomes evident on the largest scales: for instance, in the \Euclid case, $Q_4$ becomes negative when $1600 \, \hMpcc\lesssim r \lesssim 4400\, \hMpcc$ and shows a prominent peak for $r\simeq 5000 \,\hMpcc$ due to the presence of separate patches.}

In Fig.~\ref{fig:mixing_kernels}, we  {present all the elements of} the resultant mixing matrices for the two different geometries
considering the second tomographic bin $z\in(1.1,1.3)$. 
{We fix $k = 0.03 \, h \, \mathrm{Mpc}^{-1}$ and plot $\mathcal{W}_{\ell\ell'}(k,k')$
as a function of $k'$. Each panel refers to a specific combination of $\ell$ and $\ell'$.}
We plot {the results for the} \Euclid footprint in black, while {those for} the connected patch are shown in red.  
{As expected, the most significant contributions come from $k'=k$ but mixing takes place within a relatively broad range of wavenumbers. The matrix elements are always positive for $\ell=\ell'$ and show an oscillatory behaviour otherwise. These overall trends are
present for both survey footprints. However, the results for
the EWSS show high-frequency oscillations}
that are absent in the single patch case. This {modulation} can be traced back to {presence of the peak at very large separations in the $Q_\ell$ functions presented in}  %
Fig.~\ref{fig:convolution-ql}. 

\subsubsection{Comparison with the \liger mocks}
\begin{figure}
    \centering
    \includegraphics[width=1\linewidth]{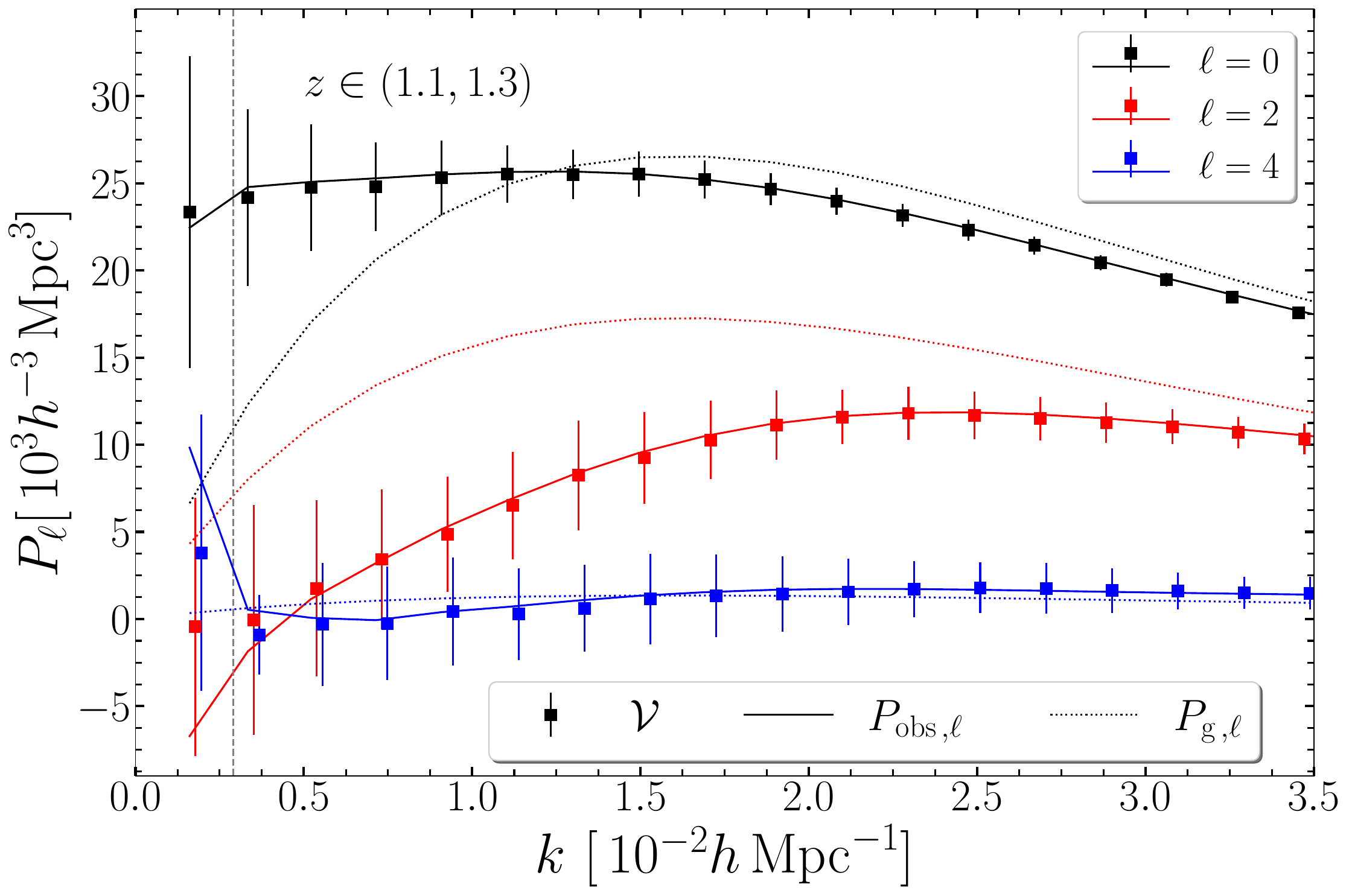}
    \caption{{
    Power-spectrum multipoles for the \vrsd mocks
    in the redshift range $z\in(1.1,1.3)$. The mean and RMS scatter
    over the 140 mocks are displayed with squares and errorbars, respectively. The solid and dotted lines show}
    the theoretical predictions %
    {obtained with
    the Kaiser model -- see Eq.~\eqref{Eq:Kaiser_P_L_LC} -- with and without accounting for the window function of the
    EWSS
    -- see Eq.~(\ref{eq:wind_fucn}) -- respectively.}
    The vertical dashed line {marks the %
    characteristic wavenumber $k_{\rm C}=2\,\pi/V^{1/3}$.}} 
    \label{fig:POWER_WC}
\end{figure}

{We finally compare }
the theoretical model that accounts for the survey window and the integral constraint
{to the spectral multipoles extracted from our mock catalogues.}
For the sake of clarity, we summarise here  how the theoretical prediction is computed: {\it (i)} we {generate the linear
matter power spectrum at }
$z = 0$ using the \texttt{CAMB} code; {\it (ii)} {we utilize Kaiser's model to obtain} the galaxy power spectrum {in redshift space and resort to Eq.~\eqref{Eq:Kaiser_P_L_LC} in order to get the  observed spectra on the past light cone};
{\it (iii)} we compute the elements of the mixing matrix using the 2PCF multipoles estimated from the random catalogue of the mocks (see Eq.~\ref{Eq:Mixing_mat}); {\it (iv)} we account for the survey window function and the integral constraint {by making use of} Eq.~\eqref{eq:wind_fucn}. 

In Fig.~\ref{fig:POWER_WC}, {we compare the model (solid lines) to the measurements extracted from the \vrsd mocks (symbols with errorbars). As a reference, we also plot the theoretical predictions obtained without accounting for the finite-volume effects (dotted lines).  Since we obtain similar results in all tomographic bins,
we only show them for the interval $z\in(1.1\,,1.3)$.
The model and the measurements are in extremely good  agreement at all 
wavenumbers $k>k_{\rm C} =2\,\pi/V^{1/3}$ (highlighted with a vertical dashed line). On larger scales, 
the quadrupole and hexadecapole moments show a small discrepancy possibly due to wide-angle corrections missed by Kaiser's model. 
A side note is in order here. The fact that Kaiser's model nicely matches our \vrsd mocks does not imply that it accounts for all 
RSD generated by peculiar velocities. In fact, the \vrsd catalogues are
constructed by setting the magnification and evolution bias to zero which
automatically cancels some wide-angle effects.
Anyway, the remarkable consistency we find validates the accuracy of the \liger mock catalogues
and demonstrates that the mixing-matrix formalism can be reliably used for surveys made of disconnected patches. 
} 

Finally, {we compare the theoretical model with the multipole moments extracted from the \obs mocks which mimic what will be observed by the actual \Euclid mission. This way} we seek to assess the significance {with which} a theoretical model that accounts only for the velocity-gradient term {can be ruled out due to the presence of additional RSD in the data. 
Given the close match between
the model and the numerical results obtained from the \vrsd mocks, we obviously expect very similar results to those presented in Sect.\ref{sec:PK_WL_RESULTS}.
We employ the likelihood-ratio test and only consider wavenumbers
$k<k_{\rm{C}}$ grouped in 11 bins covering the range $k\in[3.2,36]\times 10^{-3}\,h\,$Mpc$^{-1}$ (slightly narrower than in Sect.\ref{sec:PK_WL_RESULTS}).} 
The resulting signal-to-noise ratios {for this} ($\text{\obs-}\,P_{\rm obs}$) test are $[1.3,\, 1.2,\,1.5,\,2.9]$ {for the four tomographic bins listed in the same order as in } Table~\ref{tab:S_N_TABLE_P_K}.  
As expected, these results are in good agreement with those
deriving from the \WLtest test. This reinforces the conclusion that
it is necessary to include at least the weak lensing corrections
in order to obtain a good fit to 
the \Euclid power-spectrum multipoles 
{on scales}
$k<0.03\,h\,{\rm Mpc}^{-1}$.

\section{Summary}
\label{sec:summary}

The EWSS will measure redshifts and angular positions of nearly 30 million emission-line galaxies
over a third of the sky and in the redshift
range $0.9 \lesssim z \lesssim 1.8$ \citep{Redbook}.
If observational systematic effects are kept under control, it will provide us with the opportunity to study galaxy clustering on unprecedentedly large scales.
Theoretical considerations suggest that relativistic ``projection'' effects alter the clustering signal we measure.
In fact, galaxy observables (i.e. redshift, fluxes, and sky position)
are influenced by the inhomogeneities that light encounters as it propagates from the galaxies to the observer. %
 However, we construct maps of the galaxy distribution
by assuming that galaxies populate an unperturbed model universe
(for instance, in order to convert redshift into distances). This discrepancy leads to RSD that we need to model in order to interpret clustering data.
While the leading effect generated by the relative peculiar velocities
between the sources and the observer
\citep{Kaiser87} is regularly accounted for in all studies,
additional contributions due to Doppler, gravitational lensing, Sachs--Wolfe, and time-delay type terms have been mostly neglected. 
The importance of these relativistic corrections is expected to grow
as we investigate physical length scales approaching the Hubble radius.
In this work, we make forecasts for their impact 
on different galaxy two-point statistics that will be measured from the EWSS. Namely, the angular power spectrum, the multipole moments of the anisotropic 2PCF, and the multipole moments of the anisotropic power spectrum.

We use the \liger method (summarised schematically in Fig.~\ref{fig:liger_schematic}) to build 140 mock galaxy catalogues,
each covering the whole footprint (and redshift range) of the EWSS and
accounting for all relativistic RSD up to the first order in the cosmological perturbations (see Fig.~\ref{fig:proj_den} for one example
shown in projection). 
We study the impact of different relativistic RSD by turning on and off particular effects while producing the galaxy catalogues. 
In particular, for each of the 140 realisations, we generate the galaxy
distribution in real space (\real) and three different versions of its
redshift-space counterpart. In the \vrsd mocks, we basically only account for the spatial derivative of the peculiar velocity of the galaxies along the line of sight. In the \grsd catalogues, we add all the  
integral terms, which are dominated by the weak lensing signal
(magnification bias and convergence). Finally, the \obs mocks also include the distortions generated by a non-vanishing peculiar velocity of the observer, which we match to the observed dipole temperature anisotropy of the cosmic microwave background. The \obs catalogues 
mimic what will be obtained with \Euclid using heliocentric redshifts. 
The \real, \vrsd, and \grsd mocks are useful tools for our investigation but do not correspond to anything\footnote{The \grsd catalogues correspond to what a comoving observer would measure but this does not
coincide with the galaxy maps obtained by
correcting the galaxy redshifts measured by a non-comoving observer \elkhashabalpha.} that can be measured in reality.
Note that our study automatically includes wide-angle effects induced by correlating pairs with large separations on the celestial sphere. 
On the other hand,
we do not account for variations of the survey depth on the sky (which can generate a spurious clustering signal if not corrected for in the random catalogue) 
and redshift errors as we concentrate solely on the signal produced by relativistic RSD. 

We use the likelihood-ratio test to estimate the statistical significance with which the \Euclid measurements could reject the null hypothesis that certain relativistic projection effects can be neglected in the theoretical models.
In particular, given the summary statistics measured from 
the individual \obs mocks, we compare the likelihoods of the
models obtained by averaging the
clustering signal over  all of  the \vrsd, \grsd and \obs catalogues with no tunable parameters.
Our results can be summarised as follows.

{\it (i) Angular power spectrum}: 
The peculiar velocity of the observer noticeably increases the amplitude of the dipole ($\ell=1)$, octupole ($\ell=3$), and dotriacontapole ($\ell=5$) signal measured in the heliocentric frame
(Fig.~\ref{fig:DIFFCLSVOBS}). Statistically,
this boost can be measured with a signal-to-noise ratio of 2.1 for the redshift bin that encompasses the entire survey (Table~\ref{tab:S_N_TABLE_CLS_O_V})
The combined effect of
magnification bias and weak lensing convergence slightly increases the clustering signal for $\ell \gtrsim 50$. This contribution is particularly evident in the broad redshift bin $0.9<z<1.8$ where it is
detectable with \snrtval{5.4}. Slightly smaller statistical significance
(up to \snrtval{4.5})
is obtained by cross-correlating distant narrower bins  (see Table~\ref{tab:S_N_TABLE_CLS_G_V}). We note that the radial projection in the angular power spectrum strongly suppresses the contribution of the source peculiar velocities at small scales (large $\ell$). Consequently, the weak-lensing contribution becomes the most dominant correction at these scales, in contrast to its behaviour in the 3D statistics.%

{\it (ii) Multipole moments of the 2PCF:}
The peculiar velocity of the observer introduces small modifications
to the clustering signals and the likelihood-ratio test shows that
the \snrt for rejecting a model that assumes a comoving observer is always smaller than one.
The weak lensing contribution amplifies the quadrupole and hexadecapole moments of the 2PCF at large scales 
($r> 150 \,\hMpcc$) for $z>1.3$ (see Fig.~\ref{fig:XI_CORR_NOV}). 
The \snrt of the combined contributions due to 
lensing and $\varv_\mathrm{o}$ to the 
multipoles of the 2PCF grows with redshift and reaches
2.5 in the most distant tomographic bin we consider ($1.5<z<1.8$).
Our results are in excellent agreement with the Fisher information-based \snrt estimates provided by \cite{wp9_goran}. These authors show that neglecting lensing magnification in the models systematically shifts the constraints on the cosmological parameters derived from the \Euclid multipoles of the 2PCF. In the $\Lambda$CDM framework, the resulting bias is of the order of $0.4\,\text{--}\,0.7$ standard deviations. Even larger biases affect model-independent estimates of the growth rate of structure.

{\it (iii) Multipole moments of the power spectrum}:
The peculiar velocity of the observer %
imprints rather strong characteristic features on very large scales ($k<0.015\, h\,$Mpc$^{-1}$) which are, however, difficult to detect with the
EWSS due to the large statistical uncertainties of
the measurements. The  \snrt  is always of order one\footnote{Higher significance is found either considering wider redshift bins \citep{Elkhashab_2021} or using dedicated statistics \elkhashabalpha.} in all the tomographic bins we consider (Table~\ref{tab:S_N_TABLE_P_K}).
The weak gravitational lensing signal is mostly noticeable in the tomographic slice $1.5<z<1.8$ where it can be detected
with \snrtval{2.4}. 
In terms of the \snrt of the different effects, considering the multipole moments of the power spectrum
provides very similar results to those obtained from the 2PCF.

Since they are obtained by averaging the measurements of a given summary statistic over 140 mock catalogues,
by construction, the models considered above are unbiased and perfectly account for wide-angle effects. In actual surveys, however, observational data are fit to
theoretical predictions derived with some flavour of perturbation theory and based on assuming either the GPP or LPP approximations.
We thus repeat the likelihood-ratio test for the multipoles of the power spectrum by contrasting the mock-based predictions to a simple analytical model obtained combining
the linear matter power spectrum with Kaiser's model for RSD. In order to compare it to the mock data accounting for the  \Euclid window function, we use the mixing-matrix formalism introduced in Eqs.~(\ref{eq:wind_fucn}) and (\ref{Eq:Mixing_mat}). We find that the elements of the mixing matrix present high-frequency oscillations (Fig.~\ref{fig:mixing_kernels}) because
the EWSS is made of four disconnected patches that we
analyse simultaneously.
The window-corrected theoretical model very closely
matches the mean from the \vrsd mocks.
The striking agreement shown in Fig.~\ref{fig:POWER_WC} simultaneously demonstrates that ({\it i}) our mock catalogues are accurate,
({\it ii}) the mixing-matrix approach is
robust also for surveys composed of multiple patches, and ({\it iii})
the power spectrum multipoles extracted from the \vrsd mocks are
not particularly affected by the wide-angle effects sourced by the velocity-gradient term.
On the other hand, in the redshift bin $1.5<z<1.8$, the window-corrected Kaiser model is rejected with 
a \snrtval{2.9} by the \obs mocks that include all relativistic RSD. 
This result reinforces the conclusion that the weak lensing term cannot be disregarded at high redshift and on large scales.

\newpage

\FloatBarrier

\begin{acknowledgements}
{DB and MYE acknowledge support from the COSMOS network (\url{www.cosmosnet.it}) through the ASI (Italian Space Agency) Grants
2016-24-H.0, 2016-24-H.1-2018 and 2020-9-HH.0. MYE and DB acknowledge funding from the Italian Ministry of Education, University and Research (MIUR) through the ``Dipartimenti di eccellenza" project ``Science of the Universe".}

\AckEC
\end{acknowledgements}

\bibliographystyle{aa}
\bibliography{biblio}

\begin{thebibliography}{93}
\expandafter\ifx\csname natexlab\endcsname\relax\def\natexlab#1{#1}\fi

\bibitem[{{Bagley} {et~al.}(2020){Bagley}, {Scarlata}, {Mehta}, {Teplitz},
  {Baronchelli}, {Eisenstein}, {Pozzetti}, {Cimatti}, {Rutkowski}, {Wang}, \&
  {Merson}}]{bagley20}
{Bagley}, M.~B., {Scarlata}, C., {Mehta}, V., {et~al.} 2020, \apj, 897, 98

\bibitem[{Bahr-Kalus {et~al.}(2021)Bahr-Kalus, Bertacca, Verde, \&
  Heavens}]{Bahr-Kalus:2021jvu}
Bahr-Kalus, B., Bertacca, D., Verde, L., \& Heavens, A. 2021, \jcap, 11, 027

\bibitem[{{Bertacca}(2015)}]{Bertacca:2015}
{Bertacca}, D. 2015, Classical and Quantum Gravity, 32, 195011

\bibitem[{Bertacca(2020)}]{Bertacca_2020_Rocket_effect}
Bertacca, D. 2020, International Journal of Modern Physics D, 29, 2050085

\bibitem[{Bertacca {et~al.}(2012)Bertacca, Maartens, Raccanelli, \&
  Clarkson}]{Bertacca_2012}
Bertacca, D., Maartens, R., Raccanelli, A., \& Clarkson, C. 2012, \jcap, 10,
  025

\bibitem[{{Bertacca} {et~al.}(2018){Bertacca}, {Raccanelli}, {Bartolo},
  {Liguori}, {Matarrese}, \& {Verde}}]{Bertacca_2018}
{Bertacca}, D., {Raccanelli}, A., {Bartolo}, N., {et~al.} 2018, \prd, 97,
  023531

\bibitem[{Bertacca {et~al.}(2020)Bertacca, Ricciardone, Bellomo, Jenkins,
  Matarrese, Raccanelli, Regimbau, \& Sakellariadou}]{Bertacca_2020}
Bertacca, D., Ricciardone, A., Bellomo, N., {et~al.} 2020, \prd, 101, 103513

\bibitem[{{Beutler} {et~al.}(2019){Beutler}, {Castorina}, \&
  {Zhang}}]{Beutler+2019}
{Beutler}, F., {Castorina}, E., \& {Zhang}, P. 2019, \jcap, 03, 040

\bibitem[{Beutler \& McDonald(2021)}]{Beutler_2021}
Beutler, F. \& McDonald, P. 2021, \jcap, 11, 031

\bibitem[{Beutler {et~al.}(2014)Beutler, Saito, Seo, Brinkmann, Dawson,
  Eisenstein, Font-Ribera, Ho, McBride, Montesano, Percival, Ross, Ross,
  Samushia, Schlegel, Sánchez, Tinker, \& Weaver}]{Beutler+14}
Beutler, F., Saito, S., Seo, H.-J., {et~al.} 2014, \mnras, 443, 1065

\bibitem[{Bianchi {et~al.}(2015)Bianchi, Gil-Marín, Ruggeri, \&
  Percival}]{bianchi_measuring_2015}
Bianchi, D., Gil-Marín, H., Ruggeri, R., \& Percival, W.~J. 2015, \mnras:
  Letters, 453, L11

\bibitem[{{Bonvin} \& {Durrer}(2011)}]{Bonvin-Durrer2011}
{Bonvin}, C. \& {Durrer}, R. 2011, \prd, 84, 063505

\bibitem[{Borzyszkowski {et~al.}(2017)Borzyszkowski, Bertacca, \&
  Porciani}]{MIKO_2017}
Borzyszkowski, M., Bertacca, D., \& Porciani, C. 2017, \mnras, 471, 3899

\bibitem[{Breton {et~al.}(2022)Breton, de~la Torre, \& Piat}]{Breton+2022}
Breton, M.-A., de~la Torre, S., \& Piat, J. 2022, \aap, 661, A154

\bibitem[{{Breton} {et~al.}(2019){Breton}, {Rasera}, {Taruya}, {Lacombe}, \&
  {Saga}}]{BRETON_2022_cross_correlation}
{Breton}, M.-A., {Rasera}, Y., {Taruya}, A., {Lacombe}, O., \& {Saga}, S. 2019,
  \mnras, 483, 2671

\bibitem[{Broadhurst {et~al.}(1995)Broadhurst, Taylor, \&
  Peacock}]{Broadhurst:1994qu}
Broadhurst, T.~J., Taylor, A.~N., \& Peacock, J.~A. 1995, \apj, 438, 49

\bibitem[{Burden {et~al.}(2017)Burden, Padmanabhan, Cahn, White, \&
  Samushia}]{Burden_2017}
Burden, A., Padmanabhan, N., Cahn, R.~N., White, M.~J., \& Samushia, L. 2017,
  \jcap, 03, 001

\bibitem[{Camera {et~al.}(2015)Camera, Maartens, \&
  Santos}]{CAMERA_RELATIVISTIC}
Camera, S., Maartens, R., \& Santos, M.~G. 2015, \mnras: Letters, 451, L80

\bibitem[{Castorina \& {Di Dio}(2022)}]{Castorina_2022}
Castorina, E. \& {Di Dio}, E. 2022, \jcap, 01, 061

\bibitem[{Castorina {et~al.}(2019)Castorina, Hand, Seljak, Beutler, Chuang,
  Zhao, Gil-Marín, Percival, Ross, Choi, Dawson, Macorra, Rossi, Ruggeri,
  Schneider, \& Zhao}]{Castorina_2019}
Castorina, E., Hand, N., Seljak, U., {et~al.} 2019, \jcap, 09, 010

\bibitem[{{Castorina} \& {White}(2018)}]{Castorina-white2017}
{Castorina}, E. \& {White}, M. 2018, \mnras, 476, 4403

\bibitem[{Challinor \& Lewis(2011)}]{Challinor:2011bk}
Challinor, A. \& Lewis, A. 2011, \prd, 84, 043516

\bibitem[{Cole {et~al.}(2005)Cole, Percival, Peacock, Norberg, Baugh, Frenk,
  Baldry, Bland-Hawthorn, Bridges, Cannon, Colless, Collins, Couch, Cross,
  Dalton, Eke, De~Propris, Driver, Efstathiou, Ellis, Glazebrook, Jackson,
  Jenkins, Lahav, Lewis, Lumsden, Maddox, Madgwick, Peterson, Sutherland, \&
  Taylor}]{Cole+2005}
Cole, S., Percival, W.~J., Peacock, J.~A., {et~al.} 2005, \mnras, 362, 505

\bibitem[{{Di Dio} {et~al.}(2013){Di Dio}, Montanari, Lesgourgues, \&
  Durrer}]{Dio_2013}
{Di Dio}, E., Montanari, F., Lesgourgues, J., \& Durrer, R. 2013, \jcap, 11,
  044

\bibitem[{{Eisenstein} {et~al.}(2005){Eisenstein}, {Zehavi}, {Hogg},
  {Scoccimarro}, {Blanton}, {Nichol}, {Scranton}, {Seo}, {Tegmark}, {Zheng},
  {Anderson}, {Annis}, {Bahcall}, {Brinkmann}, {Burles}, {Castander},
  {Connolly}, {Csabai}, {Doi}, {Fukugita}, {Frieman}, {Glazebrook}, {Gunn},
  {Hendry}, {Hennessy}, {Ivezi{\'c}}, {Kent}, {Knapp}, {Lin}, {Loh}, {Lupton},
  {Margon}, {McKay}, {Meiksin}, {Munn}, {Pope}, {Richmond}, {Schlegel},
  {Schneider}, {Shimasaku}, {Stoughton}, {Strauss}, {SubbaRao}, {Szalay},
  {Szapudi}, {Tucker}, {Yanny}, \& {York}}]{Eisenstein+2005}
{Eisenstein}, D.~J., {Zehavi}, I., {Hogg}, D.~W., {et~al.} 2005, \apj, 633, 560

\bibitem[{Elkhashab {et~al.}(2021)Elkhashab, Porciani, \&
  Bertacca}]{Elkhashab_2021}
Elkhashab, M.~Y., Porciani, C., \& Bertacca, D. 2021, \mnras, 509, 1626

\bibitem[{{Euclid Collaboration: Blanchard} {et~al.}(2020){Euclid
  Collaboration: Blanchard}, {Camera}, {Carbone}, {Cardone}, {Casas}, {Clesse},
  {Ili{\'c}}, {Kilbinger}, {Kitching}, {Kunz}, {Lacasa}, {Linder}, {Majerotto},
  {Markovi{\v{c}}}, {Martinelli}, {Pettorino}, {Pourtsidou}, {Sakr},
  {S{\'a}nchez}, {Sapone}, {Tutusaus}, {Yahia-Cherif}, {Yankelevich},
  {Andreon}, {Aussel}, {Balaguera-Antol{\'\i}nez}, {Baldi}, {Bardelli},
  {Bender}, {Biviano}, {Bonino}, {Boucaud}, {Bozzo}, {Branchini}, {Brau-Nogue},
  {Brescia}, {Brinchmann}, {Burigana}, {Cabanac}, {Capobianco}, {Cappi},
  {Carretero}, {Carvalho}, {Casas}, {Castander}, {Castellano}, {Cavuoti},
  {Cimatti}, {Cledassou}, {Colodro-Conde}, {Congedo}, {Conselice}, {Conversi},
  {Copin}, {Corcione}, {Coupon}, {Courtois}, {Cropper}, {Da Silva}, {de la
  Torre}, {Di Ferdinando}, {Dubath}, {Ducret}, {Duncan}, {Dupac}, {Dusini},
  {Fabbian}, {Fabricius}, {Farrens}, {Fosalba}, {Fotopoulou}, {Fourmanoit},
  {Frailis}, {Franceschi}, {Franzetti}, {Fumana}, {Galeotta}, {Gillard},
  {Gillis}, {Giocoli}, {G{\'o}mez-Alvarez}, {Graci{\'a}-Carpio}, {Grupp},
  {Guzzo}, {Hoekstra}, {Hormuth}, {Israel}, {Jahnke}, {Keihanen}, {Kermiche},
  {Kirkpatrick}, {Kohley}, {Kubik}, {Kurki-Suonio}, {Ligori}, {Lilje}, {Lloro},
  {Maino}, {Maiorano}, {Marggraf}, {Martinet}, {Marulli}, {Massey},
  {Medinaceli}, {Mei}, {Mellier}, {Metcalf}, {Metge}, {Meylan}, {Moresco},
  {Moscardini}, {Munari}, {Nichol}, {Niemi}, {Nucita}, {Padilla}, {Paltani},
  {Pasian}, {Percival}, {Pires}, {Polenta}, {Poncet}, {Pozzetti}, {Racca},
  {Raison}, {Renzi}, {Rhodes}, {Romelli}, {Roncarelli}, {Rossetti}, {Saglia},
  {Schneider}, {Scottez}, {Secroun}, {Sirri}, {Stanco}, {Starck}, {Sureau},
  {Tallada-Cresp{\'\i}}, {Tavagnacco}, {Taylor}, {Tenti}, {Tereno},
  {Toledo-Moreo}, {Torradeflot}, {Valenziano}, {Vassallo}, {Verdoes Kleijn},
  {Viel}, {Wang}, {Zacchei}, {Zoubian}, \& {Zucca}}]{EuclidVII}
{Euclid Collaboration: Blanchard}, A., {Camera}, S., {Carbone}, C., {et~al.}
  2020, \aap, 642, A191

\bibitem[{{Euclid Collaboration: Borlaff} {et~al.}(2022){Euclid Collaboration:
  Borlaff}, {G{\'o}mez-Alvarez}, {Altieri}, {Marcum}, {Vavrek}, {Laureijs},
  {Kohley}, {Buitrago}, {Cuillandre}, {Duc}, {Gaspar Venancio}, {Amara},
  {Andreon}, {Auricchio}, {Azzollini}, {Baccigalupi},
  {Balaguera-Antol{\'\i}nez}, {Baldi}, {Bardelli}, {Bender}, {Biviano},
  {Bodendorf}, {Bonino}, {Bozzo}, {Branchini}, {Brescia}, {Brinchmann},
  {Burigana}, {Cabanac}, {Camera}, {Candini}, {Capobianco}, {Cappi}, {Carbone},
  {Carretero}, {Carvalho}, {Casas}, {Castander}, {Castellano}, {Castignani},
  {Cavuoti}, {Cimatti}, {Cledassou}, {Colodro-Conde}, {Congedo}, {Conselice},
  {Conversi}, {Copin}, {Corcione}, {Coupon}, {Courtois}, {Cropper}, {Da Silva},
  {Degaudenzi}, {Di Ferdinando}, {Douspis}, {Dubath}, {Duncan}, {Dupac},
  {Dusini}, {Ealet}, {Fabricius}, {Farina}, {Farrens}, {Ferreira}, {Ferriol},
  {Finelli}, {Flose-Reimberg}, {Fosalba}, {Frailis}, {Franceschi}, {Fumana},
  {Galeotta}, {Ganga}, {Garilli}, {Gillis}, {Giocoli}, {Gozaliasl},
  {Graci{\'a}-Carpio}, {Grazian}, {Grupp}, {Haugan}, {Holmes}, {Hormuth},
  {Jahnke}, {Keihanen}, {Kermiche}, {Kiessling}, {Kilbinger}, {Kirkpatrick},
  {Kitching}, {Knapen}, {Kubik}, {K{\"u}mmel}, {Kunz}, {Kurki-Suonio},
  {Liebing}, {Ligori}, {Lilje}, {Lindholm}, {Lloro}, {Mainetti}, {Maino},
  {Mansutti}, {Marggraf}, {Markovic}, {Martinelli}, {Martinet},
  {Mart{\'\i}nez-Delgado}, {Marulli}, {Massey}, {Maturi}, {Maurogordato},
  {Medinaceli}, {Mei}, {Meneghetti}, {Merlin}, {Metcalf}, {Meylan}, {Moresco},
  {Morgante}, {Moscardini}, {Munari}, {Nakajima}, {Neissner}, {Niemi},
  {Nightingale}, {Nucita}, {Padilla}, {Paltani}, {Pasian}, {Patrizii},
  {Pedersen}, {Percival}, {Pettorino}, {Pires}, {Poncet}, {Popa}, {Potter},
  {Pozzetti}, {Raison}, {Rebolo}, {Renzi}, {Rhodes}, {Riccio}, {Romelli},
  {Roncarelli}, {Rosset}, {Rossetti}, {Saglia}, {S{\'a}nchez}, {Sapone},
  {Sauvage}, {Schneider}, {Scottez}, {Secroun}, {Seidel}, {Serrano},
  {Sirignano}, {Sirri}, {Skottfelt}, {Stanco}, {Starck}, {Sureau},
  {Tallada-Cresp{\'\i}}, {Taylor}, {Tenti}, {Tereno}, {Teyssier},
  {Toledo-Moreo}, {Torradeflot}, {Tutusaus}, {Valentijn}, {Valenziano},
  {Valiviita}, {Vassallo}, {Viel}, {Wang}, {Weller}, {Whittaker}, {Zacchei},
  {Zamorani}, \& {Zucca}}]{Borlaf_2022}
{Euclid Collaboration: Borlaff}, A.~S., {G{\'o}mez-Alvarez}, P., {Altieri}, B.,
  {et~al.} 2022, \aap, 657, A92

\bibitem[{{Euclid Collaboration: Jelic-Cizmek} {et~al.}(2023){Euclid
  Collaboration: Jelic-Cizmek}, Sorrenti, Lepori, Bonvin, Camera, Castander,
  Durrer, Fosalba, Kunz, Lombriser, Tutusaus, Viglione, Sakr, Aghanim, Amara,
  Andreon, Baldi, Bardelli, Bodendorf, Bonino, Branchini, Brescia, Brinchmann,
  Capobianco, Carbone, Cardone, Carretero, Casas, Castellano, Cavuoti, Cimatti,
  Congedo, Conselice, Conversi, Copin, Corcione, Courbin, Courtois, Cropper,
  Degaudenzi, Giorgio, Dinis, Dubath, Dupac, Dusini, Farina, Farrens, Ferriol,
  Frailis, Franceschi, Fumana, Galeotta, Garilli, Gillis, Giocoli, Grazian,
  Grupp, Haugan, Hoekstra, Holmes, Hormuth, Hornstrup, Jahnke, Keihänen,
  Kermiche, Kiessling, Kilbinger, Kubik, Kurki-Suonio, Lilje, Lindholm, Lloro,
  Mansutti, Marggraf, Markovic, Martinet, Marulli, Massey, Medinaceli, Mei,
  Meneghetti, Merlin, Meylan, Moscardini, Munari, Niemi, Padilla, Paltani,
  Pasian, Pedersen, Percival, Pettorino, Polenta, Poncet, Popa, Raison, Rebolo,
  Renzi, Rhodes, Riccio, Romelli, Roncarelli, Rossetti, Saglia, Sapone,
  Sartoris, Schneider, Schrabback, Secroun, Seidel, Serrano, Sirignano, Sirri,
  Stanco, Starck, Surace, Tallada-Crespí, Tavagnacco, Taylor, Tereno,
  Toledo-Moreo, Torradeflot, Valentijn, Valenziano, Vassallo, Veropalumbo,
  Wang, Weller, Zamorani, Zoubian, Zucca, Biviano, Boucaud, Bozzo,
  Colodro-Conde, Ferdinando, Graciá-Carpio, Liebing, Mauri, Neissner, Scottez,
  Tenti, Viel, Wiesmann, Akrami, Allevato, Anselmi, Baccigalupi,
  Balaguera-Antolínez, Ballardini, Bruton, Burigana, Cabanac, Cappi, Carvalho,
  Castignani, Castro, {n}as Herrera, Chambers, Cooray, Coupon, Davini, de~la
  Torre, Lucia, Desprez, Domizio, Dole, Díaz-Sánchez, Vigo, Escoffier,
  Ferreira, Ferrero, Finelli, Gabarra, Ganga, García-Bellido, Giacomini,
  Gozaliasl, Guinet, Hildebrandt, Ilić, {n}oz, Joudaki, Kajava, Kansal,
  Kirkpatrick, Legrand, Loureiro, Magliocchetti, Mainetti, Maoli, Martinelli,
  Martins, Matthew, Maturi, Maurin, Metcalf, Migliaccio, Monaco, Morgante,
  Nadathur, Patrizii, Pezzotta, Popa, Porciani, Potter, Pöntinen, Reimberg,
  Rocci, Sánchez, Schneider, Schultheis, Sefusatti, Sereno, Silvestri, Simon,
  Mancini, Steinwagner, Testera, Tewes, Teyssier, Toft, Tosi, Troja, Tucci,
  Valiviita, Vergani, \& Tanidis}]{wp9_goran}
{Euclid Collaboration: Jelic-Cizmek}, G., Sorrenti, F., Lepori, F., {et~al.}
  2023, arXiv:2311.03168

\bibitem[{{Euclid Collaboration: Lepori} {et~al.}(2022){Euclid Collaboration:
  Lepori}, {Tutusaus}, {Viglione}, {Bonvin}, {Camera}, {Castander}, {Durrer},
  {Fosalba}, {Jelic-Cizmek}, {Kunz}, {Adamek}, {Casas}, {Martinelli}, {Sakr},
  {Sapone}, {Amara}, {Auricchio}, {Bodendorf}, {Bonino}, {Branchini},
  {Brescia}, {Brinchmann}, {Capobianco}, {Carbone}, {Carretero}, {Castellano},
  {Cavuoti}, {Cimatti}, {Cledassou}, {Congedo}, {Conselice}, {Conversi},
  {Copin}, {Corcione}, {Courbin}, {Da Silva}, {Degaudenzi}, {Douspis},
  {Dubath}, {Dupac}, {Dusini}, {Ealet}, {Farrens}, {Ferriol}, {Franceschi},
  {Fumana}, {Garilli}, {Gillard}, {Gillis}, {Giocoli}, {Grazian}, {Grupp},
  {Guzzo}, {Haugan}, {Holmes}, {Hormuth}, {Hudelot}, {Jahnke}, {Kermiche},
  {Kiessling}, {Kilbinger}, {Kitching}, {K{\"u}mmel}, {Kurki-Suonio}, {Ligori},
  {Lilje}, {Lloro}, {Mansutti}, {Marggraf}, {Markovic}, {Marulli}, {Massey},
  {Maurogordato}, {Melchior}, {Meneghetti}, {Merlin}, {Meylan}, {Moresco},
  {Moscardini}, {Munari}, {Nakajima}, {Niemi}, {Padilla}, {Paltani}, {Pasian},
  {Pedersen}, {Percival}, {Pettorino}, {Pires}, {Poncet}, {Popa}, {Pozzetti},
  {Raison}, {Rhodes}, {Roncarelli}, {Rossetti}, {Saglia}, {Schneider},
  {Secroun}, {Seidel}, {Serrano}, {Sirignano}, {Sirri}, {Stanco}, {Starck},
  {Tallada-Cresp{\'\i}}, {Taylor}, {Tereno}, {Toledo-Moreo}, {Torradeflot},
  {Valentijn}, {Valenziano}, {Wang}, {Weller}, {Zamorani}, {Zoubian},
  {Andreon}, {Bardelli}, {Fabbian}, {Graci{\'a}-Carpio}, {Maino}, {Medinaceli},
  {Mei}, {Renzi}, {Romelli}, {Sureau}, {Vassallo}, {Zacchei}, {Zucca},
  {Baccigalupi}, {Balaguera-Antol{\'\i}nez}, {Bernardeau}, {Biviano},
  {Blanchard}, {Bolzonella}, {Borgani}, {Bozzo}, {Burigana}, {Cabanac},
  {Cappi}, {Carvalho}, {Castignani}, {Colodro-Conde}, {Coupon}, {Courtois},
  {Cuby}, {Davini}, {de la Torre}, {Di Ferdinando}, {Farina}, {Ferreira},
  {Finelli}, {Galeotta}, {Ganga}, {Garcia-Bellido}, {Gaztanaga}, {Gozaliasl},
  {Hook}, {Ili{\'c}}, {Joachimi}, {Kansal}, {Keihanen}, {Kirkpatrick},
  {Lindholm}, {Mainetti}, {Maoli}, {Martinet}, {Maturi}, {Metcalf}, {Monaco},
  {Morgante}, {Nightingale}, {Nucita}, {Patrizii}, {Popa}, {Potter}, {Riccio},
  {S{\'a}nchez}, {Schirmer}, {Schultheis}, {Scottez}, {Sefusatti}, {Tramacere},
  {Valiviita}, {Viel}, \& {Hildebrandt}}]{EUCLID_PHOTOMETRIC_WP9}
{Euclid Collaboration: Lepori}, F., {Tutusaus}, I., {Viglione}, C., {et~al.}
  2022, \aap, 662, A93

\bibitem[{{Euclid Collaboration: Mellier} {et~al.}(2024){Euclid Collaboration:
  Mellier}, Abdurro'uf, {Acevedo Barroso}, Achúcarro, Adamek, Adam, Addison,
  Aghanim, Aguena, Ajani, Akrami, Al-Bahlawan, Alavi, Albuquerque, Alestas,
  Alguero, Allaoui, Allen, Allevato, Alonso-Tetilla, Altieri, Alvarez-Candal,
  Alvi, Amara, Amendola, Amiaux, Andika, Andreon, Andrews, Angora, Angulo,
  Annibali, Anselmi, Anselmi, Arcari, Archidiacono, Aricò, Arnaud, Arnouts,
  Asgari, Asorey, Atayde, Atek, Atrio-Barandela, Aubert, Aubourg, Auphan,
  Auricchio, Aussel, Aussel, Avelino, Avgoustidis, Avila, Awan, Azzollini,
  Baccigalupi, Bachelet, Bacon, Baes, Bagley, Bahr-Kalus, Balaguera-Antolinez,
  Balbinot, Balcells, Baldi, Baldry, Balestra, Ballardini, Ballester, Balogh,
  Bañados, Barbier, Bardelli, Baron, Barreiro, Barrena, Barriere, Barros,
  Barthelemy, Bartolo, Basset, Battaglia, Battisti, Baugh, Baumont, Bazzanini,
  Beaulieu, Beckmann, Belikov, Bel, Bellagamba, Bella, Bellini, Benabed,
  Bender, Benevento, Bennett, Benson, Bergamini, Bermejo-Climent, Bernardeau,
  Bertacca, Berthe, Berthier, Bethermin, Beutler, Bevillon, Bhargava,
  Bhatawdekar, Bianchi, Bisigello, Biviano, Blake, Blanchard, Blazek, Blot,
  Bosco, Bodendorf, Boenke, Böhringer, Boldrini, Bolzonella, Bonchi, Bonici,
  Bonino, Bonino, Bonvin, Bon, Booth, Borgani, Borlaff, Borsato, Bosco, Bose,
  Botticella, Boucaud, Bouche, Boucher, Boutigny, Bouvard, Bouwens, Bouy,
  Bowler, Bozza, Bozzo, Branchini, Brando, Brau-Nogue, Brekke, Bremer, Brescia,
  Breton, Brinchmann, Brinckmann, Brockley-Blatt, Brodwin, Brouard, Brown,
  Bruton, Bucko, Buddelmeijer, Buenadicha, Buitrago, Burger, Burigana, Busillo,
  Busonero, Cabanac, Cabayol-Garcia, Cagliari, Caillat, Caillat, Calabrese,
  Calabro, Calderone, Calura, Quevedo, Camera, Campos, Canas-Herrera, Candini,
  Cantiello, Capobianco, Cappellaro, Cappelluti, Cappi, Caputi, Cara, Carbone,
  Cardone, Carella, Carlberg, Carle, Carminati, Caro, Carrasco, Carretero,
  Carrilho, Duque, Carry, Carvalho, Carvalho, Casas, Casas, Casenove, Casey,
  Cassata, Castander, Castelao, Castellano, Castiblanco, Castignani, Castro,
  Cavet, Cavuoti, Chabaud, Chambers, Charles, Charlot, Chartab, Chary,
  Chaumeil, Cho, Chon, Ciancetta, Ciliegi, Cimatti, Cimino, Cioni, Claydon,
  Cleland, Clément, Clements, Clerc, Clesse, Codis, Cogato, Colbert, Cole,
  Coles, Collett, Collins, Colodro-Conde, Colombo, Combes, Conforti, Congedo,
  Conseil, Conselice, Contarini, Contini, Conversi, Cooray, Copin, Corasaniti,
  Corcho-Caballero, Corcione, Cordes, Corpace, Correnti, Costanzi, Costille,
  Courbin, Mifsud, Courtois, Cousinou, Covone, Cowell, Cragg, Cresci,
  Cristiani, Crocce, Cropper, Crouzet, Csizi, Cuby, Cucchetti, Cucciati,
  Cuillandre, Cunha, Cuozzo, Daddi, D'Addona, Dafonte, Dagoneau, Dalessandro,
  Dalton, D'Amico, Dannerbauer, Danto, Das, Silva, da~Silva, d'Assignies
  Doumerg, Daste, Davies, Davini, Dayal, de~Boer, Decarli, Caro, Degaudenzi,
  Degni, de~Jong, de~la Bella, de~la Torre, Delhaise, Delley, Delucchi, Lucia,
  Denniston, Paolis, Petris, Derosa, Desai, Desjacques, Despali, Desprez,
  Vicente-Albendea, Deville, Dias, Díaz-Sánchez, Diaz, Domizio, Diego,
  Ferdinando, Giorgio, Dimauro, Dinis, Dolag, Dolding, Dole, Sánchez, Doré,
  Dournac, Douspis, Dreihahn, Droge, Dryer, Dubath, Duc, Ducret, Duffy,
  Dufresne, Duncan, Dupac, Duret, Durrer, Durret, Dusini, Ealet, Eggemeier,
  Eisenhardt, Elbaz, Elkhashab, Ellien, Endicott, Enia, Erben, Vigo, Escoffier,
  Sanz, Essert, Ettori, Ezziati, Fabbian, Fabricius, Fang, Farina, Farina,
  Farinelli, Farrens, Faustini, Feltre, Ferguson, Ferrando, Ferrari,
  Ferré-Mateu, Ferreira, Ferreras, Ferrero, Ferriol, Ferruit, Filleul,
  Finelli, Finkelstein, Finoguenov, Fiorini, Flentge, Focardi, Fonseca,
  Fontana, Fontanot, Fornari, Fosalba, Fossati, Fotopoulou, Fouchez,
  Fourmanoit, Frailis, Fraix-Burnet, Franceschi, Franco, Franzetti, Freihoefer,
  Frenk, Frittoli, Frugier, Frusciante, Fumagalli, Fumagalli, Fumana, Fu,
  Gabarra, Galeotta, Galluccio, Ganga, Gao, García-Bellido, Garcia, Gardner,
  Garilli, Gaspar-Venancio, Gasparetto, Gautard, Gavazzi, Gaztanaga, Genolet,
  Santos, Gentile, George, Gerbino, Ghaffari, Giacomini, Gianotti, Gibb,
  Gillard, Gillis, Ginolfi, Giocoli, Girardi, Giri, Goh, Gómez-Alvarez,
  Gonzalez-Perez, Gonzalez, Gonzalez, Gonzalez, Beauchamps, Gozaliasl,
  Gracia-Carpio, Grandis, Granett, Granvik, Grazian, Gregorio, Grenet, Grillo,
  Grupp, Gruppioni, Gruppuso, Guerbuez, Guerrini, Guidi, Guillard, Gutierrez,
  Guttridge, Guzzo, Gwyn, Haapala, Haase, Haddow, Hailey, Hall, Hall, Hamaus,
  Haridasu, Harnois-Déraps, Harper, Hartley, Hasinger, Hassani, Hatch, Haugan,
  Häußler, Heavens, Heisenberg, Helmi, Helou, Hemmati, Henares, Herent,
  Hernández-Monteagudo, Heuberger, Hewett, Heydenreich, Hildebrandt,
  Hirschmann, Hjorth, Hoar, Hoekstra, Holland, Holliman, Holmes, Hook, Horeau,
  Hormuth, Hornstrup, Hosseini, Hu, Hudelot, Hudson, Huertas-Company, Huff,
  Hughes, Humphrey, Hunt, Huynh, Ibata, Ichikawa, Iglesias-Groth, Ilbert,
  Ilić, Ingoglia, Iodice, Israel, Israelsson, Izzo, Jablonka, Jackson,
  Jacobson, Jafariyazani, Jahnke, Jain, Jansen, Jarvis, Jasche, Jauzac,
  Jeffrey, Jhabvala, Jimenez-Teja, Muñoz, Joachimi, Johansson, Joudaki, Jullo,
  Kajava, Kang, Kannawadi, Kansal, Karagiannis, Kärcher, Kashlinsky,
  Kazandjian, Keck, Keihänen, Kerins, Kermiche, Khalil, Kiessling, Kiiveri,
  Kilbinger, Kim, King, Kirkpatrick, Kitching, Kluge, Knabenhans, Knapen,
  Knebe, Kneib, Kohley, Koopmans, Koskinen, Koulouridis, Kou, Kovács,
  Kovačić, Kowalczyk, Koyama, Kraljic, Krause, Kruk, Kubik, Kuchner, Kuijken,
  Kümmel, Kunz, Kurki-Suonio, Lacasa, Lacey, Franca, Lagarde, Lahav, Laigle,
  Marca, Marle, Lamine, Lam, Lançon, Landt, Langer, Lapi, Larcheveque, Larsen,
  Lattanzi, Laudisio, Laugier, Laureijs, Laurent, Lavaux, Lawrenson, Lazanu,
  Lazeyras, Boulc'h, Brun, Brun, Leclercq, Lee, Graet, Legrand, Leirvik, Jeune,
  Lembo, Mignant, Lepinzan, Lepori, Reun, Leroy, Lesci, Lesgourgues, Leuzzi,
  Levi, Liaudat, Libet, Liebing, Ligori, Lilje, Lin, Linde, Linder, Lindholm,
  Linke, Li, Liu, Lloro, Lobo, Lodieu, Lombardi, Lombriser, Lonare, Longo,
  López-Caniego, Lopez, Alvarez, Loureiro, Loveday, Lusso, Macias-Perez,
  Maciaszek, Maggio, Magliocchetti, Magnard, Magnier, Magro, Mahler, Mainetti,
  Maino, Maiorano, Maiorano, Malavasi, Mamon, Mancini, Mandelbaum, Manera,
  Manjón-García, Mannucci, Mansutti, Outeiro, Maoli, Maraston, Marcin,
  Marcos-Arenal, Margalef-Bentabol, Marggraf, Marinucci, Marinucci, Markovic,
  Marleau, Marpaud, Martignac, Martín-Fleitas, Martin-Moruno, Martin,
  Martinelli, Martinet, Martin, Martins, Marulli, Massari, Massey, Masters,
  Matarrese, Matsuoka, Matthew, Maughan, Mauri, Maurin, Maurogordato, McCarthy,
  McConnachie, McCracken, McDonald, McEwen, McPartland, Medinaceli, Mehta, Mei,
  Melchior, Melin, Ménard, Mendes, Mendez-Abreu, Meneghetti, Mercurio, Merlin,
  Metcalf, Meylan, Migliaccio, Mignoli, Miller, Miluzio, Milvang-Jensen,
  Mimoso, Miquel, Miyatake, Mobasher, Mohr, Monaco, Monguió, Montoro, Mora,
  Dizgah, Moresco, Moretti, Morgante, Morisset, Moriya, Morris, Mortlock,
  Moscardini, Mota, Mottet, Moustakas, Moutard, Müller, Munari, Murphree,
  Murray, Murray, Musi, Nadathur, Nagam, Nagao, Naidoo, Nakajima, Nally,
  Natoli, Navarro-Alsina, Girones, Neissner, Nersesian, Nesseris, Nguyen-Kim,
  Nicastro, Nichol, Nielbock, Niemi, Nieto, Nilsson, Noller, Norberg,
  Nouri-Zonoz, Ntelis, Nucita, Nugent, Nunes, Nutma, Ocampo, Odier, Oesch,
  Oguri, Oliveira, Onoue, Oosterbroek, Oppizzi, Ordenovic, Osato, Pacaud, Pace,
  Padilla, Paech, Pagano, Page, Palazzi, Paltani, Pamuk, Pandolfi, Paoletti,
  Paolillo, Papaderos, Pardede, Parimbelli, Parmar, Partmann, Pasian,
  Passalacqua, Paterson, Patrizii, Pattison, Paulino-Afonso, Paviot, Peacock,
  Pearce, Pedersen, Peel, Peletier, Ibanez, Pello, Penny, Percival,
  Perez-Garrido, Perotto, Pettorino, Pezzotta, Pezzuto, Philippon, Pierre,
  Piersanti, Pietroni, Piga, Pilo, Pires, Pisani, Pizzella, Pizzuti, Plana,
  Polenta, Pollack, Poncet, Pöntinen, Pool, Popa, Popa, Popp, Porciani, Porth,
  Potter, Poulain, Pourtsidou, Pozzetti, Prandoni, Pratt, Prezelus, Prieto,
  Pugno, Quai, Quilley, Racca, Raccanelli, Rácz, Radinović, Radovich,
  Ragagnin, Ragnit, Raison, Ramos-Chernenko, Ranc, Rasera, Raylet, Rebolo,
  Refregier, Reimberg, Reiprich, Renk, Renzi, Retre, Revaz, Reylé, Reynolds,
  Rhodes, Ricci, Ricci, Riccio, Ricken, Rissanen, Risso, Rix, Robin,
  Rocca-Volmerange, Rocci, Rodenhuis, Rodighiero, Monroy, Rollins, Romanello,
  Roman, Romelli, Romero-Gomez, Roncarelli, Rosati, Rosset, Rossetti, Roster,
  Rottgering, Rozas-Fernández, Ruane, Rubino-Martin, Rudolph, Ruppin,
  Rusholme, Sacquegna, Sáez-Casares, Saga, Saglia, Sahlén, Saifollahi, Sakr,
  Salvalaggio, Salvaterra, Salvati, Salvato, Salvignol, Sánchez, Sanchez,
  Sanders, Sapone, Saponara, Sarpa, Sarron, Sartori, Sartoris, Sassolas,
  Sauniere, Sauvage, Sawicki, Scaramella, Scarlata, Scharré, Schaye,
  Schewtschenko, Schindler, Schinnerer, Schirmer, Schmidt, Schmidt, Schmidt,
  Schneider, Schneider, Schneider, Schöneberg, Schrabback, Schultheis, Schulz,
  Schuster, Schwartz, Sciotti, Scodeggio, Scognamiglio, Scott, Scottez,
  Secroun, Sefusatti, Seidel, Seiffert, Sellentin, Selwood, Semboloni, Sereno,
  Serjeant, Serrano, Setnikar, Shankar, Sharples, Short, Shulevski, Shuntov,
  Sias, Sikkema, Silvestri, Simon, Sirignano, Sirri, Skottfelt, Slezak, Sluse,
  Smith, Smith, Smith, Smit, Soldano, Solheim, Sorce, Sorrenti, Soubrie,
  Spinoglio, Mancini, Stadel, Stagnaro, Stanco, Stanford, Starck, Stassi,
  Steinwagner, Stern, Stone, Strada, Strafella, Stramaccioni, Surace, Sureau,
  Suyu, Swindells, Szafraniec, Szapudi, Taamoli, Talia, Tallada-Crespí,
  Tanidis, Tao, Tarrío, Tavagnacco, Taylor, Taylor, Taylor, Teixeira, Tenti,
  Idiago, Teplitz, Tereno, Tessore, Testa, Testera, Tewes, Teyssier, Theret,
  Thizy, Thomas, Toba, Toft, Toledo-Moreo, Tolstoy, Tommasi, Torbaniuk,
  Torradeflot, Tortora, Tosi, Tosti, Trifoglio, Troja, Trombetti, Tronconi,
  Tsedrik, Tsyganov, Tucci, Tutusaus, Uhlemann, Ulivi, Urbano, Vacher, Vaillon,
  Valageas, Valdes, Valentijn, Valenziano, Valieri, Valiviita, den Broeck,
  Vassallo, Vavrek, Vega-Ferrero, Venemans, Venhola, Ventura, Kleijn, Vergani,
  Verma, Vernizzi, Veropalumbo, Verza, Vescovi, Vibert, Viel, Vielzeuf,
  Viglione, Viitanen, Villaescusa-Navarro, Vinciguerra, Visticot, Voggel, von
  Wietersheim-Kramsta, Vriend, Wachter, Walmsley, Walth, Walton, Walton,
  Wander, Wang, Wang, Weaver, Weller, Wetzstein, Whalen, Whittam, Widmer,
  Wiesmann, Wilde, Williams, Winther, Wittje, Wong, Wright, Yankelevich, Yeung,
  Yoon, Youles, Yung, Zacchei, Zalesky, Zamorani, Vitorelli, Marc, Zennaro,
  Zerbi, Zinchenko, Zoubian, Zucca, \&
  Zumalacarregui}]{euclidcollaboration2024euclidiovervieweuclid}
{Euclid Collaboration: Mellier}, Y., Abdurro'uf, {Acevedo Barroso}, J.~A.,
  {et~al.} 2024, A\&A, submitted, arXiv:2405.13491

\bibitem[{{Euclid Collaboration: Scaramella } {et~al.}(2022){Euclid
  Collaboration: Scaramella }, {Amiaux}, {Mellier}, {Burigana}, {Carvalho},
  {Cuillandre}, {Da Silva}, {Derosa}, {Dinis}, {Maiorano}, {Maris}, {Tereno},
  {Laureijs}, {Boenke}, {Buenadicha}, {Dupac}, {Gaspar Venancio},
  {G{\'o}mez-{\'A}lvarez}, {Hoar}, {Lorenzo Alvarez}, {Racca},
  {Saavedra-Criado}, {Schwartz}, {Vavrek}, {Schirmer}, {Aussel}, {Azzollini},
  {Cardone}, {Cropper}, {Ealet}, {Garilli}, {Gillard}, {Granett}, {Guzzo},
  {Hoekstra}, {Jahnke}, {Kitching}, {Maciaszek}, {Meneghetti}, {Miller},
  {Nakajima}, {Niemi}, {Pasian}, {Percival}, {Pottinger}, {Sauvage},
  {Scodeggio}, {Wachter}, {Zacchei}, {Aghanim}, {Amara}, {Auphan}, {Auricchio},
  {Awan}, {Balestra}, {Bender}, {Bodendorf}, {Bonino}, {Branchini},
  {Brau-Nogue}, {Brescia}, {Candini}, {Capobianco}, {Carbone}, {Carlberg},
  {Carretero}, {Casas}, {Castander}, {Castellano}, {Cavuoti}, {Cimatti},
  {Cledassou}, {Congedo}, {Conselice}, {Conversi}, {Copin}, {Corcione},
  {Costille}, {Courbin}, {Degaudenzi}, {Douspis}, {Dubath}, {Duncan}, {Dusini},
  {Farrens}, {Ferriol}, {Fosalba}, {Fourmanoit}, {Frailis}, {Franceschi},
  {Franzetti}, {Fumana}, {Gillis}, {Giocoli}, {Grazian}, {Grupp}, {Haugan},
  {Holmes}, {Hormuth}, {Hudelot}, {Kermiche}, {Kiessling}, {Kilbinger},
  {Kohley}, {Kubik}, {K{\"u}mmel}, {Kunz}, {Kurki-Suonio}, {Lahav}, {Ligori},
  {Lilje}, {Lloro}, {Mansutti}, {Marggraf}, {Markovic}, {Marulli}, {Massey},
  {Maurogordato}, {Melchior}, {Merlin}, {Meylan}, {Mohr}, {Moresco}, {Morin},
  {Moscardini}, {Munari}, {Nichol}, {Padilla}, {Paltani}, {Peacock},
  {Pedersen}, {Pettorino}, {Pires}, {Poncet}, {Popa}, {Pozzetti}, {Raison},
  {Rebolo}, {Rhodes}, {Rix}, {Roncarelli}, {Rossetti}, {Saglia}, {Schneider},
  {Schrabback}, {Secroun}, {Seidel}, {Serrano}, {Sirignano}, {Sirri},
  {Skottfelt}, {Stanco}, {Starck}, {Tallada-Cresp{\'\i}}, {Tavagnacco},
  {Taylor}, {Teplitz}, {Toledo-Moreo}, {Torradeflot}, {Trifoglio}, {Valentijn},
  {Valenziano}, {Verdoes Kleijn}, {Wang}, {Welikala}, {Weller}, {Wetzstein},
  {Zamorani}, {Zoubian}, {Andreon}, {Baldi}, {Bardelli}, {Boucaud}, {Camera},
  {Di Ferdinando}, {Fabbian}, {Farinelli}, {Galeotta}, {Graci{\'a}-Carpio},
  {Maino}, {Medinaceli}, {Mei}, {Neissner}, {Polenta}, {Renzi}, {Romelli},
  {Rosset}, {Sureau}, {Tenti}, {Vassallo}, {Zucca}, {Baccigalupi},
  {Balaguera-Antol{\'\i}nez}, {Battaglia}, {Biviano}, {Borgani}, {Bozzo},
  {Cabanac}, {Cappi}, {Casas}, {Castignani}, {Colodro-Conde}, {Coupon},
  {Courtois}, {Cuby}, {de la Torre}, {Desai}, {Dole}, {Fabricius}, {Farina},
  {Ferreira}, {Finelli}, {Flose-Reimberg}, {Fotopoulou}, {Ganga}, {Gozaliasl},
  {Hook}, {Keihanen}, {Kirkpatrick}, {Liebing}, {Lindholm}, {Mainetti},
  {Martinelli}, {Martinet}, {Maturi}, {McCracken}, {Metcalf}, {Morgante},
  {Nightingale}, {Nucita}, {Patrizii}, {Potter}, {Riccio}, {S{\'a}nchez},
  {Sapone}, {Schewtschenko}, {Schultheis}, {Scottez}, {Teyssier}, {Tutusaus},
  {Valiviita}, {Viel}, {Vriend}, \& {Whittaker}}]{Euclid_Wide_survey}
{Euclid Collaboration: Scaramella }, R., {Amiaux}, J., {Mellier}, Y., {et~al.}
  2022, \aap, 662, A112

\bibitem[{{Euclid Collaboration: Schirmer} {et~al.}(2022){Euclid Collaboration:
  Schirmer}, {Jahnke}, {Seidel}, {Aussel}, {Bodendorf}, {Grupp}, {Hormuth},
  {Wachter}, {Appleton}, {Barbier}, {Brinchmann}, {Carrasco}, {Castander},
  {Coupon}, {De Paolis}, {Franco}, {Ganga}, {Hudelot}, {Jullo}, {Lan{\c{c}}on},
  {Nucita}, {Paltani}, {Smadja}, {Strafella}, {Venancio}, {Weiler}, {Amara},
  {Auphan}, {Auricchio}, {Balestra}, {Bender}, {Bonino}, {Branchini},
  {Brescia}, {Capobianco}, {Carbone}, {Carretero}, {Casas}, {Castellano},
  {Cavuoti}, {Cimatti}, {Cledassou}, {Congedo}, {Conselice}, {Conversi},
  {Copin}, {Corcione}, {Costille}, {Courbin}, {Da Silva}, {Degaudenzi},
  {Douspis}, {Dubath}, {Dupac}, {Dusini}, {Ealet}, {Farrens}, {Ferriol},
  {Fosalba}, {Frailis}, {Franceschi}, {Franzetti}, {Fumana}, {Garilli},
  {Gillard}, {Gillis}, {Giocoli}, {Grazian}, {Guzzo}, {Haugan}, {Hoekstra},
  {Holmes}, {Hornstrup}, {K{\"u}mmel}, {Kermiche}, {Kiessling}, {Kilbinger},
  {Kitching}, {Kohley}, {Kunz}, {Kurki-Suonio}, {Laureijs}, {Ligori}, {Lilje},
  {Lloro}, {Maciaszek}, {Maiorano}, {Mansutti}, {Marggraf}, {Markovic},
  {Marulli}, {Massey}, {Maurogordato}, {Mellier}, {Meneghetti}, {Merlin},
  {Meylan}, {Moresco}, {Moscardini}, {Munari}, {Nakajima}, {Nichol}, {Niemi},
  {Padilla}, {Pasian}, {Pedersen}, {Percival}, {Pettorino}, {Pires}, {Poncet},
  {Popa}, {Pozzetti}, {Prieto}, {Raison}, {Rhodes}, {Rix}, {Roncarelli},
  {Rossetti}, {Saglia}, {Sartoris}, {Scaramella}, {Schneider}, {Secroun},
  {Serrano}, {Sirignano}, {Sirri}, {Stanco}, {Tallada-Cresp{\'\i}}, {Taylor},
  {Teplitz}, {Tereno}, {Toledo-Moreo}, {Torradeflot}, {Trifoglio}, {Valentijn},
  {Valenziano}, {Wang}, {Weller}, {Zamorani}, {Zoubian}, {Andreon}, {Bardelli},
  {Boucaud}, {Camera}, {Farinelli}, {Graci{\'a}-Carpio}, {Maino}, {Medinaceli},
  {Mei}, {Morisset}, {Polenta}, {Renzi}, {Romelli}, {Tenti}, {Vassallo},
  {Zacchei}, {Zucca}, {Baccigalupi}, {Balaguera-Antol{\'\i}nez}, {Biviano},
  {Blanchard}, {Borgani}, {Bozzo}, {Burigana}, {Cabanac}, {Cappi}, {Carvalho},
  {Casas}, {Castignani}, {Colodro-Conde}, {Cooray}, {Courtois}, {Crocce},
  {Cuby}, {Davini}, {de la Torre}, {Di Ferdinando}, {Escartin}, {Farina},
  {Ferreira}, {Finelli}, {Fotopoulou}, {Galeotta}, {Garcia-Bellido},
  {Gaztanaga}, {George}, {Gozaliasl}, {Hook}, {Ili{\'c}}, {Kansal},
  {Kashlinsky}, {Keihanen}, {Kirkpatrick}, {Lindholm}, {Mainetti}, {Maoli},
  {Martinelli}, {Martinet}, {Maturi}, {Mauri}, {McCracken}, {Metcalf},
  {Monaco}, {Morgante}, {Nightingale}, {Patrizii}, {Peel}, {Popa}, {Porciani},
  {Potter}, {Reimberg}, {Riccio}, {S{\'a}nchez}, {Sapone}, {Scottez},
  {Sefusatti}, {Teyssier}, {Tutusaus}, {Valieri}, {Valiviita}, {Viel}, \&
  {Hildebrandt}}]{Schirmer_2022}
{Euclid Collaboration: Schirmer}, M., {Jahnke}, K., {Seidel}, G., {et~al.}
  2022, \aap, 662, A92

\bibitem[{{Euclid Collaboration: Tanidis} {et~al.}(2024){Euclid Collaboration:
  Tanidis}, Cardone, Martinelli, Tutusaus, Camera, Aghanim, Amara, Andreon,
  Auricchio, Baldi, Bardelli, Branchini, Brescia, Brinchmann, Capobianco,
  Carbone, Carretero, Casas, Castellano, Cavuoti, Cimatti, Cledassou, Congedo,
  Conversi, Copin, Corcione, Courbin, Courtois, Da~Silvay, Degaudenzi, Dinis,
  Dubath, Dupac, Dusini, Farina, Farrens, Ferriol, Fosalba, Frailis,
  Franceschi, Fumana, Galeotta, Garilli, Gillard, Gillis, Giocoli, Grazian,
  Grupp, Guzzo, Haugan, Holmes, Hook, Hornstrup, Jahnke, Joachimi, Keihanen,
  Kermiche, Kiessling, Kunz, Kurki-Suonio, Lilje, Lindholm, Lloro, Maiorano,
  Mansutti, Marggraf, Markovic, Martinet, Marulli, Massey, Maurogordato,
  Medinaceli, Mei, Meneghetti, Meylan, Moresco, Moscardini, Munari, Niemi,
  Padilla, Paltani, Pasian, Pedersen, Percival, Pettorino, Pires, Polenta,
  Pollack, Poncet, Popa, Raison, Renzi, Rhodes, Riccio, Romelli, Roncarelli,
  Rossetti, Saglia, Sapone, Sartoris, Schirmer, Schneider, Secroun, Seidel,
  Serrano, Sirignano, Sirri, Stanco, Tallada-Crespí, Taylor, Tereno,
  Toledo-Moreo, Torradeflot, Valentijn, Valenziano, Vassallo, Veropalumbo,
  Wang, Weller, Zamorani, Zoubian, Zucca, Biviano, Boucaud, Bozzo,
  Colodro-Conde, Di~Ferdinando, Farinelli, Graciá-Carpio, Marcin, Mauri,
  Scottez, Tenti, Tramacere, Akrami, Allevato, Baccigalupi,
  Balaguera-Antolínez, Ballardini, Benielli, Bernardeau, Borgani, Borlaff,
  Burigana, Cabanac, Cappi, Carvalho, Castignani, Castro, Cañas-Herrera,
  Chambers, Cooray, Coupon, Díaz-Sánchez, Davini, de~la Torre, De~Lucia,
  Desprez, Di~Domizio, Dole, Escartin~Vigo, Escoffier, Ferreira, Ferrero,
  Finelli, Gabarra, García-Bellido, Gaztanaga, Giacomini, Gozaliasl,
  Hildebrandt, Ilić, Kajava, Kansal, Kirkpatrick, Legrand, Loureiro,
  Macias-Perez, Magliocchetti, Mainetti, Maoli, Martins, Matthew, Maurin,
  Metcalf, Migliaccio, Monaco, Morgante, Nadathur, Nucita, Pöntinen, Patrizii,
  Pezzotta, Popa, Potter, Sánchez, Sakr, Schewtschenko, Schneider, Sereno,
  Simon, Spurio~Mancini, Steinwagner, Tewes, Teyssier, Toft, Valiviita, Viel,
  \& Linke}]{EUCLID_PHOTOMETRIC_TANIDIS}
{Euclid Collaboration: Tanidis}, K., Cardone, V.~F., Martinelli, M., {et~al.}
  2024, \aap, 683, A17

\bibitem[{{Feldman} {et~al.}(1994){Feldman}, {Kaiser}, \& {Peacock}}]{FKP}
{Feldman}, H.~A., {Kaiser}, N., \& {Peacock}, J.~A. 1994, \apj, 426, 23

\bibitem[{Foglieni {et~al.}(2023)Foglieni, Pantiri, {Di Dio}, \&
  Castorina}]{Foglieni_2023}
Foglieni, M., Pantiri, M., {Di Dio}, E., \& Castorina, E. 2023, \prl, 131,
  111201

\bibitem[{Gibelyou \& Huterer(2012)}]{Gibelyou_2012}
Gibelyou, C. \& Huterer, D. 2012, \mnras, 427, 1994

\bibitem[{{G{\'o}rski} {et~al.}(2005){G{\'o}rski}, {Hivon}, {Banday},
  {Wandelt}, {Hansen}, {Reinecke}, \& {Bartelmann}}]{HEALpix}
{G{\'o}rski}, K.~M., {Hivon}, E., {Banday}, A.~J., {et~al.} 2005, \apj, 622,
  759

\bibitem[{Grimm \& Yoo(2021)}]{Grimm_2021}
Grimm, N. \& Yoo, J. 2021, \prd, 104, 083548

\bibitem[{{Hahn} \& {Abel}(2011)}]{music}
{Hahn}, O. \& {Abel}, T. 2011, \mnras, 415, 2101

\bibitem[{{Hamilton}(1998)}]{Hamilton_review}
{Hamilton}, A.~J.~S. 1998, in Astrophysics and Space Science Library, Vol. 231,
  The Evolving Universe, ed. D.~{Hamilton}, 185

\bibitem[{Hamilton(2000)}]{Hamilton00}
Hamilton, A. J.~S. 2000, \mnras, 312, 257

\bibitem[{{Hamilton} \& {Culhane}(1996)}]{Hamilton-Culhane}
{Hamilton}, A.~J.~S. \& {Culhane}, M. 1996, \mnras, 278, 73

\bibitem[{{Hartlap} {et~al.}(2007){Hartlap}, {Simon}, \&
  {Schneider}}]{Hartlap+2007}
{Hartlap}, J., {Simon}, P., \& {Schneider}, P. 2007, \aap, 464, 399

\bibitem[{Heavens \& Taylor(1995)}]{Heavens_1995}
Heavens, A.~F. \& Taylor, A.~N. 1995, \mnras, 275, 483

\bibitem[{{Hockney} \& {Eastwood}(1988)}]{Hockney-Eastwood}
{Hockney}, R.~W. \& {Eastwood}, J.~W. 1988, {Computer simulation using
  particles} ({CRC Press})

\bibitem[{Hui {et~al.}(2007)Hui, Gazta\~naga, \& LoVerde}]{Hui_2007}
Hui, L., Gazta\~naga, E., \& LoVerde, M. 2007, \prd, 76, 103502

\bibitem[{Hui {et~al.}(2008)Hui, Gazta\~naga, \& LoVerde}]{Hui_2008}
Hui, L., Gazta\~naga, E., \& LoVerde, M. 2008, \prd, 77, 063526

\bibitem[{Hui \& Greene(2006)}]{Hui_2006_perturbation}
Hui, L. \& Greene, P.~B. 2006, \prd, 73, 123526

\bibitem[{Jelic-Cizmek {et~al.}(2021)Jelic-Cizmek, Lepori, Bonvin, \&
  Durrer}]{Jelic_Cizmek_2021}
Jelic-Cizmek, G., Lepori, F., Bonvin, C., \& Durrer, R. 2021, \jcap, 04, 055

\bibitem[{Jeong {et~al.}(2012)Jeong, Schmidt, \& Hirata}]{Jeong:2011as}
Jeong, D., Schmidt, F., \& Hirata, C.~M. 2012, \prd, 85, 023504

\bibitem[{{Kaiser}(1987)}]{Kaiser87}
{Kaiser}, N. 1987, \mnras, 227, 1

\bibitem[{{Kaufman}(1967)}]{Kaufman67}
{Kaufman}, G.~M. 1967, Report N. 6710, Center for Operations Research and
  Econometrics. Catholic University of Louvain. Heverlee, Belgium.

\bibitem[{{Landy} \& {Szalay}(1993)}]{Landy-Szalay98}
{Landy}, S.~D. \& {Szalay}, A.~S. 1993, \apj, 412, 64

\bibitem[{{Laureijs} {et~al.}(2011){Laureijs}, {Amiaux}, {Arduini},
  {Augu{\`e}res}, {Brinchmann}, {Cole}, {Cropper}, {Dabin}, {Duvet}, {Ealet},
  {et~al.}}]{Redbook}
{Laureijs}, R., {Amiaux}, J., {Arduini}, S., {et~al.} 2011, arXiv:1110.3193

\bibitem[{{Lewis} \& {Bridle}(2002)}]{CAMBS}
{Lewis}, A. \& {Bridle}, S. 2002, \prd, 66, 103511

\bibitem[{Loureiro {et~al.}(2019)Loureiro, Moraes, Abdalla, Cuceu, McLeod,
  Whiteway, Balan, Benoit-Lévy, Lahav, Manera, Rollins, \&
  Xavier}]{Loureiro+2019}
Loureiro, A., Moraes, B., Abdalla, F.~B., {et~al.} 2019, \mnras, 485, 326

\bibitem[{Maciaszek {et~al.}(2022)Maciaszek, Ealet, Gillard, Jahnke, Barbier,
  Prieto, Bon, Bonnefoi, Caillat, Carle, Costille, Ducret, Fabfro, Foulon, luc
  Gimenez, Grassi, Jaquet, Lemignant, Martin, Tony, Sanchez, Jean-Claude,
  Caillat, Mathieu, Secroun, Kubik, Ferriol, Berthe, Barriere, Fontignie, luca
  Valenziano, Auricchio, Battaglia, Derosa, Farinelli, Franceschi, Medinaceli,
  Morgante, Sortino, Trifoglio, Corcione, Capobianco, Ligori, Dusini, Borsato,
  Dalcorso, Laudisio, Sirignano, Stanco, Ventura, patrizii, Chiarusi, Fornari,
  Giacomini, Margiotta, Mauri, Pasqualini, Sirri, Spurio, Tenti, Travaglini,
  Bonoli, Bortoletto, Balestra, Dalessandro, Grupp, Penka, Steinwagner,
  Hormuth, schirmer, Seidel, Padilla, Casas, Lloro, Toledo, Gomez, Colodro,
  Lizan, Diaz, Lilje, Andersen, Andersen, Sorensen, Hornstrup, Jessen, Thizy,
  Holmes, Pniel, Jhabvala, Pravdo, Seiffert, Waczynski, Laureij, Racca,
  Salvignol, Boenke, Strada, \& Mellier}]{Maciaszek_2022}
Maciaszek, T., Ealet, A., Gillard, W., {et~al.} 2022, in Space Telescopes and
  Instrumentation 2022: Optical, Infrared, and Millimeter Wave, ed. L.~E.
  Coyle, M.~D. Perrin, \& S.~Matsuura, Vol. 12180 ({SPIE}), 613

\bibitem[{Matsubara(2000{\natexlab{a}})}]{Matsubara_2000_2PCF}
Matsubara, T. 2000{\natexlab{a}}, \apj, 535, 1

\bibitem[{Matsubara(2000{\natexlab{b}})}]{Matsubara_2000}
Matsubara, T. 2000{\natexlab{b}}, \apj, 537, L77

\bibitem[{{Noorikuhani} \& {Scoccimarro}(2023)}]{Noorikuhani_Scoccimarro2023}
{Noorikuhani}, M. \& {Scoccimarro}, R. 2023, \prd, 107, 083528

\bibitem[{P{\'{a}}pai \& Szapudi(2008)}]{P_pai_2008}
P{\'{a}}pai, P. \& Szapudi, I. 2008, \mnras, 389, 292

\bibitem[{Paviot {et~al.}(2022)Paviot, de~la Torre, de~Mattia, Zhao, Bautista,
  Burtin, Dawson, Escoffier, Jullo, Raichoor, Ross, \& Rossi}]{Paviot_2022}
Paviot, R., de~la Torre, S., de~Mattia, A., {et~al.} 2022, \mnras, 512, 1341

\bibitem[{{Peacock}(1991)}]{Peacock_1991}
{Peacock}, J.~A. 1991, \mnras, 253, 1

\bibitem[{Peacock {et~al.}(2001)Peacock, Cole, Norberg, Baugh, Bland-Hawthorn,
  Bridges, Cannon, Colless, Collins, Couch, Dalton, Deeley, De~Propris, Driver,
  Efstathiou, Ellis, Frenk, Glazebrook, Jackson, Lahav, Lewis, Lumsden, Maddox,
  Percival, Peterson, Price, Sutherland, \& Taylor}]{Peacock+2001}
Peacock, J.~A., Cole, S., Norberg, P., {et~al.} 2001, Nature, 410, 169

\bibitem[{{Peebles}(1973)}]{Peebles-1973}
{Peebles}, P.~J.~E. 1973, \apj, 185, 413

\bibitem[{Peebles(1980)}]{Peebles1980}
Peebles, P. J.~E. 1980, The Large-Scale Structure of the Universe (Princeton
  University Press)

\bibitem[{Percival {et~al.}(2004)Percival, Burkey, Heavens, Taylor, Cole,
  Peacock, Baugh, Bland-Hawthorn, Bridges, Cannon, Colless, Collins, Couch,
  Dalton, De~Propris, Driver, Efstathiou, Ellis, Frenk, Glazebrook, Jackson,
  Lahav, Lewis, Lumsden, Maddox, Norberg, Peterson, Sutherland, \&
  Taylor}]{Percival_2004}
Percival, W.~J., Burkey, D., Heavens, A., {et~al.} 2004, \mnras, 353, 1201

\bibitem[{{Planck Collaboration: Aghanim} {et~al.}(2020{\natexlab{a}}){Planck
  Collaboration: Aghanim}, {Akrami}, {Ashdown}, {Aumont}, {Baccigalupi},
  {Ballardini}, {Banday}, {Barreiro}, {Bartolo}, {Basak}, {Battye}, {Benabed},
  {Bernard}, {Bersanelli}, {Bielewicz}, {Bock}, {Bond}, {Borrill}, {Bouchet},
  {Boulanger}, {Bucher}, {Burigana}, {Butler}, {Calabrese}, {Cardoso},
  {Carron}, {Challinor}, {Chiang}, {Chluba}, {Colombo}, {Combet}, {Contreras},
  {Crill}, {Cuttaia}, {de Bernardis}, {de Zotti}, {Delabrouille}, {Delouis},
  {Di Valentino}, {Diego}, {Dor{\'e}}, {Douspis}, {Ducout}, {Dupac}, {Dusini},
  {Efstathiou}, {Elsner}, {En{\ss}lin}, {Eriksen}, {Fantaye}, {Farhang},
  {Fergusson}, {Fernandez-Cobos}, {Finelli}, {Forastieri}, {Frailis},
  {Fraisse}, {Franceschi}, {Frolov}, {Galeotta}, {Galli}, {Ganga},
  {G{\'e}nova-Santos}, {Gerbino}, {Ghosh}, {Gonz{\'a}lez-Nuevo}, {G{\'o}rski},
  {Gratton}, {Gruppuso}, {Gudmundsson}, {Hamann}, {Handley}, {Hansen},
  {Herranz}, {Hildebrandt}, {Hivon}, {Huang}, {Jaffe}, {Jones}, {Karakci},
  {Keih{\"a}nen}, {Keskitalo}, {Kiiveri}, {Kim}, {Kisner}, {Knox},
  {Krachmalnicoff}, {Kunz}, {Kurki-Suonio}, {Lagache}, {Lamarre}, {Lasenby},
  {Lattanzi}, {Lawrence}, {Le Jeune}, {Lemos}, {Lesgourgues}, {Levrier},
  {Lewis}, {Liguori}, {Lilje}, {Lilley}, {Lindholm}, {L{\'o}pez-Caniego},
  {Lubin}, {Ma}, {Mac{\'\i}as-P{\'e}rez}, {Maggio}, {Maino}, {Mandolesi},
  {Mangilli}, {Marcos-Caballero}, {Maris}, {Martin}, {Martinelli},
  {Mart{\'\i}nez-Gonz{\'a}lez}, {Matarrese}, {Mauri}, {McEwen}, {Meinhold},
  {Melchiorri}, {Mennella}, {Migliaccio}, {Millea}, {Mitra},
  {Miville-Desch{\^e}nes}, {Molinari}, {Montier}, {Morgante}, {Moss}, {Natoli},
  {N{\o}rgaard-Nielsen}, {Pagano}, {Paoletti}, {Partridge}, {Patanchon},
  {Peiris}, {Perrotta}, {Pettorino}, {Piacentini}, {Polastri}, {Polenta},
  {Puget}, {Rachen}, {Reinecke}, {Remazeilles}, {Renzi}, {Rocha}, {Rosset},
  {Roudier}, {Rubi{\~n}o-Mart{\'\i}n}, {Ruiz-Granados}, {Salvati}, {Sandri},
  {Savelainen}, {Scott}, {Shellard}, {Sirignano}, {Sirri}, {Spencer},
  {Sunyaev}, {Suur-Uski}, {Tauber}, {Tavagnacco}, {Tenti}, {Toffolatti},
  {Tomasi}, {Trombetti}, {Valenziano}, {Valiviita}, {Van Tent}, {Vibert},
  {Vielva}, {Villa}, {Vittorio}, {Wandelt}, {Wehus}, {White}, {White},
  {Zacchei}, \& {Zonca}}]{planck18}
{Planck Collaboration: Aghanim}, N., {Akrami}, Y., {Ashdown}, M., {et~al.}
  2020{\natexlab{a}}, \aap, 641, A6

\bibitem[{{Planck Collaboration: Aghanim} {et~al.}(2020{\natexlab{b}}){Planck
  Collaboration: Aghanim}, {Akrami, Y.}, {Arroja, F.}, {Ashdown, M.}, {Aumont,
  J.}, {Baccigalupi, C.}, {Ballardini, M.}, {Banday, A. J.}, {Barreiro, R. B.},
  {Bartolo, N.}, {Basak, S.}, {Battye, R.}, {Benabed, K.}, {Bernard, J.-P.},
  {Bersanelli, M.}, {Bielewicz, P.}, {Bock, J. J.}, {Bond, J. R.}, {Borrill,
  J.}, {Bouchet, F. R.}, {Boulanger, F.}, {Bucher, M.}, {Burigana, C.},
  {Butler, R. C.}, {Calabrese, E.}, {Cardoso, J.-F.}, {Carron, J.}, {Casaponsa,
  B.}, {Challinor, A.}, {Chiang, H. C.}, {Colombo, L. P. L.}, {Combet, C.},
  {Contreras, D.}, {Crill, B. P.}, {Cuttaia, F.}, {de Bernardis, P.}, {de
  Zotti, G.}, {Delabrouille, J.}, {Delouis, J.-M.}, {D\'esert, F.-X.}, {Di
  Valentino, E.}, {Dickinson, C.}, {Diego, J. M.}, {Donzelli, S.}, {Dor\'e,
  O.}, {Douspis, M.}, {Ducout, A.}, {Dupac, X.}, {Efstathiou, G.}, {Elsner,
  F.}, {En\ss{}lin, T. A.}, {Eriksen, H. K.}, {Falgarone, E.}, {Fantaye, Y.},
  {Fergusson, J.}, {Fernandez-Cobos, R.}, {Finelli, F.}, {Forastieri, F.},
  {Frailis, M.}, {Franceschi, E.}, {Frolov, A.}, {Galeotta, S.}, {Galli, S.},
  {Ganga, K.}, {G\'enova-Santos, R. T.}, {Gerbino, M.}, {Ghosh, T.},
  {Gonz\'alez-Nuevo, J.}, {G\'orski, K. M.}, {Gratton, S.}, {Gruppuso, A.},
  {Gudmundsson, J. E.}, {Hamann, J.}, {Handley, W.}, {Hansen, F. K.}, {Helou,
  G.}, {Herranz, D.}, {Hildebrandt, S. R.}, {Hivon, E.}, {Huang, Z.}, {Jaffe,
  A. H.}, {Jones, W. C.}, {Karakci, A.}, {Keih\"anen, E.}, {Keskitalo, R.},
  {Kiiveri, K.}, {Kim, J.}, {Kisner, T. S.}, {Knox, L.}, {Krachmalnicoff, N.},
  {Kunz, M.}, {Kurki-Suonio, H.}, {Lagache, G.}, {Lamarre, J.-M.}, {Langer,
  M.}, {Lasenby, A.}, {Lattanzi, M.}, {Lawrence, C. R.}, {Le Jeune, M.},
  {Leahy, J. P.}, {Lesgourgues, J.}, {Levrier, F.}, {Lewis, A.}, {Liguori, M.},
  {Lilje, P. B.}, {Lilley, M.}, {Lindholm, V.}, {L\'opez-Caniego, M.}, {Lubin,
  P. M.}, {Ma, Y.-Z.}, {Mac\'{\i}as-P\'erez, J. F.}, {Maggio, G.}, {Maino, D.},
  {Mandolesi, N.}, {Mangilli, A.}, {Marcos-Caballero, A.}, {Maris, M.},
  {Martin, P. G.}, {Martinelli, M.}, {Mart\'{\i}nez-Gonz\'alez, E.},
  {Matarrese, S.}, {Mauri, N.}, {McEwen, J. D.}, {Meerburg, P. D.}, {Meinhold,
  P. R.}, {Melchiorri, A.}, {Mennella, A.}, {Migliaccio, M.}, {Millea, M.},
  {Mitra, S.}, {Miville-Desch\^enes, M.-A.}, {Molinari, D.}, {Moneti, A.},
  {Montier, L.}, {Morgante, G.}, {Moss, A.}, {Mottet, S.}, {M\"unchmeyer, M.},
  {Natoli, P.}, {N\o{}rgaard-Nielsen, H. U.}, {Oxborrow, C. A.}, {Pagano, L.},
  {Paoletti, D.}, {Partridge, B.}, {Patanchon, G.}, {Pearson, T. J.}, {Peel,
  M.}, {Peiris, H. V.}, {Perrotta, F.}, {Pettorino, V.}, {Piacentini, F.},
  {Polastri, L.}, {Polenta, G.}, {Puget, J.-L.}, {Rachen, J. P.}, {Reinecke,
  M.}, {Remazeilles, M.}, {Renault, C.}, {Renzi, A.}, {Rocha, G.}, {Rosset,
  C.}, {Roudier, G.}, {Rubi\~no-Mart\'{\i}n, J. A.}, {Ruiz-Granados, B.},
  {Salvati, L.}, {Sandri, M.}, {Savelainen, M.}, {Scott, D.}, {Shellard, E. P.
  S.}, {Shiraishi, M.}, {Sirignano, C.}, {Sirri, G.}, {Spencer, L. D.},
  {Sunyaev, R.}, {Suur-Uski, A.-S.}, {Tauber, J. A.}, {Tavagnacco, D.}, {Tenti,
  M.}, {Terenzi, L.}, {Toffolatti, L.}, {Tomasi, M.}, {Trombetti, T.},
  {Valiviita, J.}, {Van Tent, B.}, {Vibert, L.}, {Vielva, P.}, {Villa, F.},
  {Vittorio, N.}, {Wandelt, B. D.}, {Wehus, I. K.}, {White, M.}, {White, S. D.
  M.}, {Zacchei, A.}, \& {Zonca, A.}}]{planck-dipole-18}
{Planck Collaboration: Aghanim}, N., {Akrami, Y.}, {Arroja, F.}, {et~al.}
  2020{\natexlab{b}}, A\&A, 641, A1

\bibitem[{{Pozzetti} {et~al.}(2016){Pozzetti}, {Hirata}, {Geach}, {Cimatti},
  {Baugh}, {Cucciati}, {Merson}, {Norberg}, \& {Shi}}]{pozzetti16}
{Pozzetti}, L., {Hirata}, C.~M., {Geach}, J.~E., {et~al.} 2016, \aap, 590, A3

\bibitem[{Pryer {et~al.}(2022)Pryer, Smith, Booth, Blake, Eggemeier, \&
  Loveday}]{Pryer_2022}
Pryer, D., Smith, R.~E., Booth, R., {et~al.} 2022, \jcap, 08, 019

\bibitem[{{Raccanelli} {et~al.}(2014){Raccanelli}, {Bertacca}, {Dor{\'e}}, \&
  {Maartens}}]{Raccanelli_2012}
{Raccanelli}, A., {Bertacca}, D., {Dor{\'e}}, O., \& {Maartens}, R. 2014,
  \jcap, 08, 022

\bibitem[{{Raccanelli} {et~al.}(2018){Raccanelli}, {Bertacca}, {Jeong},
  {Neyrinck}, \& {Szalay}}]{raccanelli2016doppler}
{Raccanelli}, A., {Bertacca}, D., {Jeong}, D., {Neyrinck}, M.~C., \& {Szalay},
  A.~S. 2018, Physics of the Dark Universe, 19, 109

\bibitem[{Raccanelli {et~al.}(2016)Raccanelli, Montanari, Bertacca, Dor{\'{e}
  }, \& Durrer}]{Raccanelli+2016_GR_CORRECTIONS}
Raccanelli, A., Montanari, F., Bertacca, D., Dor{\'{e} }, O., \& Durrer, R.
  2016, \jcap, 05, 009

\bibitem[{Raccanelli {et~al.}(2010)Raccanelli, Samushia, \&
  Percival}]{Raccanelli_2010}
Raccanelli, A., Samushia, L., \& Percival, W.~J. 2010, \mnras, 409, 1525

\bibitem[{Reimberg {et~al.}(2016)Reimberg, Bernardeau, \&
  Pitrou}]{Reimberg+2016}
Reimberg, P., Bernardeau, F., \& Pitrou, C. 2016, \jcap, 01, 048

\bibitem[{Samushia {et~al.}(2015)Samushia, Branchini, \&
  Percival}]{Samushia_2015}
Samushia, L., Branchini, E., \& Percival, W.~J. 2015, \mnras, 452, 3704

\bibitem[{Samushia {et~al.}(2012)Samushia, Percival, \&
  Raccanelli}]{Samushia_2012}
Samushia, L., Percival, W.~J., \& Raccanelli, A. 2012, \mnras, 420, 2102

\bibitem[{{Sargent} \& {Turner}(1977)}]{Sargent_Turner_1977}
{Sargent}, W.~L.~W. \& {Turner}, E.~L. 1977, \apjl, 212, L3

\bibitem[{Scoccimarro(2015)}]{scoccimarro_fast_2015}
Scoccimarro, R. 2015, \prd, 92, 083532

\bibitem[{Szalay {et~al.}(1998)Szalay, Matsubara, \& Landy}]{Szalay+1998}
Szalay, A.~S., Matsubara, T., \& Landy, S.~D. 1998, \apj, 498, L1

\bibitem[{Szapudi(2004)}]{Szapudi_2004}
Szapudi, I. 2004, \apj, 614, 51

\bibitem[{Tansella {et~al.}(2018)Tansella, Jelic-Cizmek, Bonvin, \&
  Durrer}]{Tansella_2018}
Tansella, V., Jelic-Cizmek, G., Bonvin, C., \& Durrer, R. 2018, \jcap, 10, 032

\bibitem[{Taruya {et~al.}(2018)Taruya, Nishimichi, \& Jeong}]{Taruya+2018}
Taruya, A., Nishimichi, T., \& Jeong, D. 2018, \prd, 98, 103532

\bibitem[{Wilson {et~al.}(2017)Wilson, Peacock, Taylor, \& de~la
  Torre}]{Wilson+17}
Wilson, M.~J., Peacock, J.~A., Taylor, A.~N., \& de~la Torre, S. 2017, \mnras,
  464, 3121

\bibitem[{Yamamoto {et~al.}(2006)Yamamoto, Nakamichi, Kamino, Bassett, \&
  Nishioka}]{Yamamoto_2006}
Yamamoto, K., Nakamichi, M., Kamino, A., Bassett, B.~A., \& Nishioka, H. 2006,
  Publications of the Astronomical Society of Japan, 58, 93

\bibitem[{Yamamoto {et~al.}(1999)Yamamoto, Nishioka, \& Suto}]{Yamamoto_1999}
Yamamoto, K., Nishioka, H., \& Suto, Y. 1999, \apj, 527, 488

\bibitem[{Yamamoto {et~al.}(2000)Yamamoto, Nishioka, \&
  Taruya}]{yamamoto2000effect}
Yamamoto, K., Nishioka, H., \& Taruya, A. 2000, arXiv:0012433

\bibitem[{Yoo \& Desjacques(2013)}]{Yoo_2013}
Yoo, J. \& Desjacques, V. 2013, \prd, 88, 023502

\bibitem[{Yoo {et~al.}(2009)Yoo, Fitzpatrick, \& Zaldarriaga}]{Yoo:2009au}
Yoo, J., Fitzpatrick, A., \& Zaldarriaga, M. 2009, \prd, 80, 083514

\bibitem[{{Zaroubi} \& {Hoffman}(1996)}]{Zaroubi-Hoffman96}
{Zaroubi}, S. \& {Hoffman}, Y. 1996, \apj, 462, 25

\bibitem[{Zonca {et~al.}(2019)Zonca, Singer, Lenz, Reinecke, Rosset, Hivon, \&
  Gorski}]{Zonca2019}
Zonca, A., Singer, L., Lenz, D., {et~al.} 2019, Journal of Open Source
  Software, 4, 1298

\end{thebibliography}

\appendix
\section{Validation of the \liger method}
\label{Sec:Validation}

The calculation of the galaxy angular power spectrum with 
relativistic RSD (but excluding corrections due to $\bs{\varv}_\mathrm{o}$ and considering constant values of $b$ and $\mc{Q}$) has been integrated into the \ttt{CAMB} and \ttt{CLASS} codes \citep{Dio_2013}. 
Here, we validate the \liger method by comparing the mean of the angular power spectra estimated from our \real and \grsd mock catalogues to the output of \ttt{CAMB}.  
For simplicity, we perform the comparison for a full-sky survey and
we provide the density-weighted average of the $b$ and $\mathcal{Q}$ functions as input to \ttt{CAMB}. 
Figure~\ref{fig:CAMB_LIGER_COMP} shows that the two methods agree
to better than 1\% for $\ell \lesssim 90$ for the tomographic bin $z\in (1.1,1.3)$. The loss of power of the mock catalogues at higher multipoles is irrelevant for our study (which focuses on larger scales) and can be attributed to the coarse spatial resolution of the input 2LPT simulations (see Sect.~\ref{sec:sims}). Similar results are obtained for the other redshift bins. 

\begin{figure}
    \centering
    \includegraphics[width=1\linewidth]{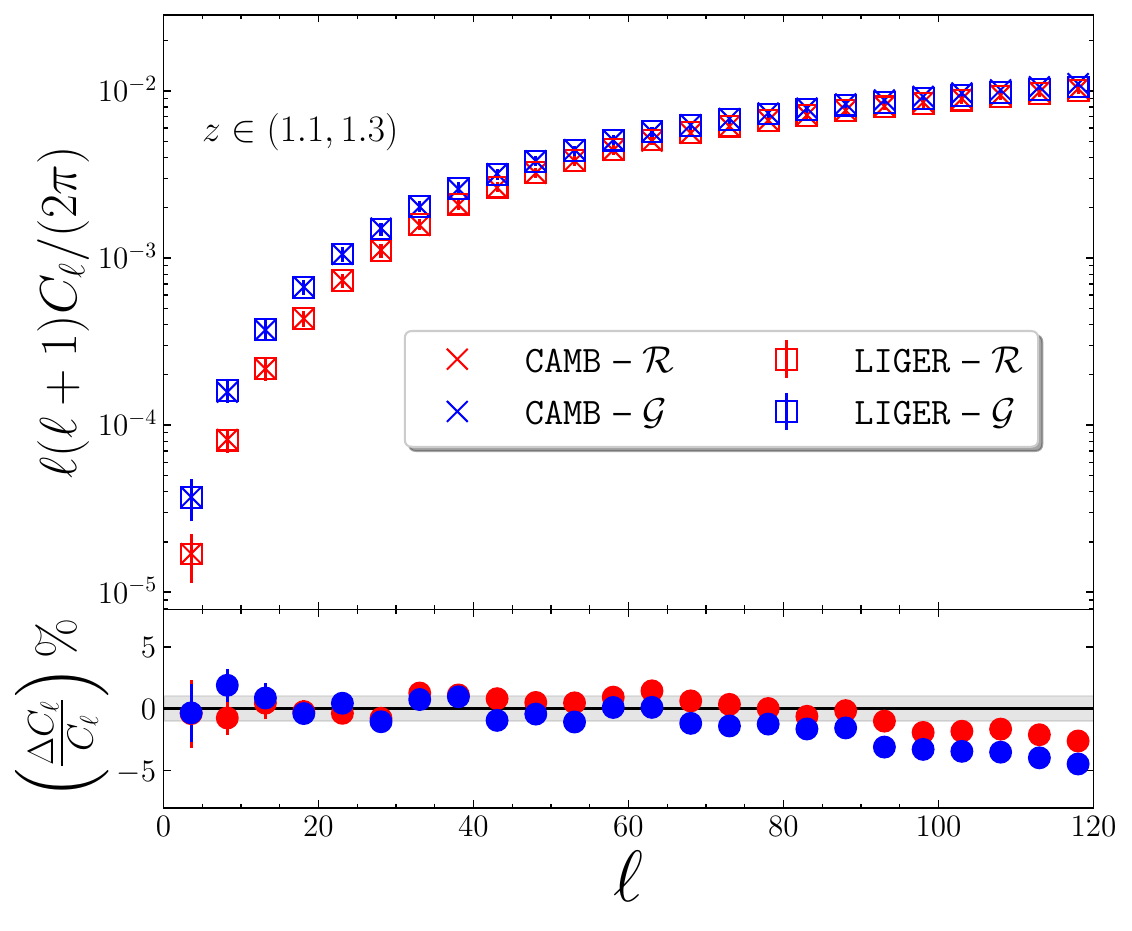}
    \caption{Top: The mean angular power spectra estimated from our mock catalogues (squares with RMS errorbars) for the \real (red) and \grsd (blue) suites are compared to the corresponding output from the \texttt{CAMB} code (crosses). All spectra are binned using intervals $\Delta \ell = 5$.     Bottom: Relative difference between the spectra (errorbars here indicate the standard error of the mean).
    }
    \label{fig:CAMB_LIGER_COMP}
\end{figure}

\end{document}